%% file: VE_NTEarXiV_3p0.tex
\documentclass{article}

\usepackage{arxiv}

\RequirePackage{amsthm,amsmath,amsfonts,amssymb, amsthm}
\usepackage[utf8]{inputenc} 
\usepackage[T1]{fontenc}    
\usepackage{hyperref}       
\usepackage{url}            
\usepackage{booktabs}       
\usepackage{amsfonts}       
\usepackage{nicefrac}       
\usepackage{microtype}      
\usepackage{ltablex}
\usepackage{subcaption}
\usepackage{newfloat}
\usepackage[tablesfirst,nolists,nomarkers]{endfloat}
\DeclareDelayedFloatFlavour{tabularx}{table}
\usepackage{graphicx}
\usepackage{natbib}
\usepackage{doi}
\usepackage[title]{appendix}
\usepackage{caption}
\usepackage{enumerate, enumitem}
\usepackage{arydshln}

\usepackage{titlesec}
\usepackage{algorithm}
\usepackage{algpseudocode}

\DeclareFloatingEnvironment[fileext=lof,name=Figure,placement=H]{intextfigure}

\titlespacing*{\section}
{0pt}{2.3ex plus 1ex minus .2ex}{2.3ex plus .2ex}
\titlespacing*{\subsection}
{0pt}{2.3ex plus 1ex minus .2ex}{2.3ex plus .2ex}

\makeatletter
\AtBeginEnvironment{tabularx}{%
	\@currsize}%
\makeatother

\newtheorem{corollary}{Corollary}

\newtheorem{lemma}{Lemma}

\DeclareMathOperator*{\argmin}{\arg\!\min}

\allowdisplaybreaks[1]

\title{Assessing vaccine effectiveness in observational studies via nested trial emulation}

\author{Justin B.~DeMonte\textsuperscript{1}, 
	Bonnie E.~Shook-Sa\textsuperscript{1,2}, 
	Michael G.~Hudgens\textsuperscript{1}}
\date{%
	\textsuperscript{1}Department of Biostatistics, Gillings School of Global Public Health, University of North Carolina at Chapel Hill, Chapel Hill, NC, USA\\%
	\textsuperscript{2}Nuffield Department of Population Health, University of Oxford, Oxford, UK\\%
	~\\
	\today
}

\renewcommand{\headeright}{}
\renewcommand{\undertitle}{}
\renewcommand{\shorttitle}{Assessing vaccine effectiveness via nested trial emulation}

\hypersetup{
	pdftitle={Assessing vaccine effectiveness in observational studies via nested trial emulation},
	pdfsubject={stat.ME, stat.AP},
	pdfauthor={Justin B.~DeMonte, Bonnie E.~Shook-Sa, Michael G.~Hudgens},
	pdfkeywords={Causal inference, Nested trial emulation, Observational studies, Vaccine effectiveness},
}
\newcounter{leftOffHere}
\begin{document}
	\maketitle
	\begin{abstract}Observational data are frequently used to evaluate real-world vaccine effectiveness (VE). 
		For vaccines such as those developed against COVID-19, VE may vary over calendar time because of changes in circulating viral variants and over time since vaccination because of waning immunity. 
		Nested trial emulation (NTE) provides a framework for estimating causal effects from observational data while reducing biases common to standard observational analyses. 
		NTE involves emulating a sequence of hypothetical trials, each initiated from a different calendar date.
		This manuscript considers a NTE inverse probability weighted estimator of vaccine effectiveness that may vary over calendar time, time since vaccination, or both. 
		The proposed approach characterizes trial-specific VE and uses those estimates to assess changes in vaccine protection across calendar time. 
		As changes in VE estimates across trials may be attributable to variation in covariate distributions across trial-eligible populations, standardization of trial-specific estimates is considered. 
		Statistical testing for evaluating heterogeneity in VE across trials is also considered. 
		The methods are used to estimate vaccine effectiveness against COVID-19 outcomes using observational data on over $110{,}000$ residents of Abruzzo, Italy during 2021.
	\end{abstract}
	
	\section{Introduction}
	\label{intro}
	As of August 2026 over 7 million deaths from coronavirus disease 2019 (COVID-19) have been reported worldwide \citep{WHO}.  
	COVID-19 vaccines were developed with unprecedented speed and proved critical to slowing the pandemic \citep{chenSlowing} and reducing mortality \citep{mesle}. 
	Initial randomized controlled trials found high efficacy of Pfizer-BioNTech and Moderna vaccines against moderate-to-severe COVID-19 \citep{BNT162b2, mRNA-1273}.   
	In many parts of the world, COVID-19 vaccines were initially deployed in late 2020 and are now widely available.  
	However, initial COVID-19 vaccine trials had relatively short observation periods (median follow-up around 2 months after completion of the vaccine regimen) and were completed before the highly transmissible Delta variant became dominant \citep{ChanVariants}. 
	
	Therefore, it is important to understand how COVID-19 vaccine effectiveness (VE) changes over extended follow-up and across changing epidemiologic conditions.  
	In particular, determining whether VE varies over time since vaccination and calendar time is essential for informing vaccine policy, vaccine development, and public health decision-making.  
	Changes along these two time scales may inform different decisions regarding booster recommendations, vaccine formulation, and deployment strategies.
	Several factors may contribute to changes in COVID-19 VE over calendar time, such as seasonality or newly emerging variants of the SARS-CoV-2 virus. 
	Likewise, COVID-19 VE may change over time since vaccination, e.g., due to waning vaccine-induced antibody levels.  
	
	Ideally, whether VE varies over time since vaccination and calendar time could be evaluated by conducting a series of randomized trials, where each trial is initiated at a different calendar date, and estimating the vaccine effect separately for each trial. 
	However, since such an approach is generally not feasible, VE studies must rely on surveillance databases or other observational data.
	One approach to analyzing such observational data could entail defining VE as one minus the hazard ratio from a covariate-adjusted Cox model \citep{jara2021effectiveness,tartof2021effectiveness, Lin, lin2022reliably,shrestha2024effectiveness,feldstein2024effectiveness}.  
	Under appropriate structural assumptions, the hazard ratio may be a well-defined causal object \citep{MartinussenHazards, WangInstrumentalVarHazRatio, ying2023defense}, and under certain assumptions the proportional hazards coefficient has a causal interpretation.
	However, the causal hazard ratio may be difficult to interpret, especially for non-statisticians such as policy makers.  
	Additionally, the adjusted hazard ratio does not in general equal the marginal hazard ratio due to noncollapsibility \citep{Daniel}.    
	In turn, estimates obtained using the Cox model approach may not accurately generalize beyond the study population at hand.
	
	Target trial emulation is one alternative approach for drawing inference from observational data.  
	Successful trial emulations minimize biases that can arise in observational analyses and produce causally interpretable results \citep{BigData, hernan2025target}.  
	The first step in a target trial emulation analysis entails developing a protocol for a hypothetical randomized trial designed to address a specific causal question.  
	Important components of the target trial protocol include eligibility criteria, the treatment regimens to be compared, and definition of ``time zero'' (the date when follow up begins).  
	Then, the observational database is prepared and analyzed to emulate the hypothetical (target) trial.   
	When some individuals in the observational database meet eligibility criteria at multiple calendar time points, nested trial emulation \citep[NTE;][]{Hernan2005firstNestedTrials} can be used to properly align time zero.  
	That is, a sequence of trials is emulated wherein each subsequent trial population is a subset of the
	previous one, containing only those individuals who continue to meet eligibility criteria on the calendar date when the subsequent trial begins.
	This manuscript describes a NTE designed to assess COVID-19 VE in Abruzzo, Italy during 2021 using a large surveillance database.
	
	Typical NTE analyses have several commonalities.  
	First, the treatment group for each trial usually consists of all new treatment users, i.e., those who initiated treatment at the start of the trial, but not before.  
	The control group contains all other eligible individuals.  
	Second, individuals may appear in the control group of multiple trials but in the treatment group of at most one trial.  
	Third, when a previously untreated individual in the control group initiates treatment, they are artificially censored \citep{RobinsFinkelstein}.  
	Likewise, individuals who initiate treatment are censored if and when they subsequently discontinue treatment.  
	This design allows for per-protocol comparisons of the longitudinal treatment strategies ``always treat'' and ``never treat'' \citep{Keogh}.  
	
	Several complications arise when applying existing NTE approaches to evaluate changes in COVID-19 VE over time. 
	First, initial COVID-19 vaccine rollout in Italy and many other countries involved multiple vaccine brands, each with their own dose schedule; such complex treatment regimens may not be easily recast in the form ``always treat.'' 
	One analytical approach involves restricting focus to one-dose treatment regimens of a single vaccine brand \citep{hulme2023challenges}.  
	The NTE approach developed in this manuscript is more general and allows for inference about the effects of recommended immunization schedules, 
	which may entail multiple vaccine types with different dosing schedules. 
	
	A second complication arises because COVID-19 VE may change over calendar time.  
	Often when NTE is used, treatment effect estimates are pooled across trials to increase statistical efficiency \citep[e.g.,][]{HernanExample, Danaei1}.  
	Such pooled estimates may lack a clear interpretation when the treatment effect is heterogeneous across trials.  
	One approach could involve targeting the treatment effect for an average trial, allowing that the treatment effect may vary across trials \citep{gervasoni2026estimands}. 
	However, in the COVID-19 setting, possible heterogeneity in VE across trials is of direct scientific interest.  
	This manuscript proposes a method for testing trial effect homogeneity (TEH) based on a comparison of area under trial-specific VE curves and illustrates how trial-specific VE estimates can be used to characterize changes in vaccine protection across calendar time.  
	
	A third complication arises because the eligible population for trials beginning at later calendar times may differ from earlier trials.  
	Such population differences may result in variation in VE across trials.  
	For example, unvaccinated individuals in early trials 
	at high risk of COVID-19 may be less likely to meet eligibility criteria for later trials due to experiencing the event.  
	If, in addition, the vaccine is protective for susceptible individuals, the resulting VE may be higher in earlier trials relative to later trials.  
	To characterize changes in VE across calendar time that cannot be attributed to variation in observed covariate distributions across the trial-eligible populations, the approach in this manuscript also considers standardizing VE across trials.  
	
	The remainder of the manuscript is organized as follows. 
	Section \ref{DataAndMotivation} introduces the conceptual framework and the motivating dataset. 
	Section \ref{methods} describes the methodological approach without standardization.  
	Section \ref{extensions} considers standardizing inferences from the sequence of emulated trials.  
	In Section \ref{DataAnalysis}, the methods are applied to estimate COVID-19 VE for a series of emulated trials using a large database of Abruzzo, Italy residents \citep{AcutiMartellucci}.  
	Section \ref{discussion} concludes with a discussion.  
	The Appendix contains additional details on the methods and data analysis, a series of simulation studies, and comparison with a Cox-model-based approach.  
	
	\section{Motivation and data description}
	\label{DataAndMotivation}
	After vaccine efficacy has been established in phase 3 randomized trials, observational data are often used to draw inference about VE in real-world settings. 
	Findings from such analyses can inform vaccine policy and public health decision making. 
	Two distinct scientific questions are often of primary interest:
	\begin{enumerate}[label=Q\arabic*]
		\item \label{rqTSV} Does VE change over time since vaccination?
		\item Does VE change over calendar time?
		\label{rqCalTime}
	\end{enumerate}
	These questions have different scientific and policy implications. 
	A decrease in VE with time since vaccination may motivate recommendations for booster doses or modifications to vaccine formulations to bolster sustained immune responses. 
	In contrast, changes in VE over calendar time may reflect changes in the epidemiologic context, such as the emergence of new viral variants, changes in population immunity, or shifts in exposure patterns. 
	For example, a decline in VE over calendar time that coincides with the emergence of novel viral strains may motivate updates to vaccine formulations, as is routinely done for influenza vaccines. 
	In this manuscript, we address questions \ref{rqTSV} and \ref{rqCalTime} in the context of the COVID-19 vaccine rollout in Italy during 2021.
	
	\subsection{Target trials and estimand}
	\label{s:estimand}
	Suppose the goal is to estimate VE against a COVID-19 outcome (e.g., SARS-CoV-2 infection, COVID-19-related hospitalization or death), allowing for the possibility that VE varies over calendar time, time since vaccination, or both. 
	Ideally, a sequence of randomized controlled trials could be conducted, each comparing an active vaccine regimen to a control regimen, and each initiated from a different calendar date. 
	Then estimates of VE from the sequence of trials could be used to answer \ref{rqTSV} and \ref{rqCalTime}. 
	For example, a series of trials initiated across various calendar times that yield similar estimates of VE would provide evidence that VE does not change over calendar time.
	
	For this ideal scenario, let $j=0, 1, ..., J$ denote trial number, where trial 0 begins February 15, 2021, and a new trial is initiated every seven days thereafter, for a total of $J+1$ trials.  
		The trial populations may overlap; persons who meet eligibility criteria on February 15 and continue to meet eligibility criteria through week $j$ would be enrolled in trials $0,1,..., j$.
	At the start of each trial, eligibility would be assessed, and eligible participants would be randomly assigned to receive an active vaccine regimen, denoted $A_j=1$, or a comparator regimen, $A_j=0$.  
	The analysis in Section~\ref{DataAnalysis} considers VE of a two-dose mRNA COVID-19 vaccine regimen.  
	Those assigned to $A_j=1$ would receive a first dose of their choice of available mRNA COVID-19 vaccine (Pfizer BioNTech or Moderna) within 7 days of enrollment and a second dose within a recommended time window (see Section~\ref{emulatedTrials} for details).  	
	Those assigned to $A_j=0$ would not receive any COVID-19 vaccine throughout the follow-up period.  
	Assume in this idealized setting that all participants fully adhere to their assigned regimen.  
	Individuals would be assessed for a COVID-19 event of interest at a series of evenly-spaced follow-up visits.  
	Assume weekly follow-up visits and let $k$ index time in weeks since the start of a given trial.  
	For further details on the target trial, see the target trial protocol (Table~\ref{protocol_p1} in the Appendix).
	
	The VE estimand corresponding to this sequence of trials can be defined using potential outcomes.  
	Let $Y^{a}_j(k)$ denote a binary potential outcome indicating severe COVID-19 or COVID-19-related death by week $k$ of trial $j$ under full adherence to treatment regimen $a$. 
	The target estimand is
	\begin{equation}
		VE_{j}(k)=1-\frac{P\{Y_j^{1}(k)=1 \mid E_j=1\}}{P\{Y_j^{0}(k)=1\mid E_j=1\}}
		\label{estimand}
	\end{equation}
	for $j=0, 1, ..., J$ and $k$ varying over a specified range of follow-up times, where $E_j$ is the indicator of eligibility for trial $j$.  
	For more discussion on the estimand, see Section~\ref{discussion}.  
	The ratio $P\{Y_j^{1}(k)=1\mid E_j=1\}/P\{Y_j^{0}(k)=1\mid E_j=1\}$ is a causal contrast comparing the counterfactual risk at time $k$ of trial $j$ under regimens $A_j=1$ and $A_j=0$.  
	Variation in $VE_j(k)$ over $j$ suggests the vaccine effect may be changing over calendar time, whereas
	variation in $VE_j(k)$ over $k$ conveys the vaccine effect changes with time since vaccination.
	
	If a series of randomized trials could be conducted, inference about (\ref{estimand}) would be straightforward.  
	However, in many settings conducting a sequence of trials is not feasible.  
	Instead, the methods described in the remainder of this section utilize observational data to draw inference about (\ref{estimand}).  
	The methods are motivated by the Abruzzo database, which is briefly introduced in Section~\ref{analyticCohort} and analyzed in Section~\ref{DataAnalysis}.
	
	\subsection{Data description and analytic cohort}
	\label{analyticCohort}  
	The Abruzzo COVID-19 VE study \citep{AcutiMartellucci} used individual data available from the Italian National Health Service on medical and demographic characteristics and COVID-19 vaccination status and outcomes.  
	The study included all persons residing or domiciled in the Abruzzo region of Italy on January 1, 2020 and without a positive SARS-CoV-2 swab prior to January 2, 2021 ($N=1{,}279{,}694$).  
	Baseline characteristics (age, sex, risk factors/comorbidities) were known for all individuals.  
	The database includes COVID-19 vaccination date and type (either Pfizer-BioNTech, Moderna, Oxford-AstraZeneca, or Janssen) for each dose received between January 2, 2021 and December 18, 2021 (up to three doses per individual).  
	Calendar date for each of the following was recorded between January 2, 2021 and February 18, 2022: first SARS-CoV-2 infection (positive reverse transcription polymerase chain reaction test from an accredited laboratory in Abruzzo), first severe COVID-19 disease (requiring hospitalization), and death (with or without positive SARS-CoV-2 swab).  
	
	To characterize how COVID-19 VE changes over calendar time and time since vaccination using the Abruzzo study data, consider a sequence of $J+1$ hypothetical target trials initiated weekly from February 15, 2021 to May 3, 2021, with each trial ending on December 18, 2021.  
	Let $l=0, 1, ..., \tau$ index calendar time, measured in weeks from February 15, where $l=J$ corresponds to the week of May 3 and $\tau+1=45$ is the administrative censoring time.  
	
	Define the analytic cohort as the set of individuals in the Abruzzo study database who meet eligibility criteria at calendar time zero, where eligibility is determined by the target trial protocol.  
	For the Abruzzo analysis presented in Section~\ref{DataAnalysis}, the target trial protocol is given in Appendix Table~\ref{protocol_p1}.    
	Variables within the analytic cohort dataset are constructed as follows.  
	
	Assume a ``study visit" (data collection time) occurs on the first day of each week during the follow-up period.  
	Variables are measured in weeks from February 15, 2021 and are determined by changes in a participant's status between study visits.  
	Let $T\in\{1,2, ...\}$ denote calendar time of the event of interest, e.g., severe disease or death due to SARS-CoV-2.  
	Similarly, let $U \in \{1, 2, ..., \tau+1\}$ denote calendar time of censoring (due to loss to follow up, as defined in the target trial protocol).  
	Let $T^*=\text{min}(T, U)$ and $\Delta=I(T < U)$, where $I(\cdot)$ denotes the indicator function.  
	If an individual remains free of the event and on-study through calendar time $\tau$, then $\Delta=0$ and ${T^*}=\tau+1$.  
	Let $\boldsymbol{X}$ denote a vector of baseline (i.e., measured at calendar time $0$) covariates.  
	
	Assume that $n_v$ vaccine brands are available (e.g., $n_v=4$ for the Abruzzo study data).  
	Let $B_l$ denote the brand of vaccine dose received at calendar time $l$, where $B_l=0$ represents no vaccine dose and values $B_l \in \{1, ..., n_v\}$ correspond to each of the available brands.  
	Throughout the manuscript, let overbars denote calendar-time histories, 
	e.g., ${\overline{B}_l}=({B_0}, {B_1}, ..., {B_l})$.  
	Vaccine dose histories ${\overline{B}_\tau}$ carry all the information needed to determine treatment ``assignments'' $A_j \in \{0,1\}$ for the emulated trials according to specifications in the target trial protocol.  
	The observed data for the $n$ individuals in the analytic cohort is denoted
	$\boldsymbol{O_i}=(T^*_i, \Delta_i, \boldsymbol{X_i}, \overline{B}_{\tau i})$ for $i=1, ..., n$.  
	$\boldsymbol{O_i},...,\boldsymbol{O_n}$ are assumed to be independent and identically distributed.  
	For notational simplicity, the subscript $i$ indexing individuals will often be omitted.
	
	\section{Methods}
	\label{methods}
	\subsection{Identifiability assumptions} 
	\label{assumptions} 
	The target parameter (\ref{estimand}) is shown in Section~\ref{identifiability} of the Appendix to be identifiable from the observable random variables under the following sufficient set of assumptions.  
	To avoid introducing extensive notation here, corresponding formal mathematical statements are given in Section~\ref{identifiability} of the Appendix.  
	\begin{enumerate}[label=\Roman*]
		\item Variables are measured without error.  
		\label{measurement_m}
		\item An individual's potential outcomes are unaffected by other individuals' treatment history (no interference).  
		\label{interference_m}
		\item Dropout is ignorable conditional on covariates. 
		\label{exchDropout_m}
		\item Treatment nonadherence is ignorable conditional on covariates.
		\label{exchNoncomp_m}
		\item Treatment assignment is exchangeable conditional on covariates.
		\label{exchTx_m}
		\item Causal consistency.
		\label{consistency_m}
		\item
		\label{positivityDropout_m}
		Positivity of remaining on-study.
		\item
		\label{positivityAdhere_m}
		Positivity of treatment adherence.
		\item
		\label{positivityTx_m}
		Positivity of treatment assignment.
		\setcounter{leftOffHere}{\value{enumi}}	
	\end{enumerate}
	The Abruzzo data were derived from comprehensive, official Italian Health Service and COVID-19 surveillance databases, suggesting that Assumption~\ref{measurement_m} is plausible.  
	The no-interference assumption (\ref{interference_m}) may be plausible in this setting because the analysis (described in detail in Section~\ref{DataAnalysis}) is restricted to individuals aged 80 years and older, who constitute less than 10\% of the population of the Abruzzo region.
	Assumptions~\ref{exchDropout_m}, \ref{exchNoncomp_m}, and \ref{exchTx_m} assert that measured covariates are sufficient to adjust for (i) selection bias due to differential loss to follow up and noncompliance and (ii) confounding.  
	The Abruzzo dataset contains a rich set of covariates including several important risk factors for COVID-19, suggesting that residual selection bias and confounding may be minimal.  
	Causal consistency (\ref{consistency_m}) is a standard assumption linking the observed and potential outcomes based on the observed treatment. 
	Positivity assumptions~(\ref{positivityDropout_m}-\ref{positivityTx_m}) are plausible in the Abruzzo trial emulation because all individuals in the target population (over 80 years of age) were eligible to receive COVID-19 vaccines starting from initial vaccine rollout, implying that individuals in the target population generally have nonzero probability of vaccine uptake and adherence. 
	
	\subsection{Marginal structural models} 
	\label{modelSpecification} 
	To facilitate interpretation and inference, additional assumptions might be considered about the possible dependence of VE on calendar time and time since vaccination.  
	Observe that (\ref{estimand}) can be re-expressed as 
	\begin{equation}
		\label{identity}
		{VE}_j(k)= 1- \dfrac{1-\prod_{m=1}^k\{1-{\lambda}_j^1(m)\}}{1-\prod_{m=1}^k\{1-{\lambda}_j^0(m)\}},
	\end{equation}
	where $\lambda_j^{a}(k)=P\{Y_j^{a}(k)=1 \mid Y_j^{a}(k-1)=0 ,E_j=1\}$ is the (discrete time) hazard of the potential outcome at time $k$ of trial $j$ under perfect adherence to vaccine regimen $A_j=a$.  
	In this section, different models for $\lambda_j^a(k)$ are considered, which we refer to as marginal structural models \citep[MSMs;][]{RobinsMSM2000}.  
	These MSMs specify how the potential outcome hazards depend on calendar time, time since vaccination, or both. 
	Recalling that time in weeks since the start of a given trial is indexed by $k$, note that time $k$ of trial $j$ corresponds to calendar time $j+k$.  
	
	Consider the MSM 
	\begin{equation}
		g\{\lambda_j^{a}(k)\} =\alpha_{0} +  \alpha_{1} a +\boldsymbol{\alpha_{2}} \boldsymbol{f_1}(k)a + \boldsymbol{\alpha_{3}} \boldsymbol{f_2}(j+k)+\boldsymbol{\alpha_4} \boldsymbol{f_3}(j+k)a
		\label{calTimeTerms}
	\end{equation} 
	for $(j,k) \in \mathcal{W}$, where $g$ is an appropriate link function for a binary outcome like the logit or probit function; $\mathcal{W}=\{(j,k):j\in \{0,1,...,J\}, k \in \{1,2,...,K_j\}\}$, $K_j=\tau-j$ is the final time point in trial $j$; and $\boldsymbol{f_1}(\cdot)$, $\boldsymbol{f_2}(\cdot)$, and $\boldsymbol{f_3}(\cdot)$ are column vectors of user-specified functions.  
	According to (\ref{calTimeTerms}), if $\boldsymbol{\alpha_2} \neq \boldsymbol{0}$ and at least one of $\boldsymbol{\alpha_3}$ or $\boldsymbol{\alpha_4}$ is non-zero, then the counterfactual hazard under $a=1$ will in general vary over both calendar time and time since vaccination.  
	Since the hazard for the outcome when unvaccinated should not depend on time since ``enrollment" in a hypothetical trial, the counterfactual hazard under $a=0$ is assumed under model (\ref{calTimeTerms}) to be a one-dimensional function of calendar time $j+k$.  
	Thus, the hazard under $a=0$ at a fixed calendar time, say $j+k=3$, is the same regardless of trial number, i.e., $\lambda_{0}^0(3)=\lambda_{1}^0(2)=\lambda_{2}^0(1)=g^{-1}\{\alpha_{0}+\boldsymbol{\alpha_{3}} \boldsymbol{f_2}(3)\}$.  
	The expression $\alpha_0+\boldsymbol{\alpha_{3}} \boldsymbol{f_2}(j+k)$ can be interpreted as a calendar-time-varying intercept representing the transformed hazard when unvaccinated at calendar time $j+k$; $\alpha_1+\boldsymbol{\alpha_{4}} \boldsymbol{f_3}(j+k)$ captures the change to the transformed hazard at calendar time $j+k$ if vaccinated; and $\boldsymbol{\alpha_{2}}\boldsymbol{f_1}(k)$ represents the change to the transformed hazard $k$ weeks after vaccination.  
	Calendar-time-varying background conditions may lead to increases or decreases in the hazard. 
	Therefore, it is important to model time flexibly, e.g., using restricted cubic splines.
	Substantive information about shifts in background epidemiologic conditions may be used to inform knot locations in spline functions of calendar time.
	
	As an alternative to (\ref{calTimeTerms}), one could consider more general MSMs.  
	For example, the outcome hazard could be modeled separately for each trial so that
	\begin{equation}
		\label{trialSpecificParams}
		g\{\lambda_j^{a}(k)\}= \alpha_{0} +  \alpha_{1j} a + \boldsymbol{\alpha_{2j}}\boldsymbol{f_1}(k)a +\boldsymbol{\alpha_{3}} \boldsymbol{f_2}(j+k)
	\end{equation}
	for $(j,k) \in \mathcal{W}$, where $\alpha_{1j} + \boldsymbol{\alpha_{2j}}\boldsymbol{f_1}(k)a$ represents the change to the hazard when vaccinated at time $k$ of trial $j$.  
	Here, as in model (\ref{calTimeTerms}), the hazard when unvaccinated is assumed to be a one-dimensional function of calendar time.  
	Note that the modeling approach in (\ref{calTimeTerms}) ``borrows'' information across trials to estimate the hazard trajectory under vaccination over both time scales.  
	If $\boldsymbol{f_1}(\cdot), \boldsymbol{f_2}(\cdot),$ and $\boldsymbol{f_3}(\cdot)$ are flexible (e.g., piecewise polynomial) functions, (\ref{calTimeTerms}) may reasonably approximate time-varying hazards in many real-world settings.
	However, (\ref{calTimeTerms}) assumes the increment to the transformed hazard when vaccinated can be decomposed into additive calendar time and time since vaccination effects.  
	This ``no interaction'' assumption may be violated if the dependence of VE on time since vaccination differs across calendar time, e.g., owing to the emergence of new viral variants.
	Adopting model (\ref{trialSpecificParams}) would circumvent 
	this 
	assumption, although the dimension of the nuisance parameters under (\ref{trialSpecificParams}) may become unwieldy when emulating more than a few trials.
	
	In some infectious disease settings (e.g., measles), MSMs that are more restrictive than (\ref{calTimeTerms}) may be appropriate. 
	For example, the antigenic profile of measles virus is stable across calendar time.  
	In turn, measles vaccines that were developed decades ago offer protection against measles viruses in circulation today \citep{Tahara, Zemella}.  
	In such a setting, one might posit the MSM
	\begin{equation}
		g\{\lambda_j^{a}(k)\} =\alpha_{0} + \alpha_{1} a + \boldsymbol{\alpha_2} \boldsymbol{f_1}(k)a + \boldsymbol{\alpha_{3}} \boldsymbol{f_2}(j+k)
		\label{timeVaryingIntercept}
	\end{equation} 
	for $(j,k) \in \mathcal{W}$, which is a special case of (\ref{calTimeTerms}) with $\boldsymbol{\alpha_4}=\boldsymbol{0}$.  
	Under the stronger assumption that hazards under both $a=0$ and $a=1$ do not depend on calendar time, but the hazard under $a=1$ may depend on time since vaccination, an even more restrictive version of (\ref{calTimeTerms}) with $\boldsymbol{\alpha_3}=\boldsymbol{\alpha_4}=\boldsymbol{0}$ could be considered.
	
	By contrast, the SARS-CoV-2 virus mutates rapidly, leading to emergence of new variants.  
	Current COVID-19 vaccines are designed to elicit an immune response to specific SARS-CoV-2 antigens and, in turn, may not protect against COVID-19 disease given exposure to a SARS-CoV-2 variant with a different antigenic structure.  
	Therefore, the application in Section \ref{DataAnalysis} utilizes models of the form (\ref{calTimeTerms}) that allow for changes in the hazard when vaccinated across calendar time and time since vaccination.
	
	\subsection{Inverse probability weighted estimator} 
	\label{IPweightedEstimator}
	This section describes an IP weighted estimator of (\ref{estimand}).  
	The estimator can be constructed in two steps.  
	The first step, described in Section~\ref{emulatedTrials}, entails converting the analytic cohort dataset described above to a NTE analysis dataset.  
	The second step, detailed in Section~\ref{inference}, involves constructing estimated IP weights and fitting a weighted regression model using the NTE analysis dataset.
	
	\subsubsection{Nested trial emulation analysis dataset}
	\label{emulatedTrials}
	Each individual in the analytic cohort contributes one set of repeated measurements to the NTE analysis dataset per trial for which they are eligible.  
	Recall that the analytic cohort consists of all individuals who meet eligibility criteria at calendar time zero.  
	Individuals remain eligible for each subsequent trial $j$ if they (i) remain free of the event of interest and (ii) are yet to receive the active treatment at calendar time $j$.  
	The follow-up period for emulated trial $j$ coincides with calendar times $j, j+1, ..., \tau$.  
	For example, an individual who meets eligibility criteria for trial 0 contributes a set of repeated measurements starting from calendar time $j=0$.  
	If the individual is also eligible for trial 1, then they contribute a set of repeated measurements starting from calendar time $j=1$, and so on.  
	
	Observed data are used to ``assign" eligible individuals to a vaccine regimen in each trial according to specifications in the target trial protocol.  
	Failure to align time zero of an emulated trial with assessment of eligibility criteria and treatment assignment can lead to bias \citep{hernan2016specifying}. 
	Individuals who are eligible for trial $j$ are ``assigned'' to the active regimen $A_j=1$ if they receive a first COVID-19 vaccine dose at calendar time $j + 1$ (i.e., if $\overline{B}_{j}=0$ and $B_{j+1}\neq 0$) and are assigned to the comparator regimen $A_j=0$ otherwise (i.e., if $\overline{B}_{j+1}= 0$).  
	If an individual is ineligible for trial $j$, then $A_j$ is undefined.  
	For example, consider an individual in the Abruzzo study with observed data $\boldsymbol{O}=(T^*=4, \Delta=1, \overline{B}_4=(0, 0, 0,1, 0),\boldsymbol{X}=\boldsymbol{x})$.  
	The individual would be enrolled in trials $0, 1,$ and $2$ with vaccine regimen assignments $A_0=0$, $A_1=0$, and $A_2=1$, respectively.
	After receiving a first COVID-19 vaccine dose, the individual would be ineligible for subsequent trials (i.e., trials $3, 4, ..., J$).  
	Figure \ref{Lexis} illustrates the hypothetical individual's contribution to trials 0, 1, and 2.  
	
	An individual who is eligible for trial $j$ contributes one record to the NTE analysis for each week they spend at risk of having an observed event in trial $j$.  
	An individual is no longer at risk in trial $j$ if they (i) have experienced the event, (ii) are lost to follow up (in the observational database), (iii) fail to adhere to their trial-$j$ treatment assignment, or (iv) reach calendar time $\tau+1$ without experiencing (i), (ii), or (iii).  
	Treatment adherence is determined according to specifications in the target trial protocol.  
	For the Abruzzo analysis, active vaccine regimens are considered that may involve multiple vaccine brands and allow grace periods for the timing of vaccine doses; the comparator strategy is ``never treat''.  
	Therefore, receipt of a first COVID-19 vaccine dose under regimen $A_j=0$ results in artificial censoring.  
	
	Continuing the example above, because the hypothetical individual received a first dose at calendar time $3$, their trial-$1$ record is artificially censored when they initiate the active regimen, as illustrated in Figure~\ref{Lexis}.   
	The individual also experienced an event just prior to the calendar week $4$ study visit.  
	Because their trial-$2$ record was at risk at this time (blue line in Figure \ref{Lexis}), they contribute an event to the analysis at time $k=2$ of trial $j=2$.  
	However, the individual was not at risk in trials $0$ and $1$ (red lines in Figure \ref{Lexis}) when the event occurred, so they do not contribute events to the analysis for trials $0$ and $1$.  
	Appendix Section~\ref{NTEdsConstruction} illustrates how the observed data vector $\boldsymbol{O}$ is converted into entries in the NTE analysis dataset.  
	
	\begin{intextfigure}
		\includegraphics[scale=.52]{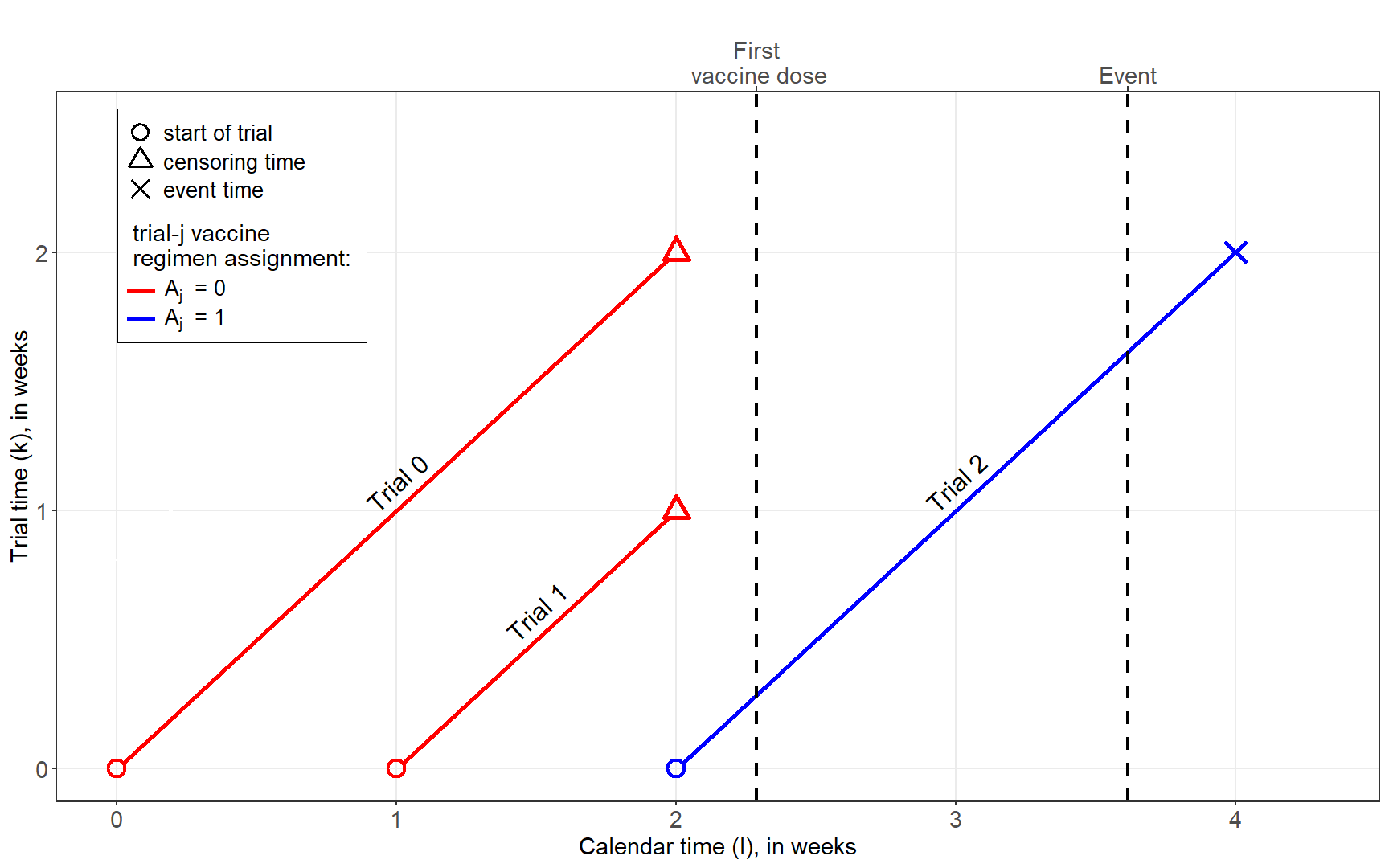}
		\caption{Lexis diagram illustrating emulated-trial-specific data for a hypothetical individual with observed data $\boldsymbol{O}=(T^*=4, \Delta=1, \overline{B}_4=(0, 0, 0, 1,0), \boldsymbol{X}=\boldsymbol{x})$.
			Circles represent entry into trial-specific cohorts, diagonal lines time spent at risk, triangles censoring, and the cross an event.  
			This individual is ``enrolled'' in trials 0, 1, and 2 in treatment groups $A_0=0$, $A_1=0$, and $A_2=1$, respectively.  
			They are ineligible for subsequent trials.  
			The trial-0 record is artificially censored at trial time $k=2$ because the individual ceases to follow their ``assigned'' trial-0 regimen at calendar time $l=2$.  
			Similarly, the trial-1 record is artificially censored at trial time $k=1$.  
			The trial-2 record is at risk and has an event at trial time $k=2$ because the individual experiences an event in the week prior to calendar time $l=4$.}
		\label{Lexis}
	\end{intextfigure}
	
	\subsubsection{Estimation procedures}
	\label{inference} 
	In general, to consistently estimate the effect of a sustained treatment or exposure from observational data with nonadherence and loss to follow up, inverse probability (IP) weighting can be used to simultaneously adjust for confounding and selection bias.  
	Below an IP weighted estimator is proposed where the weights are constructed based on estimated hazards of censoring due to nonadherence and loss to follow up.  
	The hazard of censoring due to nonadherence to vaccine regimen $A_j=a$ at time $k$ of trial $j$, denoted $\lambda_{j}^{C}(k,a,\boldsymbol{x}, \overline{z})$ is estimated by modeling vaccine uptake as a discrete time stochastic process, where $\overline{z}$ represents an observed vaccine dose history.  
	See Appendix Section~\ref{VUmethod} for details.  
	Modeling of the conditional hazard of censoring due to loss to follow up at time $k$ of trial $j$, denoted $\lambda_j^H(k,\boldsymbol{x}, \overline{z})$, is discussed in Appendix Section~\ref{IPweightEstimation}, and   
	formal definitions of $\lambda_{j}^{C}(k,a,\boldsymbol{x}, \overline{z})$ and $\lambda_j^H(k,\boldsymbol{x}, \overline{z})$ appear in Appendix Section~\ref{identifiability}.  
	
	For each record in the NTE analysis dataset with $A_j=a$, the estimated weight $\widehat{W}_j(k,a)$ is given by
	\[
	R_j(k)
	\bigg[
	\prod_{m=1}^{k} \{1-\lambda_j^H(m,\boldsymbol{X},\overline{Z}_{j+m}; \boldsymbol{\hat \xi})\} \{1-\lambda_j^C(m-1,a,\boldsymbol{X},\overline{Z}_{j+m-1}; \boldsymbol{\hat\kappa})\}
	\bigg]^{-1}
	\]
	where $R_j(k)=\{\Delta I(T^* > j+k-1) + (1-\Delta)I(T^* > j+k)\}\{1-C_j(k-1)\}E_j$ is the at-risk indicator for an observed event at time $k$ of trial $j$; 
	$C_j(k)$ is the indicator for nonadherence to assigned treatment regimen at time $k$ of trial $j$ for trial-$j$-eligible individuals (otherwise $C_j(k)$ is undefined);
	$\lambda_j^H(k,\boldsymbol{x}, \overline{z};\boldsymbol{\hat\xi})$ and $\lambda_{j}^{C}(k,a,\boldsymbol{x}, \overline{z}; \boldsymbol{\hat\kappa})$ are estimators of $\lambda_j^H(k,\boldsymbol{x}, \overline{z})$ and 
	$\lambda_{j}^{C}(k,a,\boldsymbol{x}, \overline{z})$; 
	and $\boldsymbol{\hat\xi}$ and $\boldsymbol{\hat\kappa}$ represent estimated model parameters.  
	Vaccine uptake and loss to follow up are modeled on the calendar time scale, and parameters $(\boldsymbol{\kappa},\boldsymbol{\xi})$ are estimated using the analytic cohort dataset (see Section~\ref{IPweightEstimation} for details).  
	Heuristically, weighting individuals in trial $j$ by $\widehat{W}_j(k, a)$ creates a trial-specific pseudo-population in which everyone is fully adherent to regimen $a$ \citep{RobinsFinkelstein}.  
	
	After constructing the weights, the model
	\begin{equation}  
		g[P\{Y_j(k)=1 \mid A_j=a, R_j(k)=1\}] =\boldsymbol{\alpha^\dagger} \boldsymbol{f}_{\boldsymbol{\alpha}}(j, k, a)
		\label{generalModel}
	\end{equation}
	for $(j,k) \in \mathcal{W}$ is fit to the NTE analysis data via weighted maximum likelihood with weights $\widehat{W}_{j}(k, A_j)$ where $Y_j(k)=I(T^* \leq j+k, \Delta=1)$ is the indicator of an observed event by time $k$ of trial $j$ for trial-$j$-eligible individuals (otherwise $Y_j(k)$ is undefined);
	$\boldsymbol{\alpha^\dagger}$ is a vector of unknown regression parameters; and 
	$\boldsymbol{f}_{\boldsymbol{\alpha}}(\cdot)$ is a column vector containing functions of trial number, time on trial, and the active vaccine regimen indicator.  
	The form of $\boldsymbol{f}_{\boldsymbol{\alpha}}$ should be specified according to the assumed MSM, e.g., (\ref{calTimeTerms}).  
	Let $\boldsymbol{\hat{\alpha}^\dagger}$ denote the weighted maximum likelihood estimator of the parameters in (\ref{generalModel}).  
	Let $\rho_j(k)=\log\{RR_j(k)\}$ where ${RR_j}(k)=1-{VE_j}(k)$ is the risk ratio at time $k$ of trial $j$, 
	and let $\boldsymbol{\rho_j}=\{\rho_j(1), \rho_j(2), \dots, \rho_j(K_j)\}$ and $\boldsymbol{\rho}=(\boldsymbol{\rho_0}, \boldsymbol{\rho_1}, ..., \boldsymbol{\rho_J})$.  
	A plug-in estimator for $\rho_j(k)$ is given by $\hat{\rho}_j(k)=\log[1-\prod_{m=1}^k\{1-\hat{\lambda}_j^1(m)\}]-\log[1-\prod_{m=1}^k\{1-\hat{\lambda}_j^0(m)\}]$, where $\hat{\lambda}_j^{a}(k)=g^{-1}\{\boldsymbol{\hat{\alpha}^\dagger} \boldsymbol{f}_{\boldsymbol{\alpha}}(j, k, a)\}$ for $a \in \{0,1\}$.  
	Let ${\boldsymbol{\hat{\theta}}}= (
	{\boldsymbol{\hat\kappa}}, 
	\boldsymbol{\hat\xi},
	{\boldsymbol{\hat{\alpha}^{\dagger}}}, 
	\boldsymbol{\hat{\rho}})^T$.  
	
	The estimator ${\boldsymbol{\hat{\theta}}}$ is the solution to an unbiased estimating equation vector, as shown in Section~\ref{EEs} of the Appendix.  
	It follows that, under certain regularity conditions \citep{StefanskiAndBoos}, ${\boldsymbol{\hat{\theta}}}$ is a consistent and asymptotically normal estimator of ${\boldsymbol{\theta}}=({\boldsymbol{\kappa}}, \boldsymbol{\xi}, {\boldsymbol{\alpha}}, \boldsymbol{{\rho}})^T$ if the vaccine uptake model, the loss-to-follow-up model, and the MSM for the outcome hazard are all correctly specified.  
	The empirical sandwich variance estimator, denoted $\hat{V}_n(\boldsymbol{\hat{\theta}})$, can be used to consistently estimate the asymptotic variance of $\boldsymbol{\hat{\theta}}$ and to construct pointwise Wald-type confidence intervals (CIs) for $\rho_j(k)$.  
	Upon transformation to the VE scale, a pointwise $(1-\upsilon)100\%$ CI for $VE_j(k)$ is given by $1-\exp\bigg[\log\{RR_j(k)\}\pm \Phi_{1-{\upsilon/2}} \sqrt{\hat{V}_n\{\hat{\rho}_j(k)\}/n}\bigg]$, 
	where $\Phi_{1-\upsilon/2}$ is the $(1-\upsilon/2)$th quantile of the standard normal distribution, and $\hat{V}_n\{\hat{\rho}_j(k)\}$ is the element in row $q$, column $q$ of $\hat{V}_n(\boldsymbol{\hat{\theta}})$ where $q$ denotes the index for entry $\hat{\rho}_j(k)$ in ${\boldsymbol{\hat{\theta}}}$.  
	Weighted maximum likelihood estimates can be calculated using standard software, and empirical sandwich variance estimates can be obtained from the R package \texttt{geex} \citep{Saul} or the Python library \texttt{delicatessen} \citep{Zivich}.  
	
	Instead of estimating the standard error via the empirical sandwich variance estimator, the bootstrap is often used for variance estimation in NTE analyses.  
	While the bootstrap also provides a consistent variance estimator, resampling large data sets can be computationally intensive.  
	Some NTE analyses \citep[e.g.,][]{McConeghy} discard data to achieve reasonable computation times under the bootstrap approach, leading to a loss in precision.  
	On the other hand, the M-estimation approach described above provides a computationally efficient method for obtaining valid confidence intervals in an NTE analysis.
	
	\subsection{Testing the TEH assumption}
	\label{TEH}
	Formal hypothesis testing can be used to detect heterogeneity in VE across trials.  
	Define the TEH assumption as 
	\begin{equation}
		\label{Hnaught}
		H_0:VE_0(k)=VE_1(k)=\dots =VE_J(k) \text{ for all }k\in\{1,2,...,K_J\}.   
	\end{equation}
	Departures from $H_0$, i.e., differences in VE across trials, are anticipated to be monotonic.  
	E.g., VE may decrease over calendar time as new SARS-CoV-2 variants emerge. 
	Therefore, the test statistic proposed below is intended to detect monotonic departures from $H_0$.  
	
	Let $AUC_j=\sum_{k=1}^{K_J}VE_j(k)$ denote area under the VE curve for the first $K_J$ weeks of trial $j$.  Recalling that all trials have at least $K_J$ weeks of follow-up, $AUC_j$ is a scalar summary measure that is comparable across trials.  
	Consider simple linear regression of ${AUC}_j$ on $j$, and let 
	\begin{equation}
		\label{leastSquares}
		({\beta}_0, {\beta})=\argmin_{(b_0,b)}\sum_{j=0}^{J}\{{AUC}_j-(b_0 + b j)\}^2.  
	\end{equation}
	Let $(\hat{\beta_0}, \hat{\beta})$ denote the estimator of $(\beta_0,\beta)$ obtained by solving (\ref{leastSquares}) with $AUC_j$ replaced by $\widehat{AUC}_j=\sum_{k=1}^{K_J}\widehat{VE}_j(k)$.  
	Large values of $\lvert\hat{\beta}\rvert$ provide evidence against $H_0$ for a two-sided test. 
	For testing $H_0$ against the one-sided alternative of decreasing VE across (temporally ordered) trials, large (in absolute value) negative values of $\hat \beta$ provide evidence against the null.  
	The generalized Wald test statistic $U_\beta=\hat{\beta}/\widehat{SE}(\hat{\beta})$
	may be used to test $H_0$, where $\widehat{SE}(\hat \beta)$ is computed based on the empirical sandwich variance estimator. 
	Under $H_0$, $U_\beta$ follows an approximately standard normal distribution in large samples (see Appendix Section~\ref{TEHees} for additional details).  	
	
	\section{Standardization across emulated trials}
	\label{extensions}Changes in VE across emulated trials may be attributable to variation in covariate distributions among the sequence of trial-eligible populations.  
	In NTE analyses, individuals appear in multiple trials, suggesting that covariate distributions may be similar across the emulated trials.  
	However, because prior events and prior vaccine uptake result in exclusion from later trials and, by definition, confounders predict both events and vaccine uptake, differences between covariate distributions in trial zero and trial $j>0$ may become more pronounced with increasing $j$.  
	This section describes the use of an empirical standardization procedure \citep{Keogh}, which can provide insight regarding changes in VE over calendar time that cannot be attributed to temporal variations in covariate distributions.  
	
	Consider standardizing each trial-specific VE such that the distribution of baseline covariates is the same across trials.  
	In particular, without loss of generality, suppose the trial-zero-eligible individuals are considered representative of the target population, and thus it is of interest to standardize trial-specific VE according to the trial-$0$ baseline covariate distribution.  
	Define the standardized VE for trial $j$ at time $k$ to be 
	\begin{equation}
		VE^s_j(k) = 1 - \dfrac{\int E\{Y_j^1(k)|E_j=1,\boldsymbol{X}=\boldsymbol{x}\} dF(\boldsymbol{x}|E_0=1)}
		{\int E\{Y_j^0(k)|E_j=1,\boldsymbol{X}=\boldsymbol{x}\} dF(\boldsymbol{x}|E_0=1)}
		\label{eEstimand}
	\end{equation}
	where $F(\boldsymbol{x}|E_0=1)=P(\boldsymbol{X}\leq \boldsymbol{x} \mid E_0=1 )$ and $\boldsymbol{X}$ is the vector of covariates that satisfy the identifiability assumptions in Section~\ref{assumptions}.  
	The estimand (\ref{eEstimand}) describes VE had trial $j$ been conducted in a population with the same observed covariate distribution as trial $0$.  
	
	To draw inference about (\ref{eEstimand}), a plug-in type estimator can be constructed by estimating $F(\boldsymbol{x}|E_0=1)$ with the empirical distribution of $\boldsymbol{X}$ in the analytic cohort and replacing $E\{Y^a_j(k)|E_j=1,\boldsymbol{X}=\boldsymbol{x}\}$ with a suitable estimator. 
	Specifically, consider the estimator
	\begin{equation}
		\label{eEstimator}
		\widehat{VE}^s_j(k)=1 - \dfrac{\sum_{i=1}^n \hat{E}\{Y_j^1(k)|E_j=1,\boldsymbol{X_i}\}}
		{\sum_{i=1}^n \hat{E}\{Y_j^0(k)|E_j=1,\boldsymbol{X_i}\}}
	\end{equation}
	where $\hat{E}\{Y_j^0(k)|E_j=1,\boldsymbol{X}\}$ is constructed using the following adaptation of the estimation procedures described in Section~\ref{methods}.  
	A MSM is assumed for $\lambda^a_j(k\mid \boldsymbol{X})$, i.e., the hazard of the potential outcome under $A_j=a$ conditional on covariates $\boldsymbol{X}$.  
	Letting $\boldsymbol{\gamma}$ denote the parameters of the assumed MSM, the model
	\begin{equation}  
		g[P\{Y_j(k)=1 \mid A_j=a, R_j(k)=1, \boldsymbol{X}=\boldsymbol{x}\}] =\boldsymbol{\gamma^\dagger} \boldsymbol{f}_{\boldsymbol{\gamma}}(j, k, a, \boldsymbol{x})
		\label{condlModel}
	\end{equation}
	for $(j,k) \in \mathcal{W}$ is fit to the NTE analysis data via weighted maximum likelihood with weights $\widehat{W}_{j}(k, A_j)$.  
	The estimator $\hat{E}\{Y_j^a(k)|E_j=1,\boldsymbol{X}\}$ is then constructed as 
	\begin{equation}
		\label{condRisk}
		\hat{E}\{Y_j^a(k)|E_j=1,\boldsymbol{X}\}=
		1-\prod_{m=1}^{k}\{1-\hat{\lambda}_j^{a}(m\mid \boldsymbol{X})\},
	\end{equation}
	where $\hat{\lambda}_j^{a}(k\mid \boldsymbol{X})=g^{-1}\{\boldsymbol{\hat{\gamma}^\dagger} \boldsymbol{f}_{\boldsymbol{\gamma}}(j, k, a, \boldsymbol{X})\}$ for $a \in \{0,1\}$, and $\boldsymbol{\hat{\gamma}^\dagger}$ is the weighted maximum likelihood estimator of the parameters in (\ref{condlModel}).  
	As in Section \ref{inference}, the estimator (\ref{eEstimator}) can be shown to be consistent and asymptotically normal, and Wald confidence intervals for (\ref{eEstimand}) can be constructed using the empirical sandwich variance estimator.
	
	The homogeneity assumption 
	\begin{equation}
		\label{HnaughtSeq0}
		H_0^{s}:VE^s_0(k)=\dots=VE^s_J(k) \text{ for all }k\in\{1,2,...,K_J\}
	\end{equation}
	can be tested using the procedure described in Section~\ref{TEH} with all instances of $VE_j(k)$ and $\widehat{VE}_j(k)$ replaced by $VE^s_j(k)$ and $\widehat{VE}^s_j(k)$, respectively.
	
	\section{Application to the Abruzzo, Italy data}
	\label{DataAnalysis}
	The NTE methods described above were applied to the Abruzzo data to address questions \ref{rqTSV} and \ref{rqCalTime}. 
		The application considered VE of a two-dose mRNA COVID-19 vaccine regimen, compared to remaining unvaccinated, against the composite outcome of severe COVID-19 or COVID-19-related death among individuals aged 80 years or older. 
		While the analysis focused on this particular setting, the proposed methodology is not specific to this application and can be applied to other vaccine regimens or policies, as well as to other target populations. 
		The corresponding target trial protocol is presented in Table~\ref{protocol_p1} of the Appendix.
	
	\subsection{Analysis approach}
	\label{analysisPlan}
	The active vaccine regimen was defined as two doses of (original, monovalent) Pfizer-BioNTech or Moderna vaccine, with the second dose obtained within the recommended time window \citep{CDC}.
	See \ref{dataAnalysisDetails} for more details.  
	In keeping with standard trial emulation practices \citep{hernan2016specifying}, week zero of each trial was aligned with vaccine regimen assignment and, for those with $A_j=1$, receipt of the first dose.
	The first emulated trial was initiated February 15, 2021.  
	New trials were initiated every seven days thereafter through May 3, 2021, for a total of $J+1=12$ emulated trials.  
	December 18, 2021 was chosen as the administrative censoring date because the observed vaccine regimen could not be determined from the data after this date.  
	
	It was assumed that the following set of covariates was sufficient to achieve conditional exchangeability between treatment arms and adjust for possible selection bias arising from differential censoring: baseline age, sex, hypertension, diabetes, cardiovascular disease, chronic obstructive pulmonary disease, kidney disease, and cancer.  
	Specification of the models used to estimate the IP weights is detailed in \ref{dataAnalysisDetails}.  
	The covariate-conditional outcome MSM was specified as in (\ref{calTimeTerms}) with additional terms for baseline covariates and restricted cubic spline functions for both calendar time and time since vaccination; see Appendix Section~\ref{appliedOcModelSpec} for the full model specification.
	Results in the following section were obtained using the standardization method of Section~\ref{extensions}.  
	The analysis was repeated without standardization using the method of Section~\ref{methods}; these results are presented in \ref{dataAnalysisRes_un}. 
	
	\subsection{Analysis results} 
	\label{analysisResults}
	The analytic cohort was constructed by first identifying all individuals in the Abruzzo database who met eligibility criteria (see Table~\ref{protocol_p1}) on February 15, 2021.  
	Of these 111,595 individuals, 972 ($<1\%$) were missing date of first vaccine dose and were removed from the analytic cohort, as their trial-specific eligibility could not be determined from the data.  
	The final analytic cohort comprised the remaining $n=110{,}623$ people.
	Seventy-one percent of the analytic cohort 
	($78{,}344$ individuals)
	received a first mRNA COVID-19 vaccine dose by May 9, 2021.
	Of these, 
	$76{,}286$ $(97\%)$ 
	completed a full course of mRNA COVID-19 vaccine according to the recommended dosing schedule associated with the brand of first vaccine dose received.  	
	Among completers, $69{,}036$ individuals $(90\%)$ received two doses of Pfizer and 
	$7{,}250$ $(10\%)$ received two doses of Moderna.
	
	\begin{intextfigure}
		 \begin{subfigure}[b]{0.46\textwidth}
			\centering
			\includegraphics[width=\linewidth]{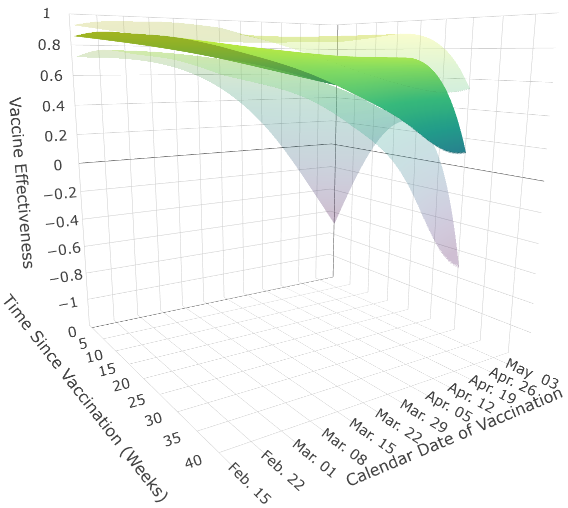}
			\caption{}
		\end{subfigure}%
		\begin{subfigure}[b]{0.5\textwidth}
			\centering
			\includegraphics[width=\linewidth]{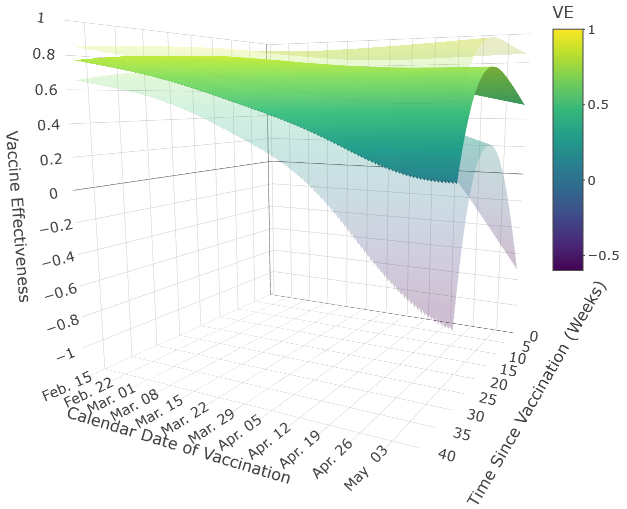}
			\caption{}
		\end{subfigure}
		\caption{Estimates of VE against severe COVID-19 or COVID-19-related death among Abruzzo residents aged 80 years or older between February 15, 2021 and December 18, 2021.  
			VE estimates are standardized to the trial-zero-eligible population.  
			The opaque surfaces depict the standardized VE point estimates and the transparent surfaces the corresponding 95\% Wald CIs.  
			Panels (a) and (b) display the same surface from two vantage points.}
		\label{contourPlots}
	\end{intextfigure}

	Figure~\ref{contourPlots} displays the estimated VE surface from two perspectives. Corresponding standardized VE point estimates and pointwise 95\% confidence intervals for selected trials and times since vaccination are reported in Table~\ref{resTable}.
	
	\begin{center}
		\renewcommand{\arraystretch}{1}
		\captionof{table}{Estimates of VE and 95\% pointwise confidence intervals (CIs) against severe COVID-19 or COVID-19-related death among Abruzzo residents aged 80 years or older ($n=110{,}623$) between February 15, 2021 and December 18, 2021.  
			VE estimates are standardized to the trial-zero-eligible population.  
			Results are presented for selected emulated trials and weeks since first dose.
			The starting date of each trial was as follows: trial $j=0$, February 15, 2021; trial $j=3$, March 8, 2021; trial $j=6$, March 29, 2021; and trial $j=9$, April 19, 2021.} 
		\label{resTable}
		\begin{tabular}{lllll}
			\hline
			Weeks since & \multicolumn{1}{c}{$\widehat{VE}_0(k )$} & \multicolumn{1}{c}{$\widehat{VE}_3(k )$} & \multicolumn{1}{c}{$\widehat{VE}_6(k )$} & \multicolumn{1}{c}{$\widehat{VE}_9(k )$} \\  first dose ($k$) & \multicolumn{4}{c}{\% (95\% CI)}  \\ \hline
			1 & 86 (71, 93) & 80 (68, 87) & 71 (48, 84) & 60 (-1, 84) \\ 
			7 & 87 (75, 93) & 82 (73, 88) & 75 (56, 86) & 67 (21, 86) \\ 
			14 & 87 (76, 93) & 82 (73, 88) & 77 (60, 87) & 72 (38, 88) \\ 
			21 & 86 (75, 92) & 81 (72, 87) & 77 (62, 86) & 73 (45, 87) \\ 
			28 & 84 (73, 90) & 79 (70, 85) & 72 (57, 81) & 61 (30, 78) \\ 
			34 & 82 (71, 89) & 74 (65, 81) & 59 (41, 71) & 25 (-21, 54) \\ 
			\hline
		\end{tabular}
	\end{center}
	
	Regarding scientific question~\ref{rqTSV}, across all emulated trials the estimated VE curve generally peaked approximately 15-20 weeks after the first vaccination. 
	This pattern is consistent with the hypothesized biological mechanism underlying protection from COVID-19 vaccination, whereby antibody concentrations following a two-dose regimen increase after the second dose and subsequently decline over time \citep{Ebinger}.
	
	Regarding scientific question~\ref{rqCalTime}, for a fixed time since vaccination, the standardized VE estimates generally decreased with trial number. 
	Because standardization was used to account for variation in measured covariate distributions across the emulated-trial populations, this pattern is consistent with a decline in VE over calendar time that is not explained by those measured covariates. 
	The hypothesis test of (\ref{HnaughtSeq0}) with $K_J=33$ yielded a one-sided $P$-value of 
	$0.07$,
	providing moderate evidence against the null hypothesis that $VE^s_j(k)$ is homogeneous across trials.  
	
	\begin{intextfigure}
		\centering
		\subfloat{\includegraphics[width=.8\textwidth]{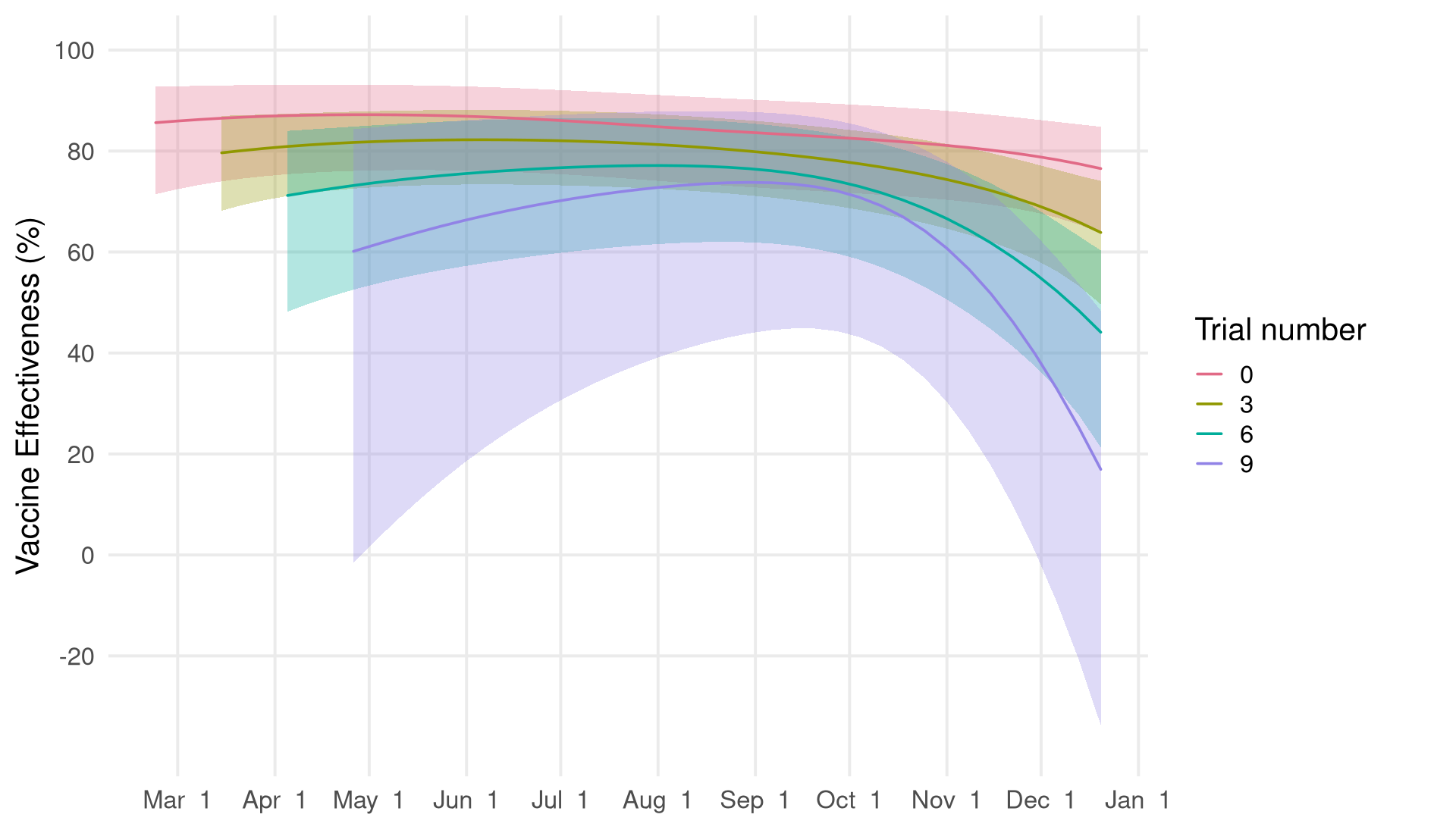}}\quad
		\subfloat{\includegraphics[width=.8\textwidth]{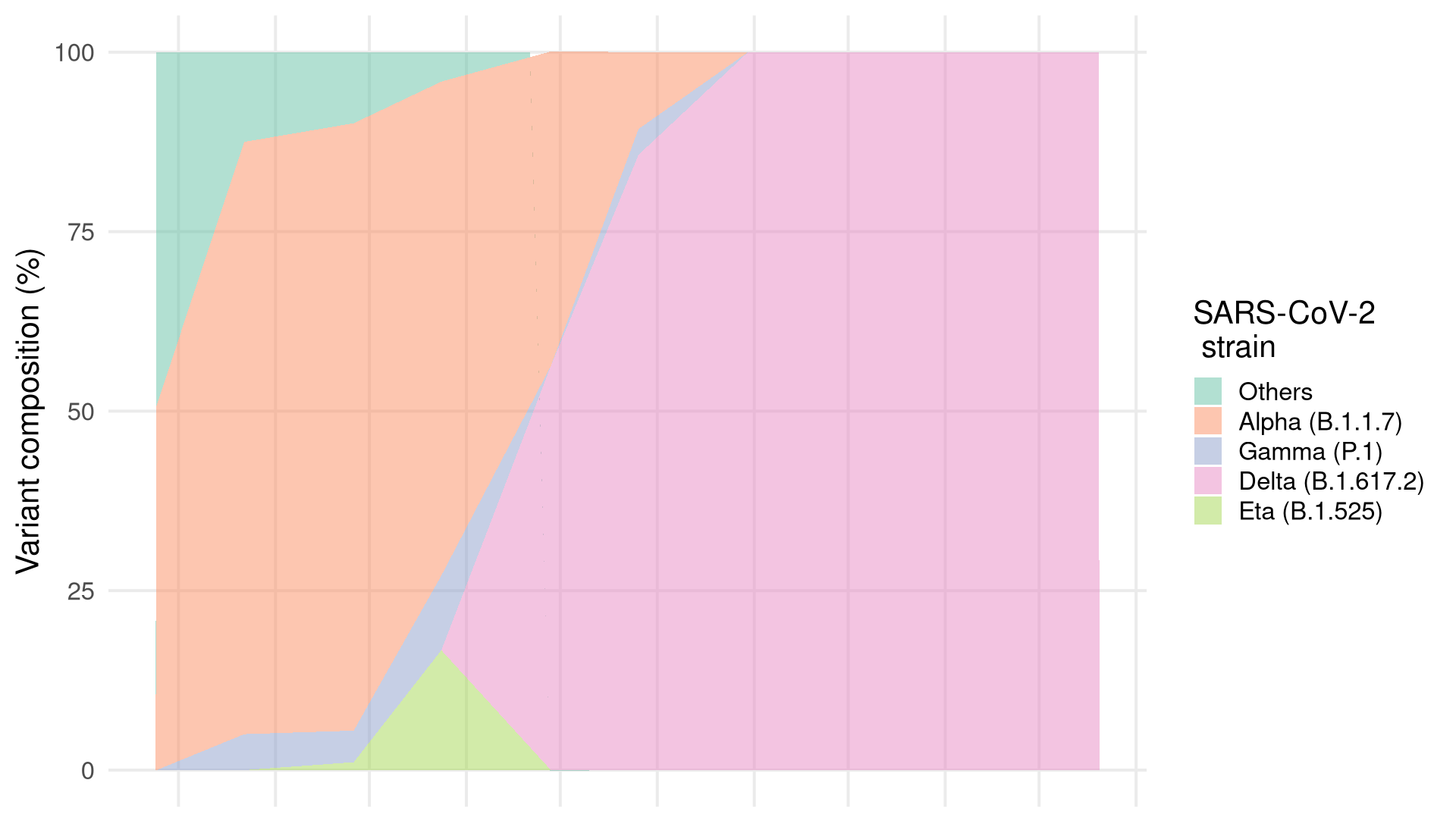}}
		\caption{(upper) Estimates of VE against severe COVID-19 or COVID-19-related death among Abruzzo residents aged 80 years or older ($n=110{,}623$) between February 15, 2021 and December 18, 2021.  
				VE estimates are standardized to the trial-zero-eligible population.  
				Results are presented for selected emulated trials.  
				VE estimates are shown by solid lines, corresponding 95\% pointwise confidence intervals (CIs) are shown by shaded bands. 
				The starting date of each trial was as follows: trial $j=0$, February 15; trial $j=3$, March 8; trial $j=6$, March 29; and trial $j=9$, April 19.
				(lower) Estimated SARS-CoV-2 variant composition \citep{voc2021} in Abruzzo during the NTE study period.}
		\label{deltaPlot}
	\end{intextfigure}
	
	To provide additional context regarding potential changes in VE over calendar time, Figure~\ref{deltaPlot} displays estimates of $VE^s_j(k)$ for selected trials (upper panel) and the circulating SARS-CoV-2 variant composition in Abruzzo (lower panel) over the same calendar time period. 
	The stacked area plot was constructed by linearly interpolating published variant prevalence estimates from regional laboratory samples collected at approximately monthly intervals \citep{voc2021}. 
	As shown in Figure~\ref{deltaPlot}, the Alpha variant predominated until mid-June, after which Delta became the dominant circulating variant for the remainder of 2021. 
	The largest declines in estimated VE occurred after the Delta variant became dominant, suggesting that the change in circulating variants may have contributed to the decrease in VE over calendar time. 
	These results are also consistent with previous studies reporting lower effectiveness of the original monovalent mRNA vaccines against Delta relative to Alpha \citep[e.g.,][]{LopezDelta}.
	
	The findings regarding scientific question~\ref{rqCalTime} should be interpreted with caution because VE estimates from later trials are generally less precise, which may be at least partially due to successive trials having smaller sample sizes. 
	The simulation study results (described in the Appendix) also demonstrate that the variance of the VE estimators tends to be larger for later trials. 
	In general, the precision of $\widehat{VE}^s_j(k)$ may depend on several factors, including the number of eligible trial participants, the proportion of trial participants who are vaccinated, the number of observed events within a trial, and the underlying true value of ${VE}^s_j(k)$.
	
	\subsection{Sensitivity analyses}
	\label{sens}
	A series of diagnostics and sensitivity analyses were conducted to examine robustness of the Abruzzo data analysis results to analytic choices.
	To assess adequacy of the vaccine uptake model specification, baseline covariate balance for each emulated trial was examined before and after weighting.  
	Several sensitivity analyses of model specifications were conducted. 
	First, to assess sensitivity to spline function specifications in the outcome MSM, information on changes in epidemiologic conditions in Abruzzo was used to choose knot locations.
	Second, the no-interaction assumption in model~(\ref{calTimeTerms}) was relaxed by including additional three-way interactions between vaccine regimen assignment, calendar time, and time since vaccination.  
	Finally, to examine the extent to which exclusion of an important predictor of the outcome may influence the final standardized estimates, the full analysis was repeated eight times, each time excluding one of the covariates listed in Section~\ref{analysisPlan}.
	Additionally, the Abruzzo data were reanalyzed using an adjusted Cox model with time-varying treatment. 
	
	Detailed methodology and results for all diagnostics and sensitivity analyses appear in Section~\ref{sensAnalyses} of the Appendix.
	Cox model analysis results appear in \ref{CoxModel}.
	After weighting, all covariates were adequately balanced in all trials except for age in certain trials.  
	The results were robust to the exclusion of individual baseline covariates and remained largely unchanged under the alternative MSM specifications described above.
	
	\subsection{Discussion of Abruzzo data analysis}
	\label{daDiscussion}
	The application in this section evaluated effectiveness of a two-dose mRNA vaccine regimen versus remaining unvaccinated among elder residents of Abruzzo.  
	It is difficult to directly compare these results with those of \citet{AcutiMartellucci} because the analyses differ in several key aspects.  
	Acuti Martellucci et al.\ estimated the effectiveness of a full course of COVID-19 vaccine versus remaining unvaccinated among individuals 60 years of age or older during the period January 31, 2021 through February 8, 2022.  
	They estimated VE against COVID-19-related death to be 94\% (95\% CI [93\%; 95\%]) and against severe COVID-19 to be 86\% (95\% CI [84\%; 88\%]).  
	For comparison, in the present analysis, estimated VE in individuals 80 or older against the composite outcome at week 44 of trial 0 (i.e., February 15, 2021 through December 18, 2021) was 
	77\% (95\% CI [64\%; 85\%]).
	Direct comparison of these estimates is challenging because of differences in study period, target population, and outcome.  
	The analyses also targeted different estimands.  
	Acuti Martellucci et al.~defined VE as one minus the adjusted odds ratio, while in this manuscript VE equals one minus the marginal risk ratio.  
	Although the odds ratio approximates the risk ratio when the outcome is rare, the adjusted odds ratio does not in general equal the marginal odds ratio due to noncollapsibility \citep{Daniel}.  
	
	The Delta (B.1.617.2) variant became dominant in Abruzzo in late June 2021, replacing Alpha (B.1.1.7), which had become the most prevalent variant in February.  
	By July 20th, estimated prevalence of Delta in Abruzzo reached $86\%$ \citep{voc2021}.  
	Most vaccinated individuals enrolled in early emulated trials would have received a second vaccine dose by June, such that vaccine-induced antibodies for these individuals would be near peak levels when Delta became dominant \citep{Ebinger}.
	On the other hand, many vaccinated individuals in later trials would have received only a single dose by June, which may have contributed to the estimated decline in VE across trials.  
	Previous findings \citep{LopezDelta} suggest two doses of the Pfizer-BioNTech original monovalent vaccine may be modestly (6\%) less effective against symptomatic COVID-19 disease with the Delta variant relative to the Alpha variant.  
	However, \citeauthor{LopezDelta} found a more substantial (12\%) loss in effectiveness when drawing that same comparison for a single dose of the Pfizer-BioNTech original monovalent vaccine. 
	
	The analysis presented in this section draws new insights from the Abruzzo study database.  
	Particularly, the analysis characterizes VE trends across calendar time and time since vaccination.  
	These results suggest that for the vaccine regimen considered (i) VE peaked approximately 15-20 weeks after the first dose, and (ii) VE declined over the calendar time of the study.  
	The combination of (i) and (ii) can result in meaningful differences in the protective effect of the vaccine, e.g., VE 1 week after the first dose in the earliest emulated trial was estimated to be 
	86\%, 
	whereas VE 34 weeks after the first dose in trial 9 was estimated to be less than 30\%.  
	Understanding changes in VE over time since vaccination is important for informing if and when additional doses following an initial vaccine series should be recommended.  
	Characterizing changes in VE over calendar time can guide decisions by policy makers and vaccine manufacturers regarding the need for updated vaccine formulations.
		
	\section{Conclusion}
	\label{discussion}
	Nested trial emulation has become increasingly popular for drawing inference about treatment effects from observational data.  
	Motivated by the need to assess the effects of COVID-19 vaccination outside of randomized trials, this paper develops NTE-based methods which allow treatment (vaccine) effects to vary on two different time scales, namely time since treatment initiation and calendar time.  
	Each year, real-world data are used to estimate the effectiveness of seasonal influenza and COVID-19 vaccine formulations with the goal of detecting loss in VE possibly due to antigenic drift and/or waning vaccine-induced antibody levels.  
	Methods in this manuscript can be applied to evaluate seasonal vaccine formulations or to any setting where treatment effects may plausibly vary across both time scales.  
	
	The methods in this manuscript can be used to evaluate vaccination policies that include multiple vaccines, as well as policies restricted to a particular vaccine or vaccine class. 
	The COVID-19 VE data analysis in Section 5 estimates the effect of a two-dose mRNA vaccination regimen with either the Pfizer-BioNTech or Moderna original monovalent vaccines. 
	Thus, the resulting VE estimates reflect the effect of a vaccination policy that allows receipt of either of these mRNA vaccines, with the distribution of the two vaccine brands determined by the policy and its implementation in the study population. 
	Moreover, such estimates should not be interpreted as the effect of any one vaccine brand and may not generalize to settings in which the distribution of vaccine brands differs substantially from the distribution in the Abruzzo study. 
	That said, the estimated efficacy of the two mRNA vaccines under consideration was similar in phase 3 trials, suggesting that these results may be robust to distribution of vaccine brand. 
	
	The target estimand in this paper $VE_j(k)$ is agnostic to early infections that may occur before a multi-dose vaccine strategy is completed.  
	Thus, results using the proposed approach can be interpreted as estimating the vaccine effect under the observed risk of early infections. 
	Alternative causal frameworks could target different scientific questions. 
	One approach could entail principal stratification, in which effects are defined within the latent subgroup of individuals who would remain uninfected during the inter-dose period regardless of whether vaccinated. 
	While such analyses focus on vaccine effects among those able to complete the dosing schedule without infection, they involve latent subpopulations and consequently rely on stronger identification assumptions than those required for the present analysis. 
	Another approach is to define a hypothetical intervention estimand that removes or prevents infection during the interval between doses, yielding the effect of a multi-dose regimen in a counterfactual setting where early infection cannot occur. 
	Although such estimands may provide complementary mechanistic insight, they address different causal questions than the policy-relevant effect considered here and generally require additional assumptions for identification.
	
	There is some conceptual overlap between the proposed NTE approach and \citet{fintzi2021assessing}, in that both methods exploit the fact that individuals vaccinated later contribute both unvaccinated and vaccinated person-time. 
	The two approaches, however, were developed for different settings. 
	\citet{fintzi2021assessing} consider randomized placebo-controlled trials with subsequent placebo crossover and aim to draw inference about VE as a function of time since vaccination. 
	Their approach leverages the randomized trial structure and, critically, assumes VE does not change over calendar time. 
	By contrast, the proposed NTE approach does not require randomization, nor does it require constant VE over calendar time. 
	Moreover, using observational data, the proposed methods can be used to assess whether VE depends on calendar time, time since vaccination, or both.
	
	There are several possible avenues for future work to build on the methods described here.  
	Alternative IP weights that are less variable and TEH tests with improved power against a broader class of alternatives could be developed.  
	Nonparametric machine learning approaches could be incorporated to relax the need for correctly specified parametric models.  
	In many infectious disease settings, an individual's outcome may be affected by other individuals' vaccination status, i.e., there may be interference.  
	Analyses that fail to account for interference may yield biased VE estimates.  
	Given a data source containing each individual's geographic location (e.g., place of residence), future research could develop NTE methods for estimating VE in the presence of interference.  
	
	\bibliographystyle{apalike}
	\bibliography{VEbib}
	
	\newpage
	\appendix
	
	\setcounter{table}{0}
	\renewcommand{\thetable}{A\arabic{table}}
	
	\renewcommand{\thefigure}{A\arabic{figure}}
	\setcounter{figure}{0}
	
	\setcounter{section}{0}
	\renewcommand{\thesection}{Appendix A\arabic{section}}
	
	\setcounter{equation}{0}
	\renewcommand{\theequation}{A.\arabic{equation}}
	
	\renewcommand{\thesubsection}{A\arabic{section}.\arabic{subsection}}
	
	\renewcommand{\thelemma}{A.\arabic{lemma}}
	\renewcommand{\thecorollary}{A.\arabic{corollary}}

\input{supplementArXiV_3p0.tex}

\end{document}

%% file: supplementArXiV_3p0.tex
\section*{Appendices}
The following Appendices contain additional details for the article ``Assessing vaccine effectiveness in observational studies via nested trial emulation'' (which is referred to as ``the main text'' in the sequel).  
\ref{webA} includes technical details for the method without standardization (Section~\ref{methods} of the main text).  
\ref{simulations} contains a series of simulations demonstrating the finite-sample performance of the proposed estimators and TEH hypothesis test under a variety of scenarios.
\ref{dataAnalysisDetails} includes additional details for the Abruzzo COVID-19 VE data analysis.  
Results of the unstandardized data analysis appear in \ref{dataAnalysisRes_un}.  
\ref{CoxModel} details a analysis of the Abruzzo data using an adjusted Cox model with time-varying treatment.  
Illustrative pseudocode for the standardized estimation procedure is provided in \ref{pseudocode}.

\section{Details of Methods in Section~\ref{methods} of the Main Text}
\label{webA}
\subsection{Identifiability}
\label{identifiability}
In this section the target estimand $VE_j(k)$ is shown to be identifiable from the observable random variables under the identifiability assumptions listed below.  
The setting with a single vaccine brand (which may require multiple doses) is considered.  
Under additional assumptions, the results can be extended to settings involving multiple brands.  
Before stating the assumptions, some new notation is introduced.  
Let $Z_l=\sum_{m=0}^l I(B_m>0)$ denote the number of COVID-19 vaccine doses received by calendar time $l$.  
Let $\mathcal{Z}^{a}_{j}(k)$ be the set of vaccine dose histories through calendar time $j+k$ that accord with initiation of vaccine regimen $a$ at calendar time $j$ and adherence to that regimen through calendar time $j+k$.  
In the Abruzzo data analysis of Section~\ref{DataAnalysis} of the main text, the comparator regimen is ``never treat'', and therefore $\mathcal{Z}^{0}_{j}(k)=\{\overline{0}\}$ for all $k \in \{1,2,..., K_j\}$; the active vaccine regimen, as defined in the target trial protocol in Table~\ref{protocol_p1}, involves remaining unvaccinated through calendar time $j$, receiving a first dose at calendar time $j+1$, and possibly receiving additional doses according to the recommended vaccine schedule of interest.  
The indicator $E_l$ of trial eligibility at calendar time $l$ is defined as $E_0=1$ (because all individuals in the analytic cohort are eligible for trial zero by construction) and $E_l=I(T^*>l, Z_l=0)$ for $l \in \{1,2,...,\tau\}$.  
The trial-specific indicator for censoring due to nonadherence to either $A_j=0$ or $A_j=1$ at time $k$ of trial $j$ is defined as 
$
C_j(k)=I(A_{j}=0,\overline{Z}_{j+k+1} \neq \overline{0} ) + I\{A_j=1, \overline{Z}_{j+k+1} \notin \mathcal{Z}^1_{j}(k+1) \}
$
if $E_j=1$ (with $C_j(-1)=0$); otherwise $C_j(k)$ is undefined.  
Let $H_l=I(T^* \leq l, \Delta=0)$ denote the indicator for loss to follow up by calendar time $l$.  
Throughout, assume the following ordering of variables within each calendar time interval: $(Z_{j+k}, H_{j+k},  Y_j(k),C_j(k))$.  

Define the IP-weight $W_j(k,a)$ as
$$I(A_j=a)R_j(k)
\bigg[
\prod_{m=1}^{k} 
\{1-\lambda^H_{j}(m,\boldsymbol{X},\overline{Z}_{j+m})\}
\{1-\lambda^C_j(m-1,a,\boldsymbol{X},\overline{Z}_{j+m-1})\}\bigg]^{-1},$$
where 
\begin{equation}
	\nonumber
	\lambda_{j}^{C}(k,a,\boldsymbol{x},\overline{z})=E\{C_j(k) \mid \boldsymbol{\mathcal{F}}^C_j( k)=(0,0,0,\overline{z}, a,\boldsymbol{x},1)\}  
\end{equation}
is the conditional hazard of censoring due to nonadherence to vaccine regimen $A_j=a$ at time $k$ of trial $j$, $\boldsymbol{\mathcal{F}}^C_j(k)=(Y_j(k),H_{j+k}, C_j(k-1), \overline{Z}_{j+k}, A_j, X, E_j)$,
\begin{equation}
	\nonumber
	\lambda^H_{j}(k,\boldsymbol{x},\overline{z})=E\{H_{j+k}\mid \boldsymbol{\mathcal{F}}^H_j(k)=(0,0,0,\overline{z}, a,\boldsymbol{x},1)\}
\end{equation}  
is the conditional hazard of loss to follow up at time $k$ of trial $j$, and $\boldsymbol{\mathcal{F}}^H_j(k)=(Y_j(k-1),H_{j+k-1}, C_j(k-1), \overline{Z}_{j+k}, A_j, X, E_j)$.

The following assumptions are sufficient to identify the target estimand $VE_j(k)$.    
\begin{enumerate}[label=\Roman*]
	\item Variables are measured without error.  
	\label{measurement}
	\item An individual's potential outcomes are unaffected by another individuals' treatment history (no interference).  
	\label{interference}
	\item $(Y^a_j(k), ..., Y^a_j(K_j)) \perp H_{j+k} \mid \boldsymbol{\mathcal{F}}^H_j(k)=(0,0,0,\overline{Z}, a, \boldsymbol{X},1) 
	$ for $a \in \{0,1\}$ (ignorable dropout).
	\label{exchDropout}
	\item $(Y^a_j(k), ..., Y^a_j(K_j)) \perp C_j(k-1) \mid  \boldsymbol{\mathcal{F}}^C_j(k-1)=(0,0,0,\overline{Z}, a, \boldsymbol{X},1)
	$ for $a \in \{0,1\}$ (ignorable nonadherence).
	\label{exchNoncomp}
	\item $(Y^a_j(1),..., Y^a_j(K_j)) \perp A_j \mid \{\boldsymbol{X},E_j=1 \}$ for $a \in \{0,1\}$ (conditional exchangeability).
	\label{exchTx}
	\item If $E_j=1$, then ${Y}_j(k)={Y}^1_j(k)I\{A_j=1, C_j(k-1)=0\}+{Y}^0_j(k)I\{A_j=0, C_j(k-1)=0\}$ for $k \in \{1,..., K_j\}$ (causal consistency).
	\label{consistency}
	\item
	\label{positivityDropout}
	For $a \in \{0,1\}$ and $k \in \{1,..., K_j\}$: 
	if $dF_{\boldsymbol{\mathcal{F}}^H_j(k)}(0,0,0,\overline{z}, a, \boldsymbol{x}, 1) > 0$,
	then $P\{H_{j+k} =0 \mid \boldsymbol{\mathcal{F}}^H_j(k)=(0,0,0,\overline{z}, a, \boldsymbol{x}, 1)\}>0$, 
	where in general $F_{\boldsymbol{G}}$ denotes the cumulative distribution function of random variable $\boldsymbol{G}$ (positivity of non-dropout).
	\item
	\label{positivityAdhere}
	For $a \in \{0,1\}$ and $k \in \{0,..., K_j-1\}$: 
	if $dF_{\boldsymbol{\mathcal{F}}^C_j(k) }(0,0,0,\overline{z}, a, \boldsymbol{x}, 1) > 0$,
	then $P\{C_j(k) =0 \mid \boldsymbol{\mathcal{F}}^C_j(k)=(0,0,0,\overline{z}, a, \boldsymbol{x}, 1)\}>0$ (positivity of adherence).
	\item
	\label{positivityTx}
	If $dF_{\boldsymbol{X} ,E_j}(\boldsymbol{x}, 1) > 0$, then $0<P\{A_j =1 \mid \boldsymbol{X}=\boldsymbol{x}, E_j=1 \}<1$ (positivity of treatment).
	\setcounter{leftOffHere}{\value{enumi}}
\end{enumerate}

The main result in this section is given by Corollary~\ref{IDcorollary} below, which shows that the estimand $VE_j(k)$ can be expressed as a functional of the distribution of the observable random variables and therefore is identifiable.  
The proof of Corollary~\ref{IDcorollary} relies on the following three lemmas. 
Assumptions \ref{measurement}-\ref{positivityTx} are made throughout.

\begin{lemma}
	\label{firstLemma}
	\begin{align*}
		E   
		[ G^a_j(k)   W_j(m,a) 
		] 
		&=E   
		[ G^a_j(k) W_j(m-1,a) \{1-Y^a_j(m-1)\}   
		] 
	\end{align*}
	for $a \in \{0,1\}$, $m \in \{2,..., k\}$, and $G^a_j(k) =h(Y^a_j(k),..., Y^a_j(K_j))$ for arbitrary function $h$. 
\end{lemma}

\begin{proof}	
	\begin{flalign}
		&E   \nonumber
		[ G^1_j(k)   W_j(m,1) 
		]&\\\nonumber
		=&E\bigg[ G^1_j(k)   W_j(m-1,1) 
		\dfrac
		{\{1-Y_j(m-1)\}(1-H_{j+m})\{1-C_j(m-1)\}}
		{\{1-\lambda^H_{j}(m,\boldsymbol{X},\overline{Z}_{j+m})\}\{1-\lambda^C_j(m-1,1,\boldsymbol{X},\overline{Z}_{j+m-1})\}}
		\bigg]&\\
		\label{iterExp1}
		=&E
		\bigg[ G^1_j(k)   W_j(m-1,1) 
		\dfrac 
		{\{1-Y_j(m-1)\}\{1-C_j(m-1)\}}
		{\{1-\lambda^C_j(m-1,1,\boldsymbol{X},\overline{Z}_{j+m-1})\}}  
		&\\\nonumber
		&
		E\bigg\{ \dfrac
		{(1-H_{j+m})}
		{\{1-\lambda^H_{j}(m,\boldsymbol{X},\overline{Z}_{j+m})\}}
		\bigg| 
		G^1_j(k), \boldsymbol{\mathcal{F}}^H_j(m)\bigg\}  \bigg]&\\
		\label{ignDropout}
		=&E
		\bigg\{ G^1_j(k)  W_j(m-1,1) \dfrac
		{\{1-Y_j(m-1)\}\{1-C_j(m-1)\}}
		{\{1-\lambda^C_j(m-1,1,\boldsymbol{X},\overline{Z}_{j+m-1})\}} \bigg\}&\\
		\label{iterExp2}
		=&E
		\bigg[ G^1_j(k)   W_j(m-1,1) 
		\{1-Y_j(m-1)\}
		&\\\nonumber
		&
		E\bigg\{ \dfrac
		{\{1-C_j(m-1)\}}
		{\{1-\lambda^C_j(m-1,1,\boldsymbol{X},\overline{Z}_{j+m-1})\}}
		\bigg| 
		G^1_j(k) ,\boldsymbol{\mathcal{F}}^C_j(m-1)\bigg\} \bigg]&\\
		\label{ignNoncomp}
		=&E[ G^1_j(k)  W_j(m-1,1) \{1-Y_j(m-1)\}]&\\
		\label{cons1}
		=&E   
		[G^1_j(k)  W_j(m-1,1) 
		\{1-Y^1_j(m-1)\}
		],
	\end{flalign}
	as desired, where (\ref{iterExp1}) holds by iterated expectation, (\ref{ignDropout}) holds 
	by Assumptions \ref{exchDropout}, \ref{positivityDropout} and by iterated expectation, 
	(\ref{iterExp2}) holds by iterated expectation, (\ref{ignNoncomp}) holds 
	by Assumptions \ref{exchNoncomp} and \ref{positivityAdhere} and by iterated expectation, and (\ref{cons1}) holds by Assumption~\ref{consistency}.  
	Proof for the $a=0$ case is analogous. 
\end{proof}

\begin{lemma}
	\label{secondLemma}
	\begin{equation*}
		E   
		[ G^a_j(k)   W_j(m,a)
		]	=
		E   
		[ G^a_j(k)  W_j(m-1,a) 
		]
	\end{equation*}
	for $a \in \{0,1\}$, $m \in \{2,...,k-1\}$, and $G_j(k) \in \{Y^a_j(k)\{1-Y^a_j(k-1)\}, 1-Y^a_j(k-1)\}$.
\end{lemma}	

\begin{proof}	
	\begin{flalign}
		\label{lemmaTheFirst}
		E   
		\{ G^a_j(k)   W_j(m,a) \}
		&=
		E   
		[ G^a_j(k)  W_j(m-1,a) \{1-Y^a_j(m-1)\}
		]\\\label{justified}
		&=
		E   
		[ G^a_j(k)  W_j(m-1,a) 
		],
	\end{flalign}
	where line (\ref{lemmaTheFirst}) holds by Lemma~\ref{firstLemma}, and line (\ref{justified}) follows because $Y^a_j(k)$ is monotonic in $k$, which implies $\{1-Y^a_j(k-1)\}=\{1-Y^a_j(k-1)\}\{1-Y^a_j(m-1)\}$ and therefore $G^a_j(k)\{1-Y^a_j(m-1)\}=G^a_j(k)$ for any $m < k-1$.   
\end{proof}

\begin{lemma}
	\label{thirdLemma}
	\begin{align*}
		\text{ (i) }E   
		[ Y^a_j(k) 
		\{1-Y^a_j(k-1)\}
		\mid E_j=1]P(E_j=1) &=E   
		\{ Y_j(k)   W_{j}(k,a) 
		\} \text{ and }\\
		\text{ (ii) }E   
		\{ 1-Y^a_j(k-1)
		\mid E_j=1\} P(E_j=1)&=E   
		\{   W_{j}(k,a) 
		\} .
	\end{align*}
\end{lemma}

\begin{proof}	
	\begin{flalign}
		\nonumber
		&E\{ Y_j(k) W_j(k,1)\}&\\
		\label{beConsistent}
		=&E\{ Y^1_j(k) W_j(k,1)\}&\\
		\label{lemma1}
		=&E   
		[ Y^1_j(k)   W_j(k-1,1) 
		\{1-Y^1_j(k-1)\}
		]		&\\
		\label{repeatedly}
		=&E   
		[ Y^1_j(k)  W_j(1,1) \{1-Y^1_j(k-1)\}
		].&\\
		\label{identical}
		=&E
		\bigg\{ Y^1_j(k)\{1-Y^1_j(k-1)\} E_j A_j  \dfrac
		{\{1-Y_j(0)\}\{1-C_j(0)\}}
		{\{1-\lambda^C_j(0,1,\boldsymbol{X},\bar{Z}_j)\}} \bigg\}&\\
		\label{iter}
		=&E
		\bigg[ Y^1_j(k)\{1-Y^1_j(k-1)\} \{1-Y_j(0)\}  E_j
		&\\\nonumber&
		E\bigg\{	\dfrac
		{A_j \{1-C_j(0)\}}
		{\{1-\lambda^C_j(0,1,\boldsymbol{X},0)\}} \bigg| Y^1_j(k),Y^1_j(k-1), \boldsymbol{X} , E_j
		\bigg\} \bigg]\\
		\label{theBayes}
		=&E
		\bigg[ Y^1_j(k)\{1-Y^1_j(k-1)\} \{1-Y_j(0)\} E_j \\\nonumber
		&
		\dfrac
		{E\{1-C_j(0)\mid A_j=1,Y^1_j(k),Y^1_j(k-1), \boldsymbol{X}, E_j \}  P\{A_j=1 \mid Y^1_j(k),Y^1_j(k-1), \boldsymbol{X}, E_j \}}
		{\{1-\lambda^C_j(0,1,\boldsymbol{X},0)\}}  \bigg]\\
		\label{whenTheMusicsOver}
		=&E[Y^1_j(k)\{1-Y^1_j(k)\} \{1-Y_j(0)\} E_j ]\\
		\label{theEnd}
		=&E[Y^1_j(k)\{1-Y^1_j(k)\} \mid E_j =1] P(E_j=1),
	\end{flalign}
	as desired, where line (\ref{beConsistent}) follows from Assumption~\ref{consistency}, (\ref{lemma1}) holds by Lemma~\ref{firstLemma}, and (\ref{repeatedly}) holds by applying Lemma~\ref{secondLemma} repeatedly for $m=k-1,...,2$.  
	Line (\ref{identical}) follows under Assumptions~\ref{exchDropout} and~\ref{positivityDropout} (as reasoned in the first three steps of the proof of Lemma~\ref{firstLemma}),  
	(\ref{iter}) holds by iterated expectation and by noting that $E_j=1$ implies $\overline{Z}_j=0$, and (\ref{theBayes}) holds by Bayes rule and Assumption~\ref{positivityTx}.  
	Line (\ref{whenTheMusicsOver}) holds because $E\{1-C_j(0)\mid A_j=1,\overline{Z}_j=0,\cdot \}=1$ by definition of $C_j(k)$ and $A_j$ and because, under Assumption~\ref{exchTx}, $P\{A_j=1 \mid Y^1_j(k)\{1-Y^1_j(k)\}, \boldsymbol{X}, E_j=1 \}=P\{A_j=1 \mid  \boldsymbol{X}, E_j=1\}=1-\lambda^C_j(0,1,\boldsymbol{X},0)$.  
	Line (\ref{theEnd}) holds because, given $E_j=1, Y_j(0)=0$.  
	Proofs for (i) with $a=0$ and (ii) with $a=0,1$ are analogous. 
\end{proof}

\begin{corollary}
	\label{IDcorollary}
	The estimand $VE_j(k)$ has the following IP-weighted g-formula representation:   
	$${VE}_j(k)
	=1- \dfrac
	{1-\prod_{m=1}^k\big(1-[E\{ Y_j(m) W_j(m,1)\}/E\{ W_j(m,1)\}]\big)} 
	{1-\prod_{m=1}^k\big(1-[E\{ Y_j(m) W_j(m,0)\}/E\{ W_j(m,0)\}]\big)} 
	$$
\end{corollary}

\begin{proof}
	\begin{equation*}
		{VE}_j(k)
		=1- \dfrac{1-\prod_{m=1}^k\{1-{\lambda}_j^{1}(m)\}}{1-\prod_{m=1}^k\{1-{\lambda}_j^{0}(m)\}},
	\end{equation*}
	where
	\begin{align}
		\lambda_j^{a}(k)
		&=\nonumber
		\dfrac
		{E[Y^{a}_j(k)\{1- Y^{a}_j(k-1) \mid E_j=1\} ]}
		{E\{1-Y^{a}_j(k-1)\mid E_j=1 \}}
	\end{align}
	for $a \in \{0,1\}$.  
	The result follows immediately from Lemma~\ref{thirdLemma}.  
\end{proof}

\subsection{Vaccine Uptake Process in the Analytic Cohort}
\label{VUmethod}
Construction of the IP weight $\widehat{W}_j(k,a)$ requires estimation of the hazards of censoring due to nonadherence and loss to follow up.  
In this manuscript, the hazard of censoring due to nonadherence is estimated by modeling vaccine uptake (VU) as a discrete-time stochastic process.  
Particularly, the hazards of nonadherence to $A_j=0$ are derived from the model-based probability of receiving a first COVID-19 vaccine dose.  	
Nonadherence to $A_j=1$ is defined in terms of the active vaccine regimen of interest, as specified in the target trial protocol.  
For a given active regimen, the hazards of nonadherence to $A_j=1$ are derived from model-based probabilities of receiving additional vaccine doses.  

The remainder of Section~\ref{VUmethod} develops the background for modeling the VU process.  
The model is constructed such that the VU process ends whenever an individual's current state implies perpetual nonadherence to the specified active regimen $A_j=1$.  
Consider a discrete-time stochastic process $\{Z_l : {l \in \mathcal{L}}\}$, where $\mathcal{L}$ is the set of all calendar time points in the study period.  
The state space $\{0, 1, ..., n_d\}$ corresponds to the number of COVID-19 vaccine doses received by calendar time $l$.  
Clearly, the states must be traversed sequentially and it is impossible to move backwards through the states.  
For any $(j,k) \in \mathcal{W}$, the transition probabilities are given by  
$p^{z^\prime,z}_{j+k}(\boldsymbol{x})$ where 
\begin{equation}
	\label{p}
	p^{z^\prime, z}_l(\boldsymbol{x})=P(Z_{l}=z \mid  {Z}_{l-1}=z^\prime, 
	{T^* \geq l, C^*_{l-2}=0},\boldsymbol{X}=\boldsymbol{x})
\end{equation}
is the conditional probability of transitioning from state $z^\prime$ to state $z$ at calendar time $l$ among individuals in the analytic cohort at risk for the outcome in at least one trial at calendar time $l$, $C^*_l=1-I({\overline{Z}_{l+1}}=\overline{0}) -I\{{\overline{Z}_{l+1}} \in \cup_{j \in \{0,1,..., M\}} {\mathcal{Z}^1_{j}(l-j+1)}\}$, and $M=\min(l,J)$.  
I.e., $C^*_l$ takes value $0$ if an individual's treatment history through calendar time $l+1$ is consistent with treatment strategies $A_j=0$ or $A_j=1$ for at least one $j \in \{0, 1, ..., M\}$ and takes value $1$ otherwise.

\subsubsection{Vaccine uptake model for one-dose regimens}
\label{VUmethodOneDose}
Consider a VU model for settings with a single, one-dose active vaccine regimen.  
For example, suppose regimen $A_j=1$ is defined as ``receive a single dose of Moderna vaccine at time zero and receive no further COVID-19 vaccine doses.''   
For this regimen, $n_d=2$ and the state space for the $Z-$process is $\{0,1,2\}$.  
States 0 and 1 are transient states, and state 2 is a terminal state because receipt of a second dose ensures perpetual nonadherence with strategy $A_j=1$.  
Consider the following first-order Markov assumption:
\begin{align}
	\label{firstOrderMarkov}
	&P(Z_{l}=z \mid {Z}_{l-1}=z^\prime, \overline{Z}_{l-2}=\overline{z}, \boldsymbol{X}=\boldsymbol{x}, R^Z_l=1)\\\nonumber
	&=
	P(Z_{l}=z \mid  {Z}_{l-1}=z^\prime, \boldsymbol{X}=\boldsymbol{x}, R^Z_l=1),
\end{align}
where $R^Z_l=I({T^* \geq l,C^*_{l-2}=0})$ is the ``at risk'' indicator for transitioning through the $Z$ process.  
In words, (\ref{firstOrderMarkov}) supposes that the probability of transitioning to state $z$ at calendar time $l$ depends at most on the state inhabited at time $l-1$ and covariates.  
Letting $D_l=Z_{l}-Z_{l-1}$, under Assumption (\ref{firstOrderMarkov}), one could posit the model
\begin{align}
	\label{oneDoseModel}
	g\{P(D_l=1 \mid Z_{l-1}=z, \boldsymbol{X}=\boldsymbol{x}, R^Z_l=1)\}
	=\boldsymbol{\kappa_0}\boldsymbol{f_{\kappa_0}}(l, \boldsymbol{x})+
	I(z=1)\boldsymbol{\kappa_1}\boldsymbol{f_{\kappa_1}}(l, \boldsymbol{x}),
\end{align}
where $\boldsymbol{f_{\kappa_0}}$ and $\boldsymbol{f_{\kappa_1}}$ are user-defined functions of calendar time and covariates.  
Model (\ref{oneDoseModel}) yields transition probabilities
\begin{align*}
	p_l^{0,0}(\boldsymbol{x}) &=1-p_l^{0,1}(\boldsymbol{x})\\
	p_l^{0,1}(\boldsymbol{x})&=g^{-1}\{\boldsymbol{\kappa_0}\boldsymbol{f_{\kappa_0}}(l, \boldsymbol{x})\}, \\
	p_l^{1,1}(\boldsymbol{x}) &=1-p_l^{1,2}(\boldsymbol{x})\\
	p_l^{1,2}(\boldsymbol{x})&=g^{-1}\{\boldsymbol{\kappa_0}\boldsymbol{f_{\kappa_0}}(l, \boldsymbol{x})+\boldsymbol{\kappa_1}\boldsymbol{f_{\kappa_1}}(l, k, \boldsymbol{x})\}.
\end{align*}
The trial-$j$ hazards of nonadherence are given by
\begin{align*}
	\lambda^C_j(k, 0,\boldsymbol{x},\overline{z})&=p^{0,1}_{j+k+1}(\boldsymbol{x}) , &{ \text{ for } k \in \{0,1,...,K_j-1\},} \\
	\lambda^C_j(0, 1,\boldsymbol{x},\overline{z})&{=1-p^{0,1}_{j+1}(\boldsymbol{x})},&\text{ and }\\
	\lambda^C_j(k, 1,\boldsymbol{x},\overline{z})&=p^{1,2}_{j+k+1}(\boldsymbol{x}) ,&{ \text{ for } k \in \{1,2,...,K_j-1\}.}
\end{align*}

\subsubsection{Vaccine uptake model for two-dose regimens with grace periods}
\label{VUmethodTwoDose}
Now consider the setting with a two-dose active vaccine regimen, e.g., ``receive two doses of Novavax vaccine, the first at time {1}, the second 3-8 weeks later, and receive no further COVID-19 vaccine doses.''  
The proposed VU model for this regimen involves a state space $\{0,1,2,3\}$, where 0, 1, and 2 are transient states and 3 is a terminal state (because receiving a third dose implies perpetual nonadherence to $A_j=1$).  
Let $k^{min}$ and $k^{max}$ be constants representing the times at which the grace period for the second dose begins and ends, respectively, measured in weeks since receipt of the first COVID-19 vaccine dose.  
For the Novavax regimen specified above, $k^{min}=3$ and $k^{max}=8$.  

When the active regimen involves a grace period for receipt of dose $z+1$, the first order Markov assumption (\ref{firstOrderMarkov}) may not hold because transition probabilities at calendar time $l$ may depend not only on the current state and covariates but additionally on the time spent in state $z^\prime$.  
Let $V_0, ..., V_{n_d}$ represent the jump times in the $Z$ process.  
Take $V_0= 0$, and note that, for $z \in \{1, 2, ..., n_d\}$, $V_z$ corresponds to the calendar time at which the $z$th dose is received.  
By convention, let $V_z=\infty$ if the person never receives a $z$th dose while on-study.  
In this setting, VU can be modeled under the following semi-Markovian assumption
\begin{align}
	\label{semiMarkov}
	&	P\{Z_l=z, V_z-V_{z^\prime} =k\mid {Z}_{l-1}=z^\prime, \overline{Z}_{l-2}=\overline{z}, V_0, ..., V_{z^\prime}, \boldsymbol{X}=\boldsymbol{x}, R^Z_l=1\}=\\\nonumber
	&	P\{Z_l=z, V_z-V_{z^\prime} =k\mid {Z}_{l-1}=z^\prime, \boldsymbol{X}=\boldsymbol{x}, R^Z_l=1\}.
\end{align}
In words, (\ref{semiMarkov}) supposes that the probability of transitioning to state $z$ $k$ weeks after entering state $z^\prime$ depends at most on the present state and covariates.  

Let
\begin{align}
	\label{q}
	&q_l^{z^\prime,z}(k\mid \boldsymbol{x})
	=P\{Z_l=z, V_z-V_{z^\prime} =k
	\mid {Z}_{l-1}=z^\prime, \boldsymbol{X}=\boldsymbol{x}, R^Z_l=1\}
\end{align}
denote the probability of transitioning to state $z$ $k$ weeks after entering state $z^\prime.$  
Under two-dose regimens with a grace period for the second dose, transitions from states $z=0, 2,3$ do not depend on sojourn time in state $z$ because there is no grace period associated with those states, i.e., $q_l^{z^\prime,z}(k\mid \boldsymbol{x})=p_l^{z^\prime,z}(\boldsymbol{x})$ for transitions $z^\prime \rightarrow z \in \{0 \rightarrow0, 0 \rightarrow1, 2 \rightarrow2, 2\rightarrow3, 3\rightarrow3\}$.  
Under Assumption~(\ref{semiMarkov}), one could posit the model
\begin{align}
	\label{2doseLRmodel}
	&g\{P(D_l=1 \mid Z_{l-1}=z,\boldsymbol{X}=\boldsymbol{x}, R^Z_l=1)\}=\\\nonumber
	&\boldsymbol{\kappa_0}\boldsymbol{f_{\kappa_0}}(l, \boldsymbol{x})+
	I(z=1)\boldsymbol{\kappa_1}\boldsymbol{f_{\kappa_1}}(l, k, \boldsymbol{x})+
	I(z=2)\boldsymbol{\kappa_2}\boldsymbol{f_{\kappa_2}}(l, \boldsymbol{x})
\end{align}
where $\boldsymbol{f_{\kappa_0}}$ and $\boldsymbol{f_{\kappa_2}}$ are user-defined functions of calendar time and covariates and $\boldsymbol{f_{\kappa_1}} $ is a user-defined function of calendar time, sojourn time in state 1 (i.e., $k$), and covariates.  
Under model (\ref{2doseLRmodel}),
\begin{align*}
	p_l^{0,0}(\boldsymbol{x}) &=1-p_l^{0,1}(\boldsymbol{x}),\\
	p_l^{0,1}(\boldsymbol{x})&=g^{-1}\{\boldsymbol{\kappa_0}\boldsymbol{f_{\kappa_0}}(l, \boldsymbol{x})\}, \\
	q_l^{1,1}(k\mid\boldsymbol{x}) &=1-q_l^{1,2}(k\mid\boldsymbol{x}),\\
	q_l^{1,2}(k\mid\boldsymbol{x})&=g^{-1}\{\boldsymbol{\kappa_0}\boldsymbol{f_{\kappa_0}}(l, \boldsymbol{x})+
	\boldsymbol{\kappa_1}\boldsymbol{f_{\kappa_1}}(l, k, \boldsymbol{x})\},\\
	p_l^{2,2}(\boldsymbol{x})&=1-p_l^{2,3}(\boldsymbol{x}),\\
	p_l^{2,3}(\boldsymbol{x})&=g^{-1}\{\boldsymbol{\kappa_0}\boldsymbol{f_{\kappa_0}}(l, \boldsymbol{x})+
	\boldsymbol{\kappa_2}\boldsymbol{f_{\kappa_1}}(l, \boldsymbol{x})\}.
\end{align*}
The trial-$j$ hazards of nonadherence are given by 
\begin{align*}
	&\lambda^C_j(k, 0,\boldsymbol{x},\overline{z})= p^{0,1}_{j+k+1}(\boldsymbol{x}) ,& \text{ for } { k \in \{0,1,...,K_j-1\}}, \\
	&\lambda^C_j(0, 1,\boldsymbol{x},\overline{z})=1-p^{0,1}_{j+1}(\boldsymbol{x}),& 
\end{align*}
and for $k \in \{1,2,...,K_j\}$,
\begin{align*}
	&\lambda^C_j(k, 1,\boldsymbol{x},\overline{z})=p^{1,2}_{j+k+1}(\boldsymbol{x}) , & \text{ if } Z_{j+k}=1 \text{ and } k<k^{min}, \\
	&\lambda^C_j(k, 1,\boldsymbol{x},\overline{z})=0 ,& \text{ if } Z_{j+k}=1 \text{ and }  k^{min} \leq k < k^{max},\\
	&\lambda^C_j(k, 1,\boldsymbol{x},\overline{z})=1- q^{1,2}_{j+k+1}(k \mid\boldsymbol{x}),& \text{ if } Z_{j+k}=1 \text{ and } k = k^{max}, \\ 
	&\lambda^C_j(k, 1,\boldsymbol{x},\overline{z})=p^{2,3}_{j+k+1}(\boldsymbol{x})  ,& \text{ if } Z_{j+k}=2.
\end{align*}  
The approach described in this subsection could be generalized to model treatment uptake for arbitrary longitudinal interventions delivered in $n_d>2$ installments, (possibly) with grace periods for each installment beyond the first.  

\subsubsection{Vaccine uptake model for recommended vaccine schedules}
\label{VUmethodRVS}
Now consider modeling VU under a recommended vaccine schedule that encompasses multiple vaccine brands, each of which may have a brand-specific grace period for receipt of the second dose.  
For example, suppose regimen $A_j=1$ is specified as ``receive (i) a single dose of Moderna or Pfizer-BioNTech at time zero or (ii) two doses of Novavax the first at time 0, the second 3-8 weeks later; and then receive no further COVID-19 vaccine doses'' \citep{CDCschedule}.  
Without loss of generality, suppose that, of the $n_v$ brands in the recommended vaccine schedule, brands $1, 2, ..., n^0_{v}$ are one-dose regimens and brands $n^0_{v}+1, ..., n_{v}$ are two-dose regimens with brand-specific grace periods for receipt of the second dose.  
For simplicity assume, for all individuals who receive a first dose of a two-dose regimen and a second dose within the recommended grace period, that the brand of the second dose matches the brand of the first dose.  
One approach for modeling VU in this setting might involve specifying a separate VU model for each of the $n_v$ brands according to the methods described in Sections~\ref{VUmethodOneDose} and~\ref{VUmethodTwoDose}.  
Alternatively, a single VU model could be specified in which certain parameters are brand-specific and other parameters are common to all brands.  

In this setting, consider the following semi-Markovian assumption: 
\begin{align}
	\label{semiMarkov2}
	&	P\{Z_l=z, V_z-V_{z^\prime} =k\mid {Z}_{l-1}=z^\prime, \overline{Z}_{l-2}=\overline{z}, V_0, ..., V_{z^\prime}, \boldsymbol{X}=\boldsymbol{x}, B_{V_{z^\prime}}=b, R^Z_l=1\}=\nonumber\\
	&	P\{Z_l=z, V_z-V_{z^\prime} =k\mid {Z}_{l-1}=z^\prime, \boldsymbol{X}=\boldsymbol{x}, B_{V_{z^\prime}}=b, R^Z_l=1\}.
\end{align}
Assumption~(\ref{semiMarkov2}) is identical to (\ref{semiMarkov}), except $B_{V_{z^\prime}}=b$ is included in the conditioning event.  
Similarly, define $p_l^{z^\prime,z}(\boldsymbol{x}, b)$ and $q_l^{z^\prime,z}(k\mid\boldsymbol{x}, b)$ according to (\ref{p}) and (\ref{q}), respectively, with the conditioning event $\boldsymbol{X}=\boldsymbol{x}$ replaced by $(\boldsymbol{X}=\boldsymbol{x}, B_{V_{z^\prime}}=b)$.  
Under Assumption~(\ref{semiMarkov2}), the following model could be specified: 
\begin{align}
	\label{RVSLRmodel}
	&g\{P(D_l=1 \mid Z_{l-1}=z, \boldsymbol{X}=\boldsymbol{x}, B_{V_z}=b, R^Z_l=1 )\}=\\\nonumber
	&\boldsymbol{\kappa_0}\boldsymbol{f_{\kappa_0}}(l, \boldsymbol{x})+
	I(z=1, b \in \{1,...,n^0_{v}\})\boldsymbol{\kappa_1}\boldsymbol{f_{\kappa_1}}(l, \boldsymbol{x})+
	\\\nonumber
	&\sum_{b^\prime=n^0_{v}+1}^{n_v}I(z=1, b=b^\prime)\boldsymbol{\kappa_{b^\prime}}\boldsymbol{f_{\kappa_{b^\prime}}}(l, k, \boldsymbol{x})+
	I(z=2, b \in \{n^0_{v}+1, ..., n_v\})\boldsymbol{\kappa_2}\boldsymbol{f_{\kappa_2}}(l, \boldsymbol{x}),
\end{align}
where $\boldsymbol{f_{\kappa_0}}, \boldsymbol{f_{\kappa_1}},$ and $\boldsymbol{f_{\kappa_2}}$ are user-specified functions of calendar time and covariates and $\boldsymbol{f_{\kappa_b}}$ for $b \in \{n^0_{v}+1, ..., n_v\}$ are brand-$b$-specific, user-defined functions of calendar time, sojourn time in state 1, and covariates.  
Under model~(\ref{RVSLRmodel}), 
\begin{align*}
	p_l^{0,0}(\boldsymbol{x},b) &=1-p_l^{0,1}(\boldsymbol{x,b}),\\	
	p_l^{0,1}(\boldsymbol{x},b)&=g^{-1}\{\boldsymbol{\kappa_0}\boldsymbol{f_{\kappa_0}}(l, \boldsymbol{x})\}, \\
	q_l^{1,1}(k\mid\boldsymbol{x},b) &=1-q_l^{1,2}(k\mid\boldsymbol{x},b),\\
	q_l^{1,2}(k\mid\boldsymbol{x},b)&=
	\begin{cases}
		g^{-1}\{\boldsymbol{\kappa_0}\boldsymbol{f_{\kappa_0}}(l, \boldsymbol{x})+
		\boldsymbol{\kappa_1}\boldsymbol{f_{\kappa_1}}(l, \boldsymbol{x})\} ,
		\hspace{3.4cm} b \in \{1,...,n^0_{v}\}\\\nonumber
		g^{-1}\{\boldsymbol{\kappa_0}\boldsymbol{f_{\kappa_0}}(l, \boldsymbol{x})+
		\boldsymbol{\kappa_{b}}\boldsymbol{f_{\kappa_{b}}}(l, k,\boldsymbol{x})\},
		\hspace{3cm} b \in \{n^0_{v}+1, ..., n_v\}
	\end{cases}\\
	p_l^{2,2}(\boldsymbol{x},b)&=1-p_l^{2,3}(\boldsymbol{x},b), \hspace{7.03cm} b \in \{n^0_{v}+1, ..., n_v\}\\
	p_l^{2,3}(\boldsymbol{x},b)&=g^{-1}\{\boldsymbol{\kappa_0}\boldsymbol{f_{\kappa_0}}(l, \boldsymbol{x})+
	\boldsymbol{\kappa_2}\boldsymbol{f_{\kappa_2}}(l, \boldsymbol{x})\}, \hspace{3.75cm} b \in \{n^0_{v}+1, ..., n_v\}.
\end{align*}
The trial-$j$ hazards of nonadherence are given by 
\begin{align*}
	&\lambda^C_j(k, 0,\boldsymbol{x},\overline{z},b)=p^{0,1}_{j+k+1}(\boldsymbol{x},b) ,& \text{ for } k \in \{0,1,...,K_j-1\}, \\
	&\lambda^C_j(0, 1,\boldsymbol{x},\overline{z},b)=1-p^{0,1}_{j+1}(\boldsymbol{x},b),& 
\end{align*}
and for $k \in \{1,2,...,K_j\}$,
\begin{align*}
	&\lambda^C_j(k, 1,\boldsymbol{x},\overline{z},b)=p^{1,2}_{j+k+1}(\boldsymbol{x},b) ,& \text{ if } Z_{j+k}=1 \text{ and } k<k^{min}_b, \\
	&\lambda^C_j(k, 1,\boldsymbol{x},\overline{z},b)=0 ,& \text{ if } Z_{j+k}=1 \text{ and }  k^{min}_b \leq k < k^{max}_b,\\
	&\lambda^C_j(k, 1,\boldsymbol{x},\overline{z},b)=1-q^{1,2}_{j+k+1}(k\mid\boldsymbol{x},b),& \text{ if } Z_{j+k}=1 \text{ and } k = k^{max}_b, \\ 
	&\lambda^C_j(k, 1,\boldsymbol{x},\overline{z},b)=p^{2,3}_{j+k+1}(\boldsymbol{x},b)  ,& \text{ if } Z_{j+k}=2
\end{align*}
where the argument $b$ is added to $\lambda^C_j$ to emphasize dependence on brand of last vaccine dose and $k^{min}_b$ and $k^{max}_b$ represent the time at which the brand-$b$ grace period begins and ends, respectively, measured in weeks since receipt of the first COVID-19 vaccine dose.  

More parsimonious model specifications could be considered.  
For example, it may be reasonable to assume that the probability of receiving additional doses after completion of a one- or two-dose regimen does not depend on vaccine brand, i.e., that the terms $I(z=1, b \in \{1,...,n^0_{v}\})\boldsymbol{\kappa_1}\boldsymbol{f_{\kappa_1}}(l, \boldsymbol{x})$ and $I(z=2,b \in \{n^0_{v}+1, ..., n_{v}\})\boldsymbol{\kappa_2}\boldsymbol{f_{\kappa_2}}(l, \boldsymbol{x})$ in~(\ref{RVSLRmodel}) could be replaced by a single term $\{I(z=1, b \in \{1,...,n^0_{v}\}) + I(z=2)\}\boldsymbol{\kappa_1}\boldsymbol{f_{\kappa_1}}(l, \boldsymbol{x})$.  
Also, the methods described above could be adapted for interventions delivered in $n_d>2$ installments, possibly with grace periods for each installment.  

\subsection{Inverse probability weight estimation}
\label{IPweightEstimation}
Construction of the IP weight $\widehat{W}_j(k,a)$ requires estimation of the hazards of censoring due to nonadherence and loss to follow up.  
Consider the following flexible parametric model for the loss to follow up hazard:
\begin{equation}
	\label{eq:IPCmodel}
	g\{\lambda^H_{j}(k,\boldsymbol{x}, \overline{z}; \boldsymbol{\xi})\}=
	\boldsymbol{\xi}\boldsymbol{f_{\xi}}(j+k, \boldsymbol{x}, \overline{z})
\end{equation}
where $\boldsymbol{\xi}$ is a vector of unknown regression parameters and $\boldsymbol{f_{\xi}}(l, \boldsymbol{x}, \overline{z})$ is a user-specified column vector containing functions of calendar time, covariates, and vaccine dose history.  
Model (\ref{eq:IPCmodel}) is fitted using the analytic cohort dataset.  
Particularly, the parameters of (\ref{eq:IPCmodel}) are estimated via maximum likelihood (ML) to obtain the estimator $\boldsymbol{\hat\xi}$, which solves the vector estimating equation $\sum_{i=1}^n\boldsymbol{\psi_{\boldsymbol{\xi}}}(\boldsymbol{O_i}; \boldsymbol{\xi})=\boldsymbol{0}$, where 
$$
\boldsymbol{\psi_{\xi}}(\boldsymbol{O};\boldsymbol{\xi})=
\sum_{l=1}^{\tau}
R^H_l
\big[H_l-
g^{-1}
\big\{
\boldsymbol{\xi}\boldsymbol{f_{\xi}}(l, \boldsymbol{X}, \overline{Z})
\big\}
\big]
\boldsymbol{f_{\xi}}(l, \boldsymbol{X}, \overline{Z})\\
$$
and $R^H_l=I({T^* \geq l,C^*_{l-1}=0})$ is the indicator of being ``at risk'' for loss to follow up at calendar time $l$.  

Similarly, parameters of the VU model are estimated using the analytic cohort dataset.  
The ML estimator $\boldsymbol{\hat\kappa}$ solves the vector estimating equation $\sum_{i=1}^n\boldsymbol{\psi_{\boldsymbol{\kappa}}}(\boldsymbol{O_i}; \boldsymbol{\kappa})=\boldsymbol{0}$, where 
$$
\boldsymbol{\psi_{\kappa}}(\boldsymbol{O};\boldsymbol{\kappa})=
\sum_{l=1}^{\tau}
\sum_{k=1}^{k_{B_{V_1}}^{max}}
R^Z_l
\big[D_l-
g^{-1}
\big\{
\boldsymbol{\kappa}\boldsymbol{f_\kappa}(\boldsymbol{X},Z_{l-1}, B_{V_1}, l, k)
\big\}
\big]
\boldsymbol{f_\kappa}(\boldsymbol{X},Z_{l-1}, B_{V_1}, l, k)
$$ and
$\boldsymbol{\kappa}\boldsymbol{f_\kappa}(\boldsymbol{X}, Z_{l-1}, B_{V_1}, l, k)$ is the linear predictor for the VU model (e.g., the right side of (\ref{RVSLRmodel})).  

\subsection[Asymptotic distribution of rho hat]{Asymptotic distribution of $\boldsymbol{\hat{\rho}}$}
\label{EEs}
Letting $i$ index the set of $n$ individuals in the analytic cohort, $\boldsymbol{\hat{\theta}^*}$ is the solution to $\sum_{i=1}^n\boldsymbol{\psi}(\boldsymbol{O_i}; \boldsymbol{\theta^*})=\boldsymbol{0}$ where
\begin{align*}
	\boldsymbol{\psi}(\boldsymbol{O}; \boldsymbol{\theta^*})&= 
	\begin{pmatrix}
		\boldsymbol{\psi_{\kappa}}(\boldsymbol{O};\boldsymbol{\kappa})\\
		\boldsymbol{\psi_{\xi}}        (\boldsymbol{O};\boldsymbol{\xi})\\
		\boldsymbol{\psi_{\lambda^0 }}(\boldsymbol{O};
		\boldsymbol{\kappa}, \boldsymbol{\xi} ,\boldsymbol{\lambda^0})\\
		\boldsymbol{\psi_{\lambda^1 }}(\boldsymbol{O};
		\boldsymbol{\kappa}, \boldsymbol{\xi} ,\boldsymbol{\lambda^1})\\
		\boldsymbol{\psi_{\rho}}(\boldsymbol{\lambda^0 ,\lambda^1})
	\end{pmatrix} ,\\
	%
	%
	%
	\boldsymbol{\psi_{\lambda^0}}(\boldsymbol{O};\boldsymbol{\kappa}, \boldsymbol{\xi}, \boldsymbol{\lambda^0})
	&=\sum_{j=0}^{J}  \sum_{k=1}^{K_j}  
	W_j(k, 0; \boldsymbol{\kappa}, \boldsymbol{\xi})
	\{Y_j(k)- \lambda_j^{0}(k)\}\\
	%
	%
	%
	\boldsymbol{\psi_{\lambda^1}}(\boldsymbol{O};\boldsymbol{\kappa}, \boldsymbol{\xi}, \boldsymbol{\lambda^1})
	&=\sum_{j=0}^{J}  \sum_{k=1}^{K_j}  
	W_j(k, 1; \boldsymbol{\kappa}, \boldsymbol{\xi})
	\{Y_j(k)- \lambda_j^{1}(k)\}\\
	\boldsymbol{\psi_{\boldsymbol{\rho}}}(\boldsymbol{\lambda^0, \lambda^1})
	&=\begin{pmatrix}
		\boldsymbol{\rho}_0^T-\boldsymbol{\nu}_0(\boldsymbol{\lambda^0_j, \lambda^1_j}) \\
		\boldsymbol{\rho}_1^T-\boldsymbol{\nu}_1(\boldsymbol{\lambda^0_j, \lambda^1_j}) \\
		\vdots \\
		\boldsymbol{\rho}_J^T-\boldsymbol{\nu}_J(\boldsymbol{\lambda^0_j, \lambda^1_j}) \\
	\end{pmatrix},\\
	\boldsymbol{\nu}_j(\boldsymbol{\lambda^0_j}, \boldsymbol{\lambda^1_j})&=\{{\nu}_j(1, \boldsymbol{\lambda^0_j}, \boldsymbol{\lambda^1_j}), {\nu}_j(2, \boldsymbol{\lambda^0_j}, \boldsymbol{\lambda^1_j}), ..., {\nu}_j(K_j, \boldsymbol{\lambda^0_j}, \boldsymbol{\lambda^1_j})\}^T,\\
	\nu_j(k,\boldsymbol{\lambda^0_j}, \boldsymbol{\lambda^1_j})
	&=\log\big[1-\prod_{m=1}^k\{1-\lambda_j^1(m)\}\big]-\log\big[1-\prod_{m=1}^k\{1-\lambda_j^0(m)\}\big],  
\end{align*}  
$\boldsymbol{\lambda^a}=(\boldsymbol{\lambda^a_0}, ..., \boldsymbol{\lambda^a_J})^T$ and $\boldsymbol{\lambda^a_j}=\{\lambda^a_j(1), ..., \lambda^a_j(K_j)\}^T$ for $a \in \{0,1\}$,
and where here and below the convention is adopted that zero times an undefined quantity equals zero.  

To see that the vector estimating equation $\boldsymbol{\psi}(\boldsymbol{O}; \boldsymbol{\theta^*})$ is unbiased, consider the following.  
The score functions for correctly specified generalized linear models are unbiased \citep{McCullaghNelder}, i.e., $E\{\boldsymbol{\psi_\kappa}(\boldsymbol{O}$;$\boldsymbol{\kappa})\}$=$\boldsymbol{0}$ and  $E\{\boldsymbol{\psi_\xi}(\boldsymbol{O}$;$\boldsymbol{\xi})\}$=$\boldsymbol{0}$.  
When the VU model (e.g., \ref{RVSLRmodel}) and the loss-to-follow-up model (\ref{eq:IPCmodel}) are correctly specified, $E\{\boldsymbol{\psi_{\lambda^1}}(\boldsymbol{O}; \boldsymbol{\kappa},\boldsymbol{\xi},\boldsymbol{\lambda^1})\}$ equals
\begin{align}
	E&\bigg[\sum_{j=0}^{J}  \sum_{k=1}^{K_j}  
	W_j(k,1)
	\{Y_j(k)- \lambda_j^{1}(k)\} \bigg]\nonumber\\
	\label{equals}
	=&\sum_{j=0}^{J} \sum_{k=1}^{K_j}  
	E\bigg(  W_j(k,1)
	\bigg[ Y_j(k)- 
	\dfrac{E\{Y_{j}(k)W_j(k,1)\}}
	{E\{ W_j(k,1)\}}
	\bigg]
	\bigg)\\
	=&\sum_{j=0}^{J} \sum_{k=1}^{K_j}  \bigg(
	E \{ W_j(k,1)
	Y_j(k) \}-
	\label{linOfExp}
	E \bigg[W_j(k,1)   \dfrac{E\{Y_{j}(k)W_j(k,1)\}}
	{E\{W_j(k,1)\}}
	\bigg]\bigg)\\
	=&\boldsymbol{0} \nonumber
\end{align}
where (\ref{equals}) holds by linearity of expectation and Corollary~\ref{IDcorollary}, and (\ref{linOfExp}) holds by linearity of expectation.  
The proof for $\boldsymbol{\psi_{\lambda^0}}(\boldsymbol{O}; \boldsymbol{\kappa},\boldsymbol{\xi},\boldsymbol{\lambda^0})$ is analogous.  

Since the estimating equation vector is unbiased, it follows that, under certain regularity conditions \citep{StefanskiAndBoos}, $\sqrt{n}(\boldsymbol{\hat{\theta}^*} - \boldsymbol{{\theta^*_{0}})} \xrightarrow{d} N\{\boldsymbol{0},V(\boldsymbol{{\theta^*_{0}}})\}$ as $n \rightarrow \infty$, where $\boldsymbol{\theta^*_{0}}$ is the true parameter value, $V({\boldsymbol{\theta^*_{0}}})=\mathbb{A}({\boldsymbol{\theta^*_{0}}})^{-1} \mathbb{B}({\boldsymbol{\theta^*_{0}}})\{\mathbb{A}(\boldsymbol{\theta^*_{0}})^{-1}\}^T$,
$\mathbb{A}({\boldsymbol{\theta^*_0}})=E\{-\dot{\boldsymbol{\psi}}(\boldsymbol{O}; \boldsymbol{\theta^*_{0}})\}$,
$\mathbb{B}({\boldsymbol{\theta^*_0}})=E[{\boldsymbol{\psi}}(\boldsymbol{O}; {\boldsymbol{\theta^*_0}}) {\boldsymbol{\psi}}(\boldsymbol{O}; \boldsymbol{\theta^*_{0}})^T]$, and ${\dot{\boldsymbol{\psi}}(\boldsymbol{O}; {\boldsymbol{\theta^*}})=\partial{{\boldsymbol{\psi}}}(\boldsymbol{O}; {\boldsymbol{\theta^*}})}/\partial{\boldsymbol{\theta^*}}$.

The vector of estimating functions can be written in terms of the MSM parameters as follows: 
\begin{align*}
	\boldsymbol{\psi}(\boldsymbol{O}; \boldsymbol{\theta})&= 
	\begin{pmatrix}
		\boldsymbol{\psi_{\kappa}}(\boldsymbol{O};\boldsymbol{\kappa})\\
		\boldsymbol{\psi_{\xi}}        (\boldsymbol{O};\boldsymbol{\xi})\\
		\boldsymbol{\psi_{\alpha }}(\boldsymbol{O};
		\boldsymbol{\kappa}, \boldsymbol{\xi} ,\boldsymbol{\alpha})\\
		\boldsymbol{\psi_{\rho}}(\boldsymbol{\alpha})
	\end{pmatrix} ,\\
	%
	%
	%
	\boldsymbol{\psi_{\alpha}}(\boldsymbol{O};\boldsymbol{\kappa}, \boldsymbol{\xi}, \boldsymbol{\alpha})
	&=\sum_{j=0}^{J}  \sum_{k=1}^{K_j}  
	W_j(k, A_j; \boldsymbol{\kappa}, \boldsymbol{\xi})
	\{Y_j(k)- \lambda_j^{A_j}(k; \boldsymbol{\alpha})\}\boldsymbol{f}_{\boldsymbol{\alpha}}(j, k, A_j),\\
	%
	%
	\boldsymbol{\psi_{\boldsymbol{\rho}}}(\boldsymbol{\alpha})
	&=\begin{pmatrix}
		\boldsymbol{\rho}_0^T-\boldsymbol{\nu}^{*}_0(\boldsymbol{\alpha}) \\
		\boldsymbol{\rho}_1^T-\boldsymbol{\nu}^{*}_1(\boldsymbol{\alpha}) \\
		\vdots \\
		\boldsymbol{\rho}_J^T-\boldsymbol{\nu}^{*}_J(\boldsymbol{\alpha}) \\
	\end{pmatrix},\\
	\boldsymbol{\nu}^{*}_j(\boldsymbol{\alpha})&=\{{\nu}^{*}_j(1, \boldsymbol{\alpha}), {\nu}^{*}_j(2, \boldsymbol{\alpha}), ..., {\nu}^{*}_j(K_j, \boldsymbol{\alpha})\}^T,\\
	\nu^{*}_j(k,\boldsymbol{\alpha})
	&=\log\big[1-\prod_{m=1}^k\{1-\lambda_j^1(m)\}\big]-\log\big[1-\prod_{m=1}^k\{1-\lambda_j^0(m)\}\big],  \\
	\lambda_j^a(k)
	&=g^{-1}\{\boldsymbol{\alpha}\boldsymbol{f}_{\boldsymbol{\alpha}}(j, k, a)\},
\end{align*}
where $\boldsymbol{\hat \theta}=(\boldsymbol{\hat\kappa},\boldsymbol{\hat\xi},\boldsymbol{\hat{\alpha}^\dagger}, \boldsymbol{\hat\rho})$ is the solution to $\sum_{i=1}^n\boldsymbol{\psi}(\boldsymbol{O_i}; \boldsymbol{\kappa},\boldsymbol{\xi},\boldsymbol{\alpha}, \boldsymbol{\rho})=\boldsymbol{0}$.  
When the VU model (e.g., \ref{RVSLRmodel}), the loss-to-follow-up model (\ref{eq:IPCmodel}), and the marginal structural model (MSM) for the hazard of the outcome are all correctly specified, $\boldsymbol{\hat\theta}$ is consistent for $\boldsymbol{\theta_0}$ and asymptotically normal, where $\boldsymbol{\theta_0}$ is the true parameter value.  

\subsection[Asymptotic distribution of beta hat]{Asymptotic distribution of $\hat{\beta}$}
\label{TEHees}
Estimating equations for $\boldsymbol{\hat{\theta}^{\dagger}}=( \boldsymbol{\hat\kappa}, \boldsymbol{\hat\xi}, \boldsymbol{\hat{\alpha}^*}, \hat\beta)$ are given by
\begin{flalign*}
	&&\boldsymbol{\psi}(\boldsymbol{O};\boldsymbol{\theta^{\dagger}})&=
	\begin{pmatrix}
		\boldsymbol{\psi_\kappa}(\boldsymbol{O};\boldsymbol{\kappa}) \\
		\boldsymbol{\psi_\xi}   (\boldsymbol{O};\boldsymbol{\xi}) \\
		\boldsymbol{\psi_\alpha}(\boldsymbol{O};\boldsymbol{\kappa},\boldsymbol{\xi}, \boldsymbol{\alpha}) \\
		\psi_\beta{({\boldsymbol\alpha})}
	\end{pmatrix},& \\
	\text{where} &&&&\\
	&&\psi_\beta{\boldsymbol({\boldsymbol\alpha})}&
	=\beta-\dfrac{\sum_{j=0}^J(j  - \bar{j}) ({AUC}_j-\overline{AUC}_j )}{ \sum_{j=0}^J (j  - \bar{j})^2},&
\end{flalign*}
the overbar denotes an average (taken across $j$), ${AUC}_j = \sum_{k=1}^{K_J} VE_j(k)$; $VE_j(k)$ is a function of $\{\lambda_j^0(k), \lambda_j^1(k)\}$ given by (\ref{identity}) in the main text, and $\{\lambda_j^0(k),\lambda_j^1(k)\}$ depend on $\boldsymbol{\alpha}$ through the specified MSM for the hazard of the outcome.
It follows from Section~\ref{EEs} that $E\{\boldsymbol{\psi_\kappa}(\boldsymbol{O};\boldsymbol{\kappa}) \}$=$\boldsymbol{0}$, 
$E\{\boldsymbol{\psi_\xi}(\boldsymbol{O};\boldsymbol{\xi}) \}$=$\boldsymbol{0}$, and 
$E\{\boldsymbol{\psi_\alpha}(\boldsymbol{O};\boldsymbol{\kappa},\boldsymbol{\xi},\boldsymbol{\alpha})  \}$=$\boldsymbol{0}$ under $H_{0}$, provided that the corresponding models are correctly specified.  
To see that $E\{\psi_\beta(\boldsymbol\alpha)\}=0$, note that under $H_{0}$, $\beta=0$ and $\overline{AUC}_j={AUC}_j$ for $j \in \{0,1,...,J\}$.  
Since $\boldsymbol{\hat{\theta}^{\dagger}}$ is the solution to an unbiased estimating equation, under certain regularity conditions, $\sqrt{n}(\boldsymbol{\hat{\theta}^{\dagger} - \boldsymbol{{\theta}^{\dagger}_{0})}} \xrightarrow{d} N\{\boldsymbol{0},V(\boldsymbol{{\theta}_{0}^{\dagger}})\}$ as $n \rightarrow \infty$, where $\boldsymbol{\theta^{\dagger}}=( \boldsymbol{\kappa}, \boldsymbol{\xi}, \boldsymbol{\alpha}, \beta)$ and $\boldsymbol{\theta_{0}^{\dagger}}$ is the true parameter value under $H_{0}$.  
The empirical sandwich estimator $\hat{V}_n(\boldsymbol{\hat{\theta}^{\dagger}})$ is a consistent estimator for $V(\boldsymbol{{\theta}_{0}^{\dagger}})$.  

\subsection{Constructing the NTE dataset: A worked example}
\label{NTEdsConstruction}
	This section provides a small worked example illustrating construction of the NTE analysis dataset.  
	As described in the main text, the NTE analysis dataset is derived according to specifications in the target trial protocol.  
	The worked example in this section is based on the target trial protocol for the applied Abruzzo data analysis (see Table~\ref{protocol_p1} below).  
	To keep the example small, suppose the goal is to emulate $J+1=3$ trials and there are $\tau=4$ time points of follow-up.  
	As in \ref{dataAnalysisDetails} below, the vaccine dose variable $B_l$ is coded as 0) no vaccine dose, 1) Pfizer-BioNTech, 2) Moderna, 3) Oxford-AstraZeneca, 4) Janssen.  
	
	Table~\ref{rawData} displays observed data for a small, hypothetical analytic cohort.   
	Recall that 
	each individual in the analytic cohort contributes one set of longitudinal measurements to the NTE analysis dataset per trial for which they are eligible.  
	The corresponding NTE dataset for the hypothetical analytic cohort in Table~\ref{rawData} is displayed in Table~\ref{NTEdata}. 

\begin{table}
	\centering  \renewcommand{\arraystretch}{1}
	\caption{Example analytic cohort dataset.  
		ID is an individual's unique numerical identifier;
		$l$ represents calendar time and
		$T^*$ is the (possibly right-censored) observation time, both measured in weeks since February 15, 2021;
		$\Delta$ is the indicator for an observed event; 
		$\boldsymbol{X}$ are baseline covariates (measured on February 15, 2021); 
		$B_l$ is the brand of vaccine dose received during calendar week $l$; 
		and $E_l$ is the eligibility indicator for trials $l=0,1,2$.
		The solid lines separate data for individuals.}
	\label{rawData}
	\begin{tabular}{ccccccc}
		\hline
		ID &$l$ & $T^*$ & $\Delta$ &$\boldsymbol{X}$   &  $B_l$  &  $E_l$\\ 
		\hline
		1 & 0      & 4 &1 &$\boldsymbol{x}$ &0 & 1\\
		1 & 1      & 4 &1 &$\boldsymbol{x}$ &0 & 1\\
		1 & 2      & 4 &1 &$\boldsymbol{x}$ &0 & 1\\
		1 & 3      & 4 &1 &$\boldsymbol{x}$ &1 & -\\
		1 & 4      & 4 &1 &$\boldsymbol{x}$ &0 & -\\
		\hline
		2 & 0      & 5 &0 &$\boldsymbol{x}$ &0 & 1\\
		2 & 1      & 5 &0 &$\boldsymbol{x}$ &1 & 0\\
		2 & 2      & 5 &0 &$\boldsymbol{x}$ &0 & 0\\
		2 & 3      & 5 &0 &$\boldsymbol{x}$ &0 & -\\
		2 & 4      & 5 &0 &$\boldsymbol{x}$ &3 & -\\
		\hline
		3 & 0      & 2 &1 &$\boldsymbol{x}$ &0 & 1\\
		3 & 1      & 2 &1 &$\boldsymbol{x}$ &0 & 1 \\
		3 & 2      & 2 &1 &$\boldsymbol{x}$ &0 & 0\\
		3 & 3      & 2 &1 &$\boldsymbol{x}$ &0 & -\\
		3 & 4      & 2 &1 &$\boldsymbol{x}$ &0 & -\\
		\hline
	\end{tabular}	
\end{table}

	Individual 1 gets a first COVID-19 vaccine dose (Pfizer-BioNTech) at calendar time $l=3$ (i.e., $\overline{B}_2=0$, $B_3=1$) and experiences the event at calendar time $l=4$ (i.e., $T^*=4, \Delta=1$).  
	Therefore, they are ``enrolled'' in trials 0, 1, and 2 in treatment groups $A_0=0$, $A_1=0$, and $A_2=1$, respectively.  
	Individual 1 has eligibility indicators $\overline{E}_2=(1, 1, 1)$, and in turn, they contribute longitudinal measurements to the NTE analysis dataset for trials 1, 2, and 3 (see Table~\ref{NTEdata}).
	Their trial-0 record is artificially censored at trial time $k=2$ because the individual ceases to follow their ``assigned" trial-0 regimen (``remain unvaccinated for COVID-19 through December 18, 2021'') at calendar time $l=2$.  
	For the same reason, their trial-1 record is artificially censored at trial time $k=1$.  
	Their trial-2 record is at risk at calendar time $l=4$ (when they experience the event), and therefore they contribute an observed event at time $k=2$ of trial 2.
	
	Individual 2 gets a first COVID-19 vaccine dose (Pfizer-BioNTech) at calendar time $l=1$ (i.e., ${B}_0=0$, $B_1=1$), and does not experience the event by calendar time $\tau$ (i.e., $T^*=\tau+1=5, \Delta=0$).  
	Therefore, they are enrolled in trial 0 only, in treatment group $A_0=1$.  
	They are ineligible for subsequent trials, because at calendar times 1 and 2 they meet the exclusion criterion of having any dose of any COVID-19 vaccine prior to enrollment.
	Individual 2 has eligibility indicators $\overline{E}_2=(1, 0, 0)$, and they contribute longitudinal measurements to the NTE analysis dataset for trial 0 only (see Table~\ref{NTEdata}).
	They receive a dose of non-mRNA vaccine (AstraZeneca) at calendar time $l=4$ (i.e., $B_4=3$), so they are artificially censored at trial time $k=3$ due to noncompliance with their assigned (active) trial-0 treatment regimen.
	
	Individual 3 does not receive any COVID-19 vaccine doses through calendar time $\tau=4$, and they experience the event at calendar time $l=2$ (i.e., $T^*=2, \Delta=1$).  
	Therefore, they are enrolled in trials 0 and 1 only, in treatment groups $A_0=0$ and $A_1=0$, respectively.  
	They are ineligible for trial 3 due to meeting the exclusion criterion of having experienced the event prior to enrollment. 
	Individual 3 has eligibility indicators $\overline{E}_2=(1, 1, 0)$, and they contribute longitudinal measurements to the NTE analysis dataset for trials 0 and 1 only (see Table~\ref{NTEdata}).
	Because they do not get vaccinated at any time during the study period, they remain compliant with their assigned (comparator) treatment regimen at all times during trials 0 and 1, and they are never artificially censored.  
	Their trial-0 record is at risk at calendar time $l=2$ (when they experience the event), and therefore they contribute an observed event at time $k=2$ of trial 0.
	Their trial-1 record is also at risk at that same calendar time, and therefore they also contribute an observed event at time $k=1$ of trial 1.  

\begin{table}
	\centering \renewcommand{\arraystretch}{1}
	\caption{Contribution to the NTE analysis dataset for individuals in the hypothetical analytic cohort.  
		ID is an individual's unique numerical identifier;
		$j$, $l$, and $k$ represent trial number, calendar time, and trial time, respectively; 
		$A_j$ represents the trial-specific vaccine regimen assignment; 
		$C_j(k)$ is an indicator for nonadherence to $A_j$ by time $k$ of trial $j$; 
		$Y_j(k)$ is the indicator of an observed event by time $k$ of trial $j$; 
		and $R_j(k)$ is the indicator for being at risk of an observed event at time $k$ of trial $j$. 
		Dashed lines separate data for emulated trials.
		Solid lines separate data for individuals.}
	\label{NTEdata}
	\begin{tabular}{ccccccccc}
		\hline
		ID&$j$ & $l$ & $k$ &  $A_j$ & $C_{j}(k)$ & $Y_j(k)$& $R_j(k)$&$\boldsymbol{X}$\\ 
		\hline
		1& 0 & 0 & 0 &0  & 0 & 0  & 1 & $\boldsymbol{x}$ \\ 
		1&0 & 1 & 1 &0  & 0 & 0  & 1 & $\boldsymbol{x}$ \\ 
		1&0 & 2 & 2 &0  & 1 & 0  & 1 & $\boldsymbol{x}$ \\ 
		1&0 & 3 & 3 &0  & 1 & 0  & 0 & $\boldsymbol{x}$ \\ 
		1&0 & 4 & 4 &0  & 1 & 0  & 0 & $\boldsymbol{x}$\\ 
		\hdashline
		1&1 & 1 & 0 &0  & 0 & 0  & 1 & $\boldsymbol{x}$ \\ 
		1&1 & 2 & 1 &0  & 1 & 0  & 1 & $\boldsymbol{x}$ \\ 
		1&1 & 3 & 2 &0  & 1 & 0  & 0 & $\boldsymbol{x}$\\ 
		1&1 & 4 & 3 &0  & 1 & 0  & 0 & $\boldsymbol{x}$\\ 
		\hdashline
		1&2 & 2 & 0 &1  & 0 & 0  & 1 & $\boldsymbol{x}$ \\ 
		1&2 & 3 & 1 &1  & 0 & 0  & 1 & $\boldsymbol{x}$ \\ 
		1&2 & 4 & 2 &1  & 0 & 1  & 1 & $\boldsymbol{x}$\\ 
		\hline
		2& 0 & 0 & 0 &1  & 0 & 0  & 1 & $\boldsymbol{x}$ \\ 
		2&0 & 1 & 1 &1  & 0 & 0  & 1 & $\boldsymbol{x}$ \\ 
		2&0 & 2 & 2 &1  & 0 & 0  & 1 & $\boldsymbol{x}$ \\ 
		2&0 & 3 & 3 &1  & 1 & 0  & 1 & $\boldsymbol{x}$ \\ 
		2&0 & 4 & 4 &1  & 1 & 0  & 0 & $\boldsymbol{x}$\\ 
		\hline
		3& 0 & 0 & 0 &0  & 0 & 0  & 1 & $\boldsymbol{x}$ \\ 
		3&0 & 1 & 1 &0  & 0 & 0  & 1 & $\boldsymbol{x}$ \\ 
		3&0 & 2 & 2 &0  & 0 & 1  & 1 & $\boldsymbol{x}$ \\ 
		3&0 & 3 & 3 &0  & 0 & 1  & 0 & $\boldsymbol{x}$ \\ 
		3&0 & 4 & 4 &0  & 0 & 1  & 0 & $\boldsymbol{x}$\\ 
		\hdashline
		3&1 & 1 & 0 &0  & 0 & 0  & 1 & $\boldsymbol{x}$ \\ 
		3&1 & 2 & 1 &0  & 0 & 1  & 1 & $\boldsymbol{x}$ \\ 
		3&1 & 3 & 2 &0  & 0 & 1  & 0 & $\boldsymbol{x}$\\ 
		3&1 & 4 & 3 &0  & 0 & 1  & 0 & $\boldsymbol{x}$\\ 
		\hline
	\end{tabular}	
\end{table}

\section{Simulation Studies}
\label{simulations}
\subsection{Unstandardized estimator}
\label{sims_un}
Simulation studies were conducted to examine the finite sample performance of the VE estimator and TEH test discussed in Section \ref{methods} of the main text.  
An observational cohort was simulated with $\tau=20$ time points of follow up and $n=50{,}000$ individuals.  
Motivated by the Abruzzo data, baseline covariates were simulated as follows.  
Age was generated according to $X_1 \sim FN(0, 7) + 80$, where $FN(\text{mean}, \text{standard deviation})$ is the folded normal distribution.  
Sex ($X_2$) and comorbidity status ($X_3$) were generated from $\text{Bern}$[logit$^{-1}\{-0.42 - 0.047 (X_1-\tilde{X_1})\}$] and $\text{Bern}$[logit$^{-1}\{0.44 + 0.009 (X_1-\tilde{X_1}) + 0.37 X_2 \}]$, respectively, where $\text{Bern}(p)$ is the Bernoulli distribution with mean $p$, and $\tilde{X_1}=85.6$ is the population mean of $X_1$.  
Baseline vaccine regimen assignments were generated as $A_0 \sim\text{Bern}\{p_A(0)\}$, where $p_A(l)=$ logit$^{-1}\{-2.64+0.25l-0.022l^2 -0.052 (X_1-\tilde{X_1}) + 0.03 X_2 -0.048 X_3\}$.  
For $l=1, ..., \tau$, vaccine regimen assignments were generated as $A_l \sim \text{Bern}\{p_A(l)\}$ if $A_{l-1}=0$ and $A_{l-1}$ is undefined otherwise.  

A set of potential event times $\{T^{A_l=1}: l \in \mathcal{L}\} \cup \{T^{A_0=0}\}$ was generated for each person in the simulated cohort, where $T^{A_l=1}$ is the potential event time associated with initiating the active vaccine regimen at calendar time $l$, and $T^{A_0=0}$ is the potential event time associated with not initiating the active regimen at any of these time points.  
Potential event times were initialized to $\infty$.  
Then, for each calendar time $l$ in $\{1,..., \tau\}$, if $T^{A_j=1} > l-1$ then $T^{A_j=1}$ was updated according to 
\begin{flalign*}
	&P(T^{A_j=1}=l \mid T^{A_j=1}>l-1, \boldsymbol{X})=
	\text{logit}^{-1}[-4 
	-0.013 (X_1-\tilde{X_1}) -0.26 X_2 + 0.425 X_3
	\\
	& +\alpha_{1}l 	
	+ \alpha_{2}l^2 
	+I(l > j) \{
	-2.5
	+ \alpha_{3}l
	+ \alpha_{4}l^2 
	+ \alpha_{5} (l-j)  
	+ \alpha_{6} (l-j)^2 
	\}]. 
\end{flalign*} 
Potential event time $T^{A_0=0}$ was generated analogously, with true conditional hazard $P(T^{A_0=0}=l \mid T^{A_0=0}>l-1, \boldsymbol{X})=\text{logit}^{-1}\{-4 -0.013 (X_1-\tilde{X_1}) -0.26 X_2 + 0.425 X_3+\alpha_{1}l 	
+ \alpha_{2}l^2 \}$.
Observed event times were computed according to $T = \sum_{l=0}^{\tau} I(A_l=1, A_{l-1}=0) T^{A_l=1} + I(\overline{A}_l=\overline{0})T^{A_0=0}$, where $A_{-1}\equiv0$.  

Three scenarios were considered.  
In scenario 1, $\boldsymbol{\alpha}=(\alpha_1, ..., \alpha_6)=(0,0,0,0, .02, .005)$ so that the true VE was homogeneous across trials and decreased over time since vaccination.  
In scenarios 2 and 3, the true hazard varied over calendar time.  
Particularly, in scenario 2, $\boldsymbol{\alpha}=(-.01, -.003, .02, .006, 0 ,0)$ such that the hazard when vaccinated depended on calendar time but not time since vaccination.
In scenario 3, $\boldsymbol{\alpha}=(-.01, -.003, .02, .006, .02, .005)$ so that the hazard when vaccinated depended on both time scales.  
True values for $VE_j(k)$ for select $j$ and $k$ were approximated by computing 
$$1-\dfrac{\sum_{i=1}^N I(T^{A_j=1}_i\leq j+k, E_{j,i}=1)}{\sum_{i=1}^N I(T^{A_0=0}_i\leq j+k, E_{j,i}=1)}$$ in a very large cohort of $N=10^7$ individuals generated according to the data generating process (DGP) described above.

For each scenario, $3{,}000$ replications were conducted, and $J+1=13$ trials were emulated from each simulated dataset.  
For each trial, individuals were excluded if and only if, prior to enrollment, they (i) initiated active vaccine regimen or (ii) experienced an event.  
Each simulated data set was analyzed using two different MSM specifications.  
For the first analysis, the hazard was modeled according to (\ref{timeVaryingIntercept}) in the main text with $\boldsymbol{f_1}(t)=\boldsymbol{f_2}(t)=t+t^2$, i.e., it was assumed that the hazard when vaccinated did not depend on calendar time but could depend on time since vaccination.  
For the second analysis, the hazard model was specified according to (\ref{calTimeTerms}) in the main text with $\boldsymbol{f_1}(t)=\boldsymbol{f_2}(t)=\boldsymbol{f_3}(t)=t+t^2$, i.e., it was assumed that the hazard when vaccinated could depend on both calendar time and time since vaccination.  
IP weights were estimated using correctly specified models.  
For each analysis, $\widehat{VE}_5(k)$ for select $k$ and $\widehat{VE}_j(5)$ for select $j$ were calculated along with estimated standard errors and corresponding 95\% CIs.  
Additionally, a one-sided generalized Wald test of (\ref{Hnaught}) in the main text was conducted for the model (\ref{calTimeTerms}) analysis.  

Point and variance estimators generally performed as anticipated in the simulation study.  
Table \ref{simResults} presents simulation study results by scenario and analysis model.  
When VE was homogeneous across trials (Scenario 1), bias was low and pointwise CI coverage was near the nominal level for both modeling approaches, as expected.  
In scenarios 2 and 3, the true VE differed across trials and the model (\ref{calTimeTerms}) analysis continued to exhibit low bias and near-nominal pointwise CI coverage overall.  
On the other hand, the model (\ref{timeVaryingIntercept}) analysis produced biased estimates with below-nominal coverage in scenarios 2 and 3, as expected due to model misspecification.  
Results suggest that an incorrect choice of time scale for modeling vaccine effects can have substantial impact on the resulting inference.  
For a fixed $k=5$, empirical and average estimated standard errors for the correctly specified model (\ref{calTimeTerms}) analysis tended to increase with trial number $j$, although standard errors were highest for trial $j=0$. 

The null hypothesis (\ref{Hnaught}) was true by design in scenario 1.  
One-sided generalized Wald tests of the null were rejected at the $0.05$ significance level in $4\%$ of scenario 1 replications.  
The null hypothesis was false in scenarios 2 and 3 and was rejected in $100\%$ of the replications.  

Additional simulations were conducted to evaluate performance of the methods when models were specified to include flexible functions of time variables.  
Simulations were carried out under the same DGP and scenarios described in Section~\ref{sims_un}.  
For analyzing the simulated data, the following variables were transformed using restricted cubic spline (RCS) bases with four knots at the 5th, 35th, 65th, and 95th percentiles \citep{HarrellTextbook}: calendar time in the model for vaccine uptake; both calendar time and time since vaccination in the weighted outcome hazard model.  

Results for the additional simulations are presented in Table~\ref{AddlSims_p3}.  
Empirical bias was tolerable and CI coverage was close to the nominal level in all scenarios. 
For each replication and each scenario, a generalized Wald test was conducted for the TEH assumption $H_{0}$ in (\ref{Hnaught}).  
The null hypothesis $H_{0}$ was true by design in scenario 1.  One-sided generalized Wald tests of $H_{0}$ were rejected at the $0.05$ significance level in $4\%$ of scenario 1 replications.  
The null hypothesis was false by design in scenarios 2 and 3 and was rejected in $100\%$ of the replications.  
Under these simulation conditions, results suggest that the methods performed well when model specification included flexible functions of time.  

\begin{table}
	\centering \renewcommand{\arraystretch}{1}
	\caption{Simulation study results by scenario across $3{,}000$ replications, $\tau=20$ time points of follow-up, and $n=50{,}000$.} 
	\label{simResults}
	\begin{tabular}{clccccccccc}
		\hline
		& & & \multicolumn{4}{c}{Model (\ref{timeVaryingIntercept})} &\multicolumn{4}{c}{Model (\ref{calTimeTerms})} \\
		&  & True & & &  & 95\% CI &  &  &  & 95\% CI\\
		&  & value &  & ESE$^\ddagger$ & ASE$^\ddagger$ & coverage &  & ESE$^\ddagger$ & ASE$^\ddagger$ & coverage \\
		Scn. & Estimand &(\%)  & Bias & $\times10^2$ & $\times10^2$ & (\%) & Bias & $\times10^2$ & $\times10^2$& (\%) \\
		\hline
		1 & $VE_0(5)$ & 90.2 & 0.0 & 5.3 & 5.3 & 95 & 0.0 & 10.0 & 9.9 & 95 \\ 
		& $VE_3(5)$ & 90.1 & 0.0 & 5.3 & 5.3 & 95 & 0.0 & 5.5 & 5.5 & 95 \\ 
		& $VE_6(5)$ & 90.2 & 0.0 & 5.3 & 5.3 & 95 & 0.0 & 6.5 & 6.5 & 95 \\ 
		& $VE_{9}(5)$ & 90.4 & -0.2 & 5.3 & 5.3 & 93 & -0.2 & 8.7 & 8.6 & 95 \\ 
		& $VE_{12}(5)$ & 90.1 & 0.1 & 5.3 & 5.3 & 95 & 0.1 & 10.3 & 10.2 & 95 \\ 
		\hline
		& $VE_5(1)$ & 91.7 & -0.3 & 7.8 & 7.8 & 93 & -0.3 & 8.0 & 8.0 & 94 \\ 
		& $VE_5(4)$ & 90.6 & -0.1 & 5.8 & 5.8 & 95 & -0.1 & 6.3 & 6.3 & 95 \\ 
		& $VE_5(8)$ & 88.7 & 0.0 & 4.1 & 4.0 & 94 & 0.0 & 4.8 & 4.7 & 94 \\ 
		& $VE_5(12)$ & 85.5 & 0.0 & 3.2 & 3.2 & 94 & 0.0 & 3.6 & 3.6 & 95 \\ 
		& $VE_5(15)$ & 81.3 & 0.0 & 2.9 & 2.8 & 94 & 0.0 & 3.2 & 3.1 & 94 \\ 
		\hline
		2 & $VE_0(5)$ & 90.1 & -6.0 & 5.0 & 5.0 & 0 & 0.0 & 10.1 & 9.9 & 95 \\ 
		& $VE_3(5)$ & 87.7 & -3.4 & 5.0 & 5.0 & 0 & 0.0 & 5.5 & 5.5 & 95 \\ 
		& $VE_6(5)$ & 83.2 & 1.2 & 5.1 & 5.1 & 68 & 0.0 & 5.8 & 5.8 & 95 \\ 
		& $VE_{9}(5)$ & 74.5 & 10.0 & 5.1 & 5.1 & 0 & -0.3 & 7.5 & 7.4 & 95 \\ 
		& $VE_{12}(5)$ & 55.8 & 28.8 & 5.1 & 5.1 & 0 & 0.3 & 8.9 & 8.8 & 95 \\ 
		\hline
		& $VE_5(1)$ & 88.6 & -0.4 & 7.5 & 7.6 & 93 & -0.3 & 7.4 & 7.5 & 94 \\ 
		& $VE_5(4)$ & 86.1 & -0.7 & 5.6 & 5.6 & 86 & -0.1 & 5.7 & 5.8 & 95 \\ 
		& $VE_5(8)$ & 81.4 & -0.5 & 4.0 & 4.0 & 89 & -0.1 & 4.4 & 4.4 & 95 \\ 
		& $VE_5(12)$ & 73.8 & 1.8 & 3.4 & 3.3 & 44 & 0.0 & 3.6 & 3.5 & 95 \\ 
		& $VE_5(15)$ & 64.8 & 6.2 & 3.2 & 3.2 & 0 & 0.0 & 3.2 & 3.2 & 95 \\ 
		\hline
		3 & $VE_0(5)$ & 88.9 & -7.2 & 4.6 & 4.6 & 0 & 0.0 & 9.0 & 8.9 & 95 \\ 
		& $VE_3(5)$ & 86.2 & -4.4 & 4.6 & 4.6 & 0 & 0.0 & 5.1 & 5.1 & 95 \\ 
		& $VE_6(5)$ & 80.9 & 0.9 & 4.6 & 4.6 & 83 & 0.1 & 5.2 & 5.2 & 95 \\ 
		& $VE_{9}(5)$ & 71.1 & 10.7 & 4.7 & 4.6 & 0 & -0.2 & 6.7 & 6.6 & 94 \\ 
		& $VE_{12}(5)$ & 49.9 & 31.9 & 4.7 & 4.6 & 0 & 0.3 & 7.9 & 7.8 & 95 \\ 
		\hline
		& $VE_5(1)$ & 88.3 & -0.7 & 6.7 & 6.8 & 87 & -0.2 & 6.6 & 6.7 & 94 \\ 
		& $VE_5(4)$ & 84.8 & -1.3 & 5.1 & 5.1 & 66 & -0.1 & 5.2 & 5.2 & 95 \\ 
		& $VE_5(8)$ & 76.1 & -1.4 & 3.5 & 3.5 & 64 & -0.1 & 3.9 & 3.9 & 95 \\ 
		& $VE_5(12)$ & 55.9 & 2.4 & 2.8 & 2.8 & 47 & 0.1 & 2.9 & 2.9 & 95 \\ 
		& $VE_5(15)$ & 21.1 & 14.2 & 2.7 & 2.7 & 0 & 0.1 & 2.6 & 2.6 & 95 \\ 
		\hline \multicolumn{11}{l}{Abbreviations: Scn., scenario; ASE, average estimated standard error; ESE, empirical } \\
		\multicolumn{11}{l}{standard error; CI, confidence interval.}\\
		\multicolumn{11}{l}{$^\ddagger$Empirical standard error and average estimated standard error were computed on the}\\
		\multicolumn{11}{l}{log risk ratio scale.  All other quantities were computed on the VE scale.}\end{tabular}
\end{table}

\begin{table}
	\centering
	\renewcommand{\arraystretch}{1}
	\caption{Additional simulation study results by scenario across $3{,}000$ replications, $\tau=20$ time points of follow-up, and $n=50{,}000$.  All time functions in the analytical models were specified using RCSs.}
	\label{AddlSims_p3}
	\begin{tabular}{clcccccc}
		\toprule
		&&True &  & &&  95\% CI    \\
		&&value  & &  ESE$^\ddagger$ &ASE$^\ddagger$ &  coverage  \\
		Scenario &Estimand& (\%)&Bias& $\times 10^2$  & $\times 10^2$ & (\%) \\
		\hline
		1 & $VE_0(5)$ & 90.2 & 0.0 & 11.6 & 11.5 & 95 \\ 
		& $VE_3(5)$ & 90.1 & -0.1 & 5.9 & 5.9 & 94 \\ 
		& $VE_6(5)$ & 90.2 & -0.1 & 7.4 & 7.3 & 94 \\ 
		& $VE_{9}(5)$ & 90.4 & -0.2 & 9.2 & 9.2 & 94 \\ 
		& $VE_{12}(5)$ & 90.1 & 0.1 & 11.1 & 11.1 & 95 \\ 
		\hline
		& $VE_5(1)$ & 91.7 & -0.2 & 12.4 & 12.2 & 94 \\ 
		& $VE_5(4)$ & 90.6 & -0.2 & 7.7 & 7.6 & 93 \\ 
		& $VE_5(8)$ & 88.7 & 0.0 & 5.0 & 5.0 & 95 \\ 
		& $VE_5(12)$ & 85.5 & 0.1 & 3.7 & 3.6 & 94 \\ 
		& $VE_5(15)$ & 81.3 & -0.1 & 3.3 & 3.2 & 95 \\ 
		\hline
		2 & $VE_0(5)$ & 90.1 & 0.1 & 11.5 & 11.5 & 95 \\ 
		& $VE_3(5)$ & 87.7 & -0.1 & 5.8 & 5.8 & 95 \\ 
		& $VE_6(5)$ & 83.2 & 0.0 & 6.6 & 6.6 & 95 \\ 
		& $VE_{9}(5)$ & 74.5 & -0.1 & 8.0 & 8.0 & 95 \\ 
		& $VE_{12}(5)$ & 55.8 & 0.1 & 9.9 & 9.8 & 95 \\ 
		\hline
		& $VE_5(1)$ & 88.6 & -0.5 & 11.1 & 10.9 & 92 \\ 
		& $VE_5(4)$ & 86.1 & -0.2 & 7.0 & 6.9 & 94 \\ 
		& $VE_5(8)$ & 81.4 & 0.0 & 4.7 & 4.7 & 95 \\ 
		& $VE_5(12)$ & 73.8 & -0.1 & 3.6 & 3.5 & 95 \\ 
		& $VE_5(15)$ & 64.8 & 0.0 & 3.2 & 3.2 & 95 \\ 
		\hline
		3 & $VE_0(5)$ & 88.9 & 0.1 & 10.9 & 10.7 & 95 \\ 
		& $VE_3(5)$ & 86.2 & -0.4 & 5.3 & 5.3 & 92 \\ 
		& $VE_6(5)$ & 80.9 & -0.4 & 6.0 & 6.0 & 94 \\ 
		& $VE_{9}(5)$ & 71.1 & -0.4 & 7.3 & 7.3 & 94 \\ 
		& $VE_{12}(5)$ & 49.9 & -0.2 & 8.7 & 8.6 & 95 \\ 
		\hline
		& $VE_5(1)$ & 88.3 & -0.3 & 10.6 & 10.5 & 94 \\ 
		& $VE_5(4)$ & 84.8 & -0.6 & 6.4 & 6.4 & 91 \\ 
		& $VE_5(8)$ & 76.1 & -0.1 & 4.1 & 4.1 & 95 \\ 
		& $VE_5(12)$ & 55.9 & 0.6 & 2.9 & 2.9 & 92 \\ 
		& $VE_5(15)$ & 21.1 & -0.3 & 2.6 & 2.6 & 94 \\ 
		\hline \multicolumn{7}{l}{Abbreviations: Scn., scenario; ASE, average estimated standard } \\
		\multicolumn{7}{l}{error; ESE, empirical standard error; CI, confidence interval.} \\
		\multicolumn{7}{l}{$^\ddagger$Empirical standard error and average estimated standard error}\\
		\multicolumn{7}{l}{were computed on the log risk ratio scale.  All other quantities}\\
		\multicolumn{7}{l}{were computed on the VE scale.}
	\end{tabular}
\end{table}

\subsection{Hypothesis testing under nonmonotonic departures from TEH}
\label{nonmonoSims}
	The proposed hypothesis test is intended to detect monotonic departures from TEH.  
	The SARS-CoV-2 virus mutates rapidly, and over time its antigenic structure can become increasingly dissimilar from the antigenic structure for which the vaccine was designed.  
	In turn, VE may decrease with calendar time, suggesting a monotonic departure from TEH.  
	However, nonmonotonic changes could result from seasonal trends or variant waves, e.g., emergence of a new variant followed by resurgence of the original strain.  
	To investigate the proposed test's performance under nonmonotonic departures from TEH, additional simulation scenarios were considered. 
	
	The simulation design was identical to that of Section~\ref{sims_un} except the true counterfactual hazards implied nonmonotonic changes in VE across trials.  
	Potential event times $T^{A_j=0}$ and $T^{A_j=1}$ for $j \in \{0,1,...,\tau\}$ were generated as described in Section~\ref{sims_un}, with true conditional hazards 
	$P(T^{A_0=0}=l \mid T^{A_0=0}>l-1, \boldsymbol{X})=\text{logit}^{-1}\{-\alpha_0 -0.013 (X_1-\tilde{X_1}) -0.26 X_2 + 0.425 X_3+\alpha_{1}l + \alpha_{2}l^2 \}$ and
	\begin{flalign*}
		&P(T^{A_j=1}=l \mid T^{A_j=1}>l-1, \boldsymbol{X})=
		\text{logit}^{-1}[-\alpha_0 
		-0.013 (X_1-\tilde{X_1}) -0.26 X_2 + 0.425 X_3
		\\
		& +\alpha_{1}l 	
		+ \alpha_{2}l^2 
		+I(l > j) \{
		-3
		+ \alpha_3
		+\alpha_{4}l^2 
		+ .02 (l-j)  
		+ .005 (l-j)^2 
		\}].
	\end{flalign*}
	In scenario 4, $(\alpha_0, \alpha_1, \alpha_2, \alpha_3, \alpha_4)=(-5, .2, -.01, .5, -.03)$, simulating a strong seasonality effect, such that the true hazards (both when unvaccinated and when vaccinated) increase to a peak around time point 10 and then decline over the remainder of the study period.  
	In scenario 5, $(\alpha_0, \alpha_1, \alpha_2, \alpha_3, \alpha_4)=(-4, -.2, .01, .5, -.03)$, simulating the emergence of some new viral strain which evades vaccine-induced immunity and is less likely to cause severe disease, followed by resurgence of the original strain.  
	Under scenario 5, the true hazard when unvaccinated decreases over the first half and increases over the second half of the study period, while the true hazard when vaccinated increases over the first half and decreases over the second half of the study period.  
	Two-sided generalized Wald tests of the null were rejected at the $0.05$ significance level in $81\%$ of scenario 4 replications and $48\%$ of scenario 5 replications.
	Under scenarios 4 and 5, power of the TEH test was diminished relative to the monotonic departure scenarios (2 and 3) above.  
	However, note that in Italy during 2021 the SARS-CoV-2 virus did not exhibit these simulated viral mutation patterns \citep{voc2021,ItalyDPC}. 

\subsection{Standardized estimator}
\label{stdzSims}
Simulation studies were conducted to evaluate performance of the standardized estimator $\widehat{VE}^s_j(k)$ in finite samples.  
If there is a subset of covariates that both modify VE and predict ``participation'' in a given trial $j$, then in general $VE_j(k) \neq {VE^s_j(k)}$ for $k \in \{1, ..., K_j\}$ \citep{DahabrehTutorial}.  
However, if effect modification (on the VE scale) is only slight, as in the DGP described in Section~\ref{sims_un}, the estimands $VE_j(k)$ and ${VE^s_j(k)}$ can be approximately equal across trials.  
Therefore simulations were conducted for a DGP under which covariates modify VE more substantially and in turn true values for $VE_j(k)$ and ${VE^s_j(k)}$ diverge as trial number $j$ increases.
Specifically, the simulation DGP in this section differs from that of Section~\ref{sims_un} in that the true outcome hazard model includes an interaction term between continuous covariate {$X_1$} and $A_j$.  

A cohort with $n=50{,}000$ individuals and $\tau=20$ time points of follow up was simulated.  
Covariates $X_1,X_2,X_3$ (representing age at calendar time zero, sex, and comorbidity status, respectively) and vaccine regimen assignments $A_j$ were generated as described in Section~\ref{sims_un}.  
Potential event times $T^{A_j=0}$ and $T^{A_j=1}$ for $j \in \{0,1,...,\tau\}$ were generated as described in Section~\ref{sims_un}, with true conditional hazards 
$P(T^{A_0=0}=l \mid T^{A_0=0}>l-1, \boldsymbol{X})=\text{logit}^{-1}\{-4 -0.013 (X_1-\tilde{X_1}) -0.26 X_2 + 0.425 X_3+\gamma_{1}l 	
+ \gamma_{2}l^2 \}$ and
\begin{flalign*}
	&P(T^{A_j=1}=l \mid T^{A_j=1}>l-1, \boldsymbol{X})=
	\text{logit}^{-1}[-4 
	-0.013 (X_1-\tilde{X_1}) -0.26 X_2 + 0.425 X_3
	\\
	& +\gamma_{1}l 	
	+ \gamma_{2}l^2 
	+I(l > j) \{
	-2.5
	+ \gamma_{3}l
	+ \gamma_{4}l^2 
	+ \gamma_{5} (l-j)  
	+ \gamma_{6} (l-j)^2 
	+ 0.20 (X_1-\tilde{X_1})
	\}].
\end{flalign*}  
As in Section~\ref{sims_un}, three scenarios were considered where the outcome hazard when vaccinated depended on time since vaccination but not calendar time (scenario 1), calendar time but not time since vaccination (scenario 2), and both calendar time and time since vaccination (scenario 3). 
True values for $\boldsymbol{\gamma}=(\gamma_1, ..., \gamma_6)$ were as follows.  
In scenario 1, $\boldsymbol{\gamma}=(0,0,0,0, .02, .005)$; in scenario 2, $\boldsymbol{\gamma}=(-.01, -.003, .02, .006, 0 ,0)$, and in scenario 3, $\boldsymbol{\gamma}=(-.01, -.003, .02, .006, .02, .005)$.  
The resulting treatment effect on the VE scale remained approximately constant across calendar time under Scenario 1 and varied over both time scales in Scenarios 2 and 3.    
Observed event times were computed as described in Section~\ref{sims_un}.  

The simulated dataset was replicated $3{,}000$ times, and thirteen trials were emulated from each replication.  
The vaccine uptake model and outcome MSM were specified to accord with the data generating process.  
Estimates of ${VE}^s_j(5)$ for select $j$ were obtained using (i) the unstandardized estimator $\widehat{VE}_j(k)$ of Section~\ref{methods} of the main text and (ii) the standardized estimator $\widehat{VE}^s_j(k)$ of Section~\ref{extensions} of the main text.  
True values for $VE^s_j(5)$ for select $j$ were approximated by computing 
$$1-\dfrac{\sum_{i=1}^N P(T^{A_j=1} \leq j+k\mid E_j=1, X_i)}{\sum_{i=1}^N P(T^{A_0=0} \leq j+k\mid E_j=1, X_i)}$$ in a very large cohort of $N=10^7$ individuals generated according to the DGP described above.  

Results of this simulation study are presented in Table \ref{StdznSimResults}.  
As expected, empirical bias of $\widehat{VE}^s_j(5)$ was low and pointwise CI coverage was near the nominal level under all three scenarios.  
On the other hand, $\widehat{VE}_j(5)$, which does not standardize across trial-eligible populations, was biased for ${VE}^s_j(5)$ and 95\% confidence interval coverage was poor, particularly as trial number $j$ increased.  
As trial number $j$ increases, the trial-$j$-eligible population becomes increasingly dissimilar from the trial-zero-eligible population because some individuals who appeared in trial zero do not meet criteria for subsequent trials.  
The bias of $\widehat{VE}_j(5)$ was more pronounced under scenarios 2 and 3, where the true hazard under vaccination changed with calendar time.   
As observed in the unstandardized estimator simulation studies, for a fixed $k=5$, empirical and average estimated standard errors for the standardized estimator tended to increase with trial number $j$, although standard errors were highest for trial $j=0$. 

\begin{table}
	\centering \renewcommand{\arraystretch}{1}
	\caption{Simulation study results by scenario for standardization methods.  Results are based on $3{,}000$ replications of a simulated cohort with $\tau=20$ time points of follow-up for $n=50{,}000$ individuals.} 
	\label{StdznSimResults}
	\begin{tabular}{rlrrrrrrrrr}
		\hline
		& & & \multicolumn{4}{c}{Unstandardized} & \multicolumn{4}{c}{Standardized}  \\
		\cmidrule(lr){4-7}  \cmidrule(lr){8-11} &  & True & & & & 95\% CI && & & 95\% CI \\
		&  & Value & & ESE$^\ddagger$ & ASE$^\ddagger$ & coverage  &  &   ESE$^\ddagger$ & ASE$^\ddagger$   & coverage  \\
		Scn. & Estimand & (\%) & Bias & $\times 10^2$ & $\times 10^2$ & (\%) & Bias& $\times 10^2$ & $\times 10^2$ & (\%)  \\
		\hline
		1 & $VE^s_0(5)$ & 85.4 & 0.1 & 9.2 & 9.2 & 94 & -0.1 & 9.0 & 9.0 & 95 \\ 
		& $VE^s_3(5 )$ & 85.4 & -1.0 & 5.1 & 5.0 & 75 & 0.0 & 5.0 & 5.0 & 95 \\ 
		& $VE^s_6(5 )$ & 85.4 & -2.2 & 5.8 & 5.7 & 34 & 0.0 & 5.7 & 5.6 & 95 \\ 
		& $VE^s_{9}(5 )$ & 85.4 & -3.4 & 7.7 & 7.6 & 23 & 0.0 & 7.5 & 7.4 & 94 \\ 
		& $VE^s_{12}(5 )$ & 85.4 & -4.7 & 9.1 & 9.0 & 15 & 0.0 & 9.0 & 8.8 & 94 \\ 
		\hline
		2 & $VE^s_0(5)$ & 85.3 & 0.2 & 9.2 & 9.2 & 95 & -0.1 & 9.0 & 9.0 & 94 \\ 
		& $VE^s_3(5 )$ & 81.9 & -1.3 & 5.0 & 5.0 & 73 & 0.0 & 5.0 & 5.0 & 94 \\ 
		& $VE^s_6(5 )$ & 75.1 & -3.8 & 5.1 & 5.1 & 24 & 0.0 & 5.0 & 5.1 & 95 \\ 
		& $VE^s_{9}(5)$ & 62.1 & -8.7 & 6.5 & 6.5 & 12 & 0.0 & 6.3 & 6.4 & 96 \\ 
		& $VE^s_{12}(5 )$ & 36.1 & -19.5 & 7.7 & 7.8 & 8 & -0.1 & 7.5 & 7.6 & 95 \\ 
		\hline
		3 & $VE^s_0(5)$ & 83.5 & 0.3 & 8.4 & 8.3 & 94 & -0.1 & 8.2 & 8.2 & 95 \\ 
		& $VE^s_3(5 )$ & 79.6 & -1.4 & 4.7 & 4.6 & 71 & 0.0 & 4.6 & 4.6 & 94 \\ 
		& $VE^s_6(5 )$ & 72.0 & -4.3 & 4.7 & 4.7 & 15 & 0.0 & 4.6 & 4.6 & 95 \\ 
		& $VE^s_{9}(5)$ & 57.3 & -9.7 & 5.9 & 5.9 & 7 & 0.0 & 5.7 & 5.8 & 96 \\ 
		& $VE^s_{12}(5 )$ & 27.9 & -20.9 & 6.9 & 7.0 & 5 & -0.2 & 6.7 & 6.8 & 95 \\ 
		\hline \multicolumn{11}{l}{Abbreviations: Scn., scenario; ASE, average estimated standard error; ESE, empirical } \\
		\multicolumn{11}{l}{standard error; CI, confidence interval.} \\
		\multicolumn{11}{l}{$^\ddagger$Empirical standard error and average estimated standard error were computed on the}\\
		\multicolumn{11}{l}{log risk ratio scale.  All other quantities were computed on the VE scale.}\end{tabular}
\end{table}

\section{Additional details about the Abruzzo data analysis}
\label{dataAnalysisDetails}
This section provides additional details on the Abruzzo data analysis in Section~\ref{DataAnalysis} of the main text.  
Precise model specifications are detailed below, and the target trial protocol for the Abruzzo analysis appears in Table~\ref{protocol_p1}.
In the following, let $X_1$ represent (continuous) age at baseline (i.e., calendar time $l=0$), $X_2$ (binary) sex at birth, and $X_3, ..., X_8$ (binary, baseline) hypertension, diabetes, cardiovascular disease, chronic obstructive pulmonary disease, kidney disease, and cancer status, respectively. 
The vaccine dose variable $B_l$ was coded as 0) no vaccine dose, 1) Pfizer-BioNTech, 2) Moderna, 3) Oxford-AstraZeneca, 4) Janssen.  
The constants $k^{min}_1=3$, $k^{max}_1=6$, $k^{min}_2=4$, and $k^{max}_2=6$ were defined according to CDC guidelines for timing of the second dose of the original monovalent Pfizer-BioNTech and Moderna vaccines \citep{CDC}.  

In each emulated trial, the active vaccine regimen was ``receive a first mRNA COVID-19 vaccine dose within seven days of enrollment, complete the full course of vaccine according to recommended guidelines for the brand of the first vaccine dose, and receive no further COVID-19 vaccine doses.''  
	The comparator regimen was ``remain unvaccinated for COVID-19 through December 18, 2021''.
	An individual was no longer compliant with either strategy if they (i) got any non-mRNA COVID-19 vaccine dose; (ii) received a third dose of any COVID-19 vaccine; (iii) got a second mRNA dose of a different brand than the first dose; (iv) received a second mRNA dose prior to the start of the recommended time window (for the brand of their first COVID-19 vaccine dose); or (v) reached the end of the recommended time window without getting a second dose.  
	In symbols, the noncompliance indicator was $C^*_l= 
	\max[
	I(B_{V_1}>2, Z_l>0), 
	I(Z_l>2), 
	I(Z_l=2, B_{V_1} \neq  B_{V_2}),
	\{I(Z_l=2, B_{V_1}=1, l-V_1<k^{min}_{1}) + I(Z_l=2, B_{V_1}=2, l-V_1<k^{min}_{2})\},
	\{I(Z_l=1, B_{V_1}=1, l-V_1>k^{max}_1)   + I(Z_l=1, B_{V_1}=2, l-V_1>k^{max}_2)\}
	]$.
	The trial-specific compliance indicator for the outcome hazard model was $C_{j}(k)=A_jC^*_{j+k}+(1-A_j)Z_{j+k+1}$ if $E_j=1$ (otherwise $C_{j}(k)$ was undefined).

\subsection{Censoring model specification}
\label{appliedCensModelSpec}  
The VU model was specified as follows.
Note that in Italy, additional COVID-19 vaccine doses for those who completed an initial vaccine series first became available in Fall, 2021 (specifically at week $l=32$ in the Abruzzo NTE study period).  
Therefore, it was impossible for fully-vaccinated individuals to go off regimen prior to time $l=32$.  
In other words, the hazard of noncompliance is a known constant ($0$) and does not need to be modeled whenever $l<32, Z_l=2,$ and $B_{V_1} \in \{1,2\}$.  
Therefore, the ``at-risk'' indicator for the VU model in the Abruzzo data analysis was given by $R^Z_l=I(T^* \geq l,C^*_{l-2}=0)  \{1-I(l<32, Z_l=2, B_{V_1} \in \{1,2\})\}$.
	
	The following VU model was assumed
	\begin{align} 
		\label{VUmodel_analysis}
		&P\{D_l=1 \mid Z_{l-1}=z, \boldsymbol{X}=\boldsymbol{x}, B_{V_z}=b, R^Z_l=1\}= \text{logit}^{-1}\{\boldsymbol{\kappa_0}\boldsymbol{f_{\kappa_0}}(l, \boldsymbol{x})+\\\nonumber
		&
		{\kappa_1}I(z=1)+{\kappa_2}I(z=2)+
		{\kappa_3}\{I(z=1, b=1, k<k^{min}_{1}) + I(z=1, b=2, k<k^{min}_{2})\} + \\\nonumber
		&{\kappa_4}\{I(z=1, b=1, k^{min}_{1}\leq k \leq k^{max}_{1}) + I(z=1, b=2, k^{min}_{2}\leq k \leq k^{max}_{2})\},
\end{align}
where $\boldsymbol{f_{\kappa_0}}=\{1,\boldsymbol{spl}(l), \boldsymbol{spl}(x_1-\tilde{x}_1), x_2, ..., x_8,
	\boldsymbol{spl}(l) \times \boldsymbol{spl}(x_1-\tilde{x}_1), \boldsymbol{spl}(l) \times x_2, \boldsymbol{spl}(l) \times x_3, ..., \boldsymbol{spl}(l) \times x_8
	\}^T$, here and below $\tilde{x}_1$ is the sample mean of $x_1$ and $\boldsymbol{spl}(\cdot)$ is a RCS function with knots at the 5th, 35th, 65th, and 95th percentiles \citep{HarrellTextbook}, and the product of two spline functions contains all possible products between component spline terms.  
The parameters $\boldsymbol{\kappa_0}$ represent the effect of calendar time and covariates on the probability of an incremental change in vaccine status; ${\kappa_1}$ and ${\kappa_2}$ are shared (across vaccine brands) intercepts for prior vaccination status $1$ and $2$, respectively; $\kappa_3$ is a parameter associated with getting a second dose of a two-dose vaccine brand prior to the start of the brand-specific grace period; and $\kappa_4$ is a parameter associated with getting a second dose of a two-dose vaccine brand during the brand-specific grace period.
The model for censoring due to loss to follow up was specified according to (\ref{eq:IPCmodel}) with $\boldsymbol{f_{\xi}}(l,\boldsymbol{x}, \overline{z})=(1,\boldsymbol{spl}(l), I(z_l=1),I(z_l=2), \boldsymbol{spl}(x_1-\tilde{x}_1), x_2, x_3, ..., x_8 )^T$.  

\subsection{Outcome hazard model specification}
\label{appliedOcModelSpec}
The MSM for the covariate-conditional outcome hazards was specified by adapting~(\ref{calTimeTerms}) in the main text to include terms for baseline covariates.  
	Specifically, the outcome MSM was specified as 
	\begin{equation}
		\label{appliedOcModel}
		{\lambda}_j^{a}(m\mid \boldsymbol{x})=\text{logit}^{-1}\{\boldsymbol{\gamma} \boldsymbol{f_\gamma}(j,k,a,\boldsymbol{x})\}, 
	\end{equation}
	where $\boldsymbol{f_\gamma}(j,k,a,\boldsymbol{x})$=$(1$, $\boldsymbol{spl}(j+k)$, $\boldsymbol{spl}(x_1-\tilde{x}_1)$, $x_2$, ..., $x_8$, $a$, $a\boldsymbol{spl}(j+k)$, $a\boldsymbol{spl}(k))^T$.  	
	All percentiles used to set the spline knots were computed using the NTE analysis dataset. 

A series of sensitivity analyses were conducted to evaluate adequacy of model specifications detailed above.  
	See Section~\ref{sensAnalyses} for more details.        	

\begin{tabularx}{\linewidth}{
		>{\hsize=.32\hsize}X
		>{\hsize=1.32\hsize}X
		>{\hsize=1.34\hsize}X
	}
	\caption{Specification and emulation of a sequence of trials to assess effectiveness of a full course of COVID-19 vaccine between February 15, 2021 and December 18, 2021.}
	\label{protocol_p1}
	\\
	\toprule
	\textbf{}      
	& \textbf{Target trial specification}   
	& \textbf{Target trial emulation} \\
	\midrule
	\endfirsthead
	\toprule
	\textbf{}      
	& \textbf{Target trial specification}   
	& \textbf{Target trial emulation} \\
	\endhead
	\multicolumn{3}{l}{\footnotesize Abbreviation: NHS, National Health Service}\\
	\multicolumn{3}{c}{\footnotesize(Table continued on next page)}
	\endfoot
	\bottomrule
	\endlastfoot
	Inclusion criteria
	&Resident of or domiciled in Abruzzo region of Italy on January~1, 2020
	&Same\\
	\cline{2-3}
	&Aged 80 years or older on February~15, 2021
	&Same\\
	\cline{2-3}
	&Alive at time of enrollment 
	&Alive at time of enrollment according to NHS data\\
	\hline
	Exclusion criteria
	&Positive SARS-CoV-2 swab prior to February~15, 2021
	&Positive SARS-CoV-2 swab documented in NHS data prior to February~15, 2021\\
	\cline{2-3}	
	&Severe COVID-19 disease (as diagnosed by a specialist physician and requiring hospitalization) prior to enrollment
	&Severe COVID-19 disease (as diagnosed by a specialist physician and requiring hospitalization) documented in NHS data prior to enrollment\\
	\cline{2-3}
	&Received any dose of any COVID-19 vaccine prior to enrollment 
	&Any dose of any COVID-19 vaccine documented in NHS data prior to enrollment \\
	\hline
	Vaccine  regimens		
	& The active regimen is ``receive a first dose of mRNA COVID-19 vaccine (original monovalent Pfizer-BioNTech or Moderna) within seven days of enrollment, complete the full course of vaccine according to recommended guidelines \citep{CDC} for the brand of the first vaccine dose, and receive no further COVID-19 vaccine doses.''  The comparator regimen is ``remain unvaccinated for COVID-19 through December 18, 2021''.  
	&Same, except recommended time windows for the second dose of two-dose regimens are coarsened to weeks to match the time unit of analyses.\\
	\hline
	Treatment assignment	
	& Each week during the enrollment period (February 15, 2021 to May 9, 2021), a pragmatic randomized trial will be initiated.  On the first day of each trial, eligible individuals will be enrolled and assigned at random (with equal probability) to active regimen or comparator.  Those assigned to active regimen may choose the mRNA vaccine brand they receive (original monovalent Pfizer-BioNTech or Moderna, as available).  They will be instructed to receive a first dose of their chosen vaccine within seven days and then follow the corresponding dosing schedule (as described above).  
	& On February 15, 2021, all eligible individuals will be ``enrolled" in a hypothetical trial.  Those who receive a first COVID-19 vaccine dose during the first week of the trial will be classified as receiving active regimen; all others will be classified as receiving comparator.  A series of identical hypothetical trials will be initiated on the first day of each subsequent week through May. 9, 2021. Individuals may appear in multiple trials, provided they meet eligibility criteria.  \\
	\hline
	Outcome  
	& Severe COVID-19 disease (as diagnosed by a specialist physician and requiring hospitalization) or COVID-19-related death 
	& Severe COVID-19 disease (as diagnosed by a specialist physician and requiring hospitalization) or death with a SARS-CoV-2 positive swab documented in NHS data\\
	\hline
	Follow up
	&Eligibility will be assessed and treatment will be randomly assigned on the first day of each trial.  Participants are followed until the first of the following:
	
	1. Experience of an event
	
	2. Discontinuation of assigned vaccine regimen
	
	3. Death without a SARS-CoV-2 positive swab
	
	4. December 18, 2021
	& Same, except:
	
	1. Discontinuation is defined in terms of the regimen an individual was observed to initiate at the start of an emulated trial. 
	
	2. At-risk status will be evaluated at a series of weekly hypothetical ``study visits".  
	\\
	\hline
	Nonad-herence
	&Participants who discontinue their assigned treatment strategy will be censored.  Participants’ observations will be censored on the day of the first occurrence of any of the following:
	
	1. Receipt of any non-mRNA COVID-19 vaccine dose
	
	2. Receipt of first dose of any COVID-19 vaccine (for participants assigned to comparator)
	
	3. Failure to complete a second dose by the end of the recommended time window (for participants assigned to active regimen)
	
	4. Receipt of an additional dose of COVID-19 vaccine following completion of an initial vaccine series (i.e., a third COVID-19 vaccine dose)
	& Same, except censoring will be handled separately per person-trial, nonadherence is defined in terms of the regimen an individual was observed to initiate at the start of an emulated trial, and censoring status will be updated at weekly hypothetical study visits. \\
	\hline
	Causal contrast
	& VE, defined as one minus the counterfactual risk ratio comparing counterfactual risk of severe COVID-19 or death had all individuals remained adherent to the active regimen with that same risk had all individuals remained adherent to the comparator regimen
	&Same\\
	\hline
	Analysis plan
	&Analyses will be analogous to those described in the main text.  Since exchangeability of treatment groups is expected due to randomization, no adjustment for measured confounders is needed.  Inverse probability weighting can be used to adjust for selection bias arising due to differential censoring (nonadherence and loss to follow up).  Estimates and pointwise 95\% CIs for VE across calendar time and time since vaccination will be calculated and reported for each trial.
	& Analyses will be conducted according to methods described in the main text.  Inverse probability weighting will be used to adjust for confounding and differential censoring due to nonadherence and loss to follow up. All time variables will be coarsened to weeks.  Specific analytical decisions are detailed in \ref{dataAnalysisDetails}.  Estimates and pointwise 95\% CIs for VE across calendar time and time since vaccination will be calculated and reported for each trial.  A test of the TEH assumption, as defined in the main text, will be conducted.  In order to assess changes in VE over calendar time that cannot be attributed to temporal variations in covariate distributions, analyses will be repeated using the standardization procedures described in Section~\ref{extensions} of the main text.  
\end{tabularx}
\subsection{Diagnostics and sensitivity analyses}	
\label{sensAnalyses}
This section details a series of diagnostics and sensitivity analyses designed to explore how analytic choices could influence results.

\subsubsection{Assessment of baseline covariate balance for each emulated trial}
\label{Sens_VUmodel}
\begin{figure}
		\centering
		\includegraphics[width=\textwidth]{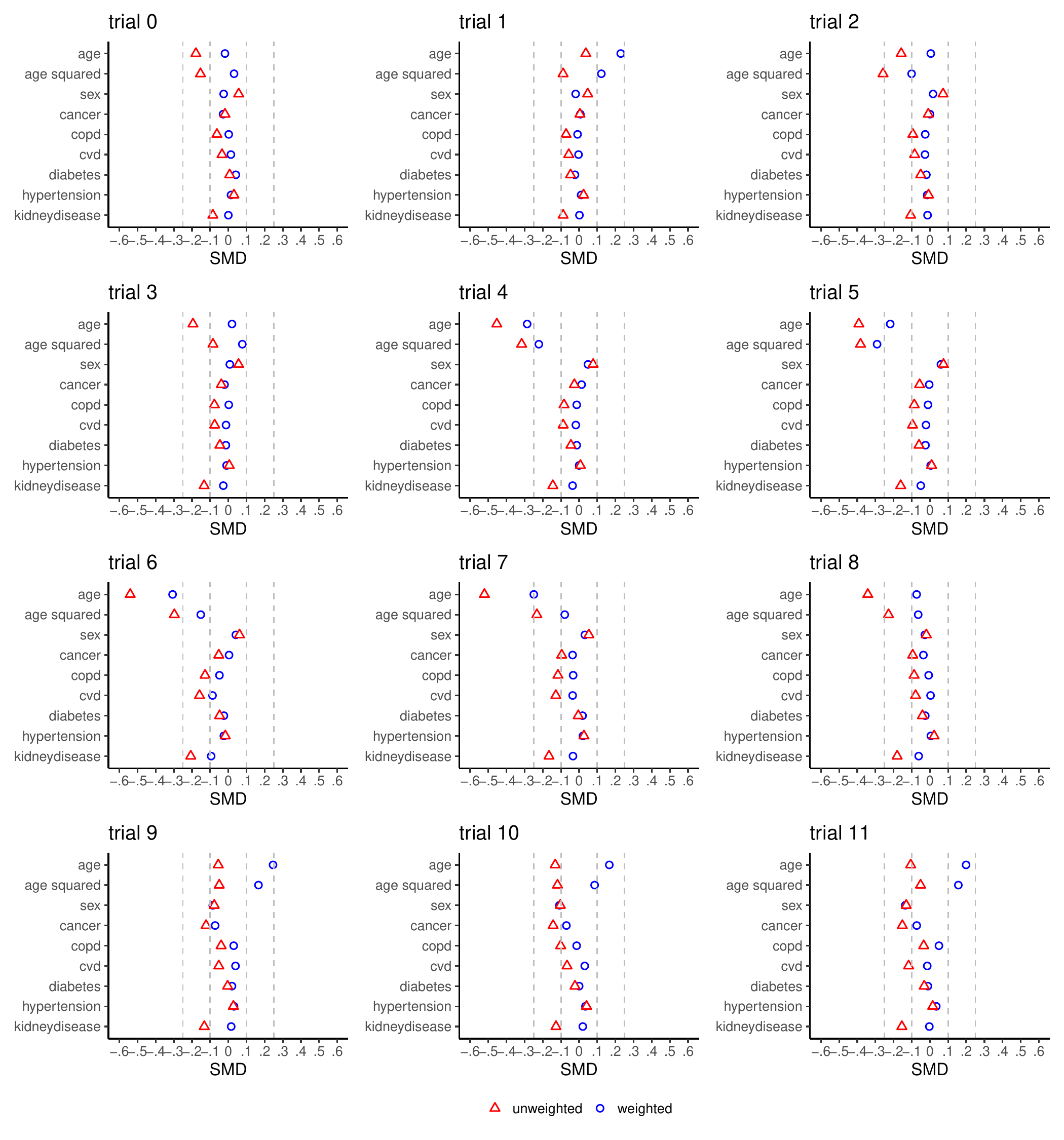}
		\caption{Standardized mean difference (SMD) of covariates (comparing active and comparator groups) before and after weighting for each emulated trial.}
		\label{LovePlots_contin}
	\end{figure}

To assess adequacy of the VU model specification, baseline covariate balance was assessed for each emulated trial.  
	Specifically, within each trial, standardized mean differences (comparing the active and comparator arms) were computed for each baseline covariate before and after weighting.  
	Standardized mean differences were defined and computed as in Section~4.1.1 of \citet{AustinStuart}, except that the weighted/unweighted sample variance for binary variables was computed as $p(1-{p})$, where ${p}$ is the weighted/unweighted sample proportion \citep{Greifer2026}.  
	Figure~\ref{LovePlots_contin} displays standardized mean difference (SMD) of covariates before and after weighting for each emulated trial.
	
\begin{figure}
		\centering
		\includegraphics[scale=.55]{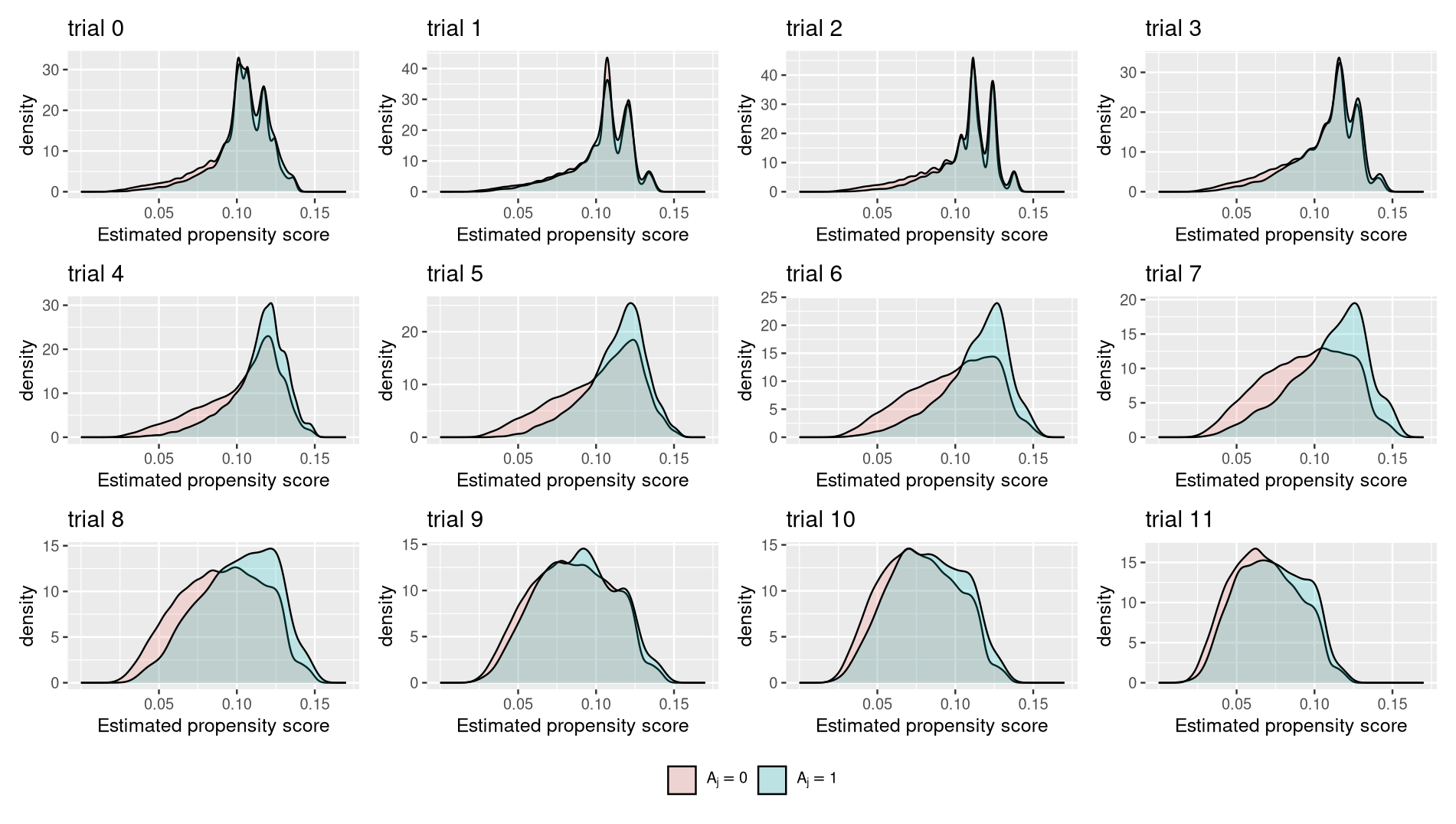}
		\caption{Density plots for estimated propensity score by trial and treatment group}
		\label{propScoreDensities}
	\end{figure}

	After weighting, all covariates were adequately balanced in all trials except for age and age squared in certain trials.  
	Propensity score distributions for the treated and untreated (Figure~\ref{propScoreDensities}) generally overlapped, suggesting that the positivity of treatment assignment assumption (\ref{positivityTx_m}) is tenable. 
	
\begin{figure}
	\centering
	\includegraphics[width=\textwidth]{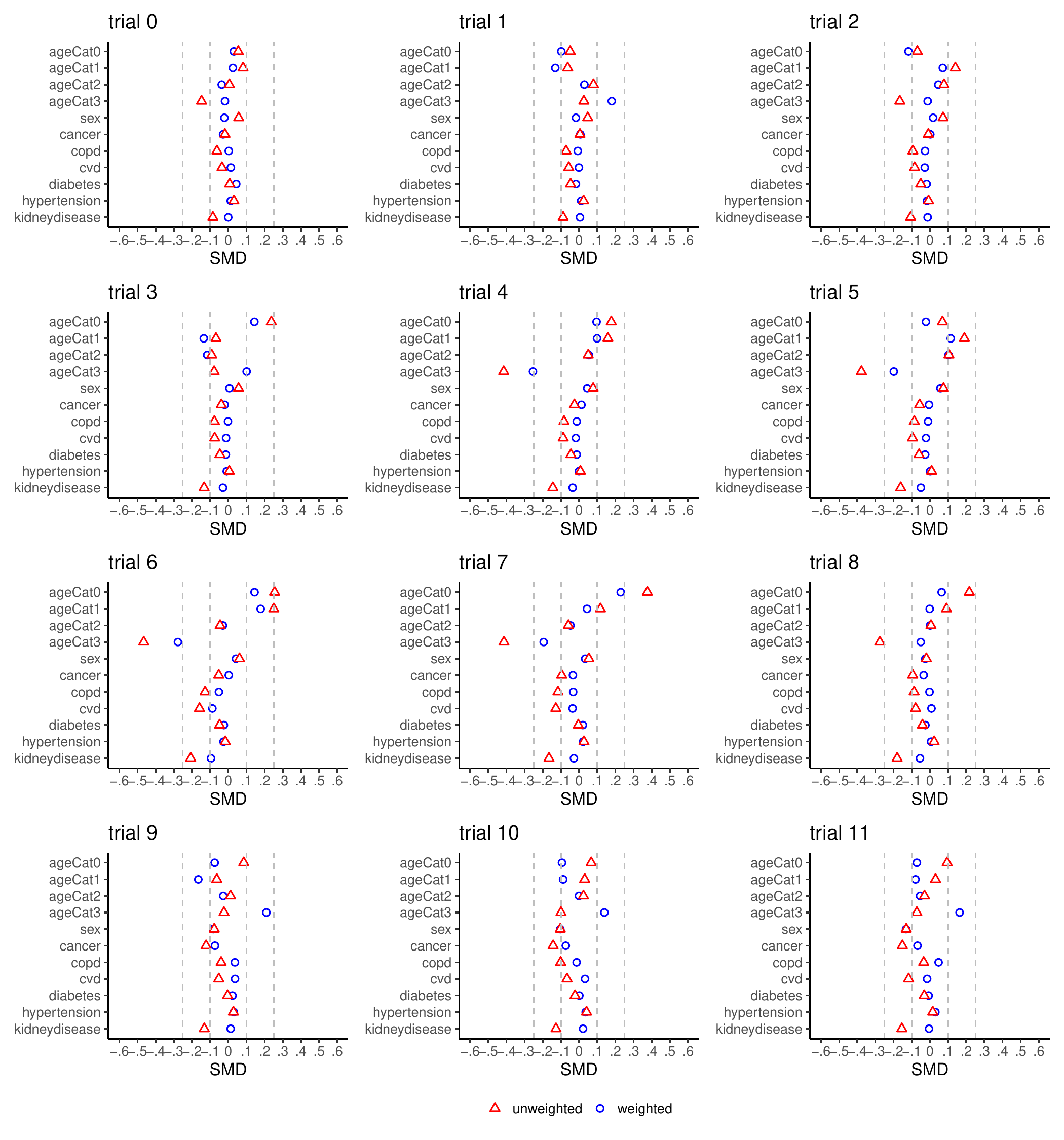}
	\caption{Standardized mean difference (SMD) of covariates (comparing active and comparator groups) before and after weighting for each emulated trial.
		The vaccine uptake model was specified with age as a categorical variable.}
	\label{LovePlots_cat}
\end{figure}

	To explore the extent to which very old individuals may influence (i) residual imbalance in age for certain trials and (ii) final standardized VE estimates, two sensitivity analyses were considered.
	First, age was treated as categorical with categories $X_1 < 82, 82\leq X_1 < 85, 85 \leq X_1 < 89,$ and $X_1 \geq 89$. 
	The VU model was specified as in~(\ref{VUmodel_analysis}) with 
	$\boldsymbol{f_{\kappa_0}}=\{1,\boldsymbol{spl}(l), I(82\leq x_1 < 85), I(85 \leq x_1 < 89), I(x_1 \geq 89), x_2, ..., x_8,
	\boldsymbol{spl}(l) \times I(82\leq x_1 < 85), 
	\boldsymbol{spl}(l) \times I(85 \leq x_1 < 89),
	\boldsymbol{spl}(l) \times I(x_1 \geq 89),
	\boldsymbol{spl}(l) \times x_2, \boldsymbol{spl}(l) \times x_3, ..., \boldsymbol{spl}(l) \times x_8
	\}^T$.
	With estimated weights obtained from this VU model, baseline covariate balance improved (see Figure~\ref{LovePlots_cat}), although absolute standardized mean differences for age remained higher, relative to other covariates, in certain trials.   
	Final VE point estimates and 95\% CIs (Table~\ref{sens_VUcatAge}) were similar to those of the primary analysis.

\begin{table}
	\centering  \renewcommand{\arraystretch}{1}
	\caption{Sensitivity analysis results for selected emulated trials and weeks since first
		dose. 
		The vaccine uptake model was specified with age as a categorical variable.} 
	\label{sens_VUcatAge}
	\begin{tabular}{lllll}
		\hline
		Weeks since & \multicolumn{1}{c}{$\widehat{VE}_0(k )$} & \multicolumn{1}{c}{$\widehat{VE}_3(k )$} & \multicolumn{1}{c}{$\widehat{VE}_6(k )$} & \multicolumn{1}{c}{$\widehat{VE}_9(k )$} \\  first dose ($k$) & \multicolumn{4}{c}{\% (95\% CI)}  \\ \hline
		1 & 85 (72, 93) & 80 (69, 87) & 73 (53, 84) & 64 (13, 85) \\ 
		7 & 87 (76, 93) & 82 (73, 88) & 76 (59, 86) & 69 (30, 87) \\ 
		14 & 87 (77, 93) & 83 (74, 88) & 78 (63, 87) & 74 (44, 88) \\ 
		21 & 86 (76, 92) & 82 (73, 88) & 78 (64, 86) & 74 (50, 87) \\ 
		28 & 84 (75, 90) & 80 (71, 86) & 73 (60, 82) & 62 (36, 78) \\ 
		34 & 83 (74, 89) & 76 (67, 82) & 61 (44, 72) & 27 (-16, 54) \\ 
		\hline
	\end{tabular}
\end{table}

\begin{figure}
	\centering
	\includegraphics[width=\textwidth]{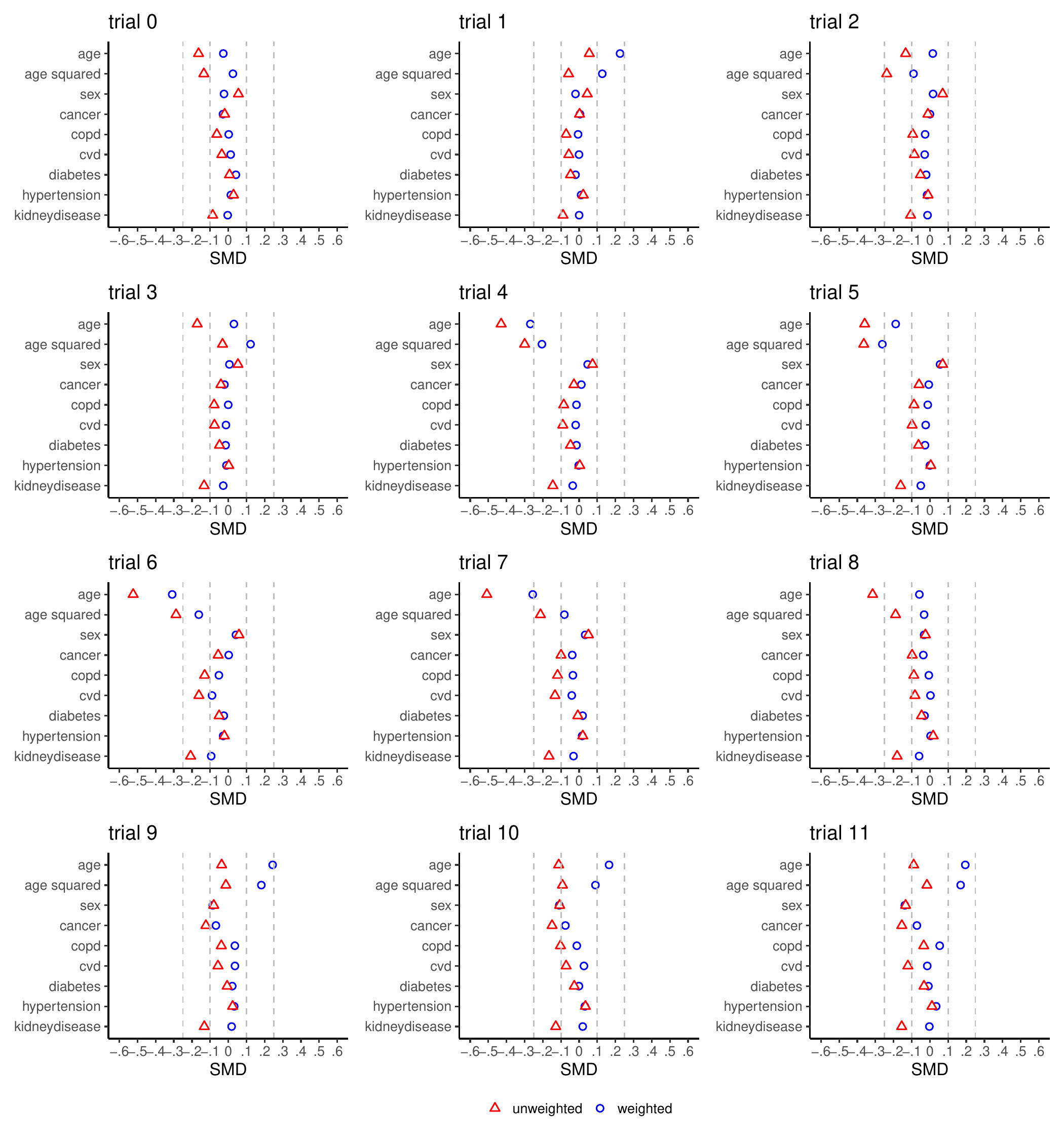}
	\caption{Standardized mean difference (SMD) of covariates (comparing active and comparator groups) before and after weighting for each emulated trial.
		The target population was altered to include only individuals aged 80-100 years.}
	\label{LovePlots_restrict}
\end{figure}

	Second, the target population was restricted to individuals 80-100 years of age.  
	This restriction ensured that the age ranges for the treated and the untreated were essentially identical in every trial.  
	The VU model was specified as in (\ref{VUmodel_analysis}).
	Baseline covariate balance (Figure~\ref{LovePlots_restrict}) after weighting and final VE point estimates and 95\% CIs (Table~\ref{sens_VUrestrict}) were similar to those of the primary analysis.

\begin{table}
	\centering  \renewcommand{\arraystretch}{1}
	\caption{Sensitivity analysis results for selected emulated trials and weeks since first dose. 
		The target population was altered to include only individuals aged 80-100 years.} 
	\label{sens_VUrestrict}
	\begin{tabular}{lllll}
		\hline
		Weeks since & \multicolumn{1}{c}{$\widehat{VE}_0(k )$} & \multicolumn{1}{c}{$\widehat{VE}_3(k )$} & \multicolumn{1}{c}{$\widehat{VE}_6(k )$} & \multicolumn{1}{c}{$\widehat{VE}_9(k )$} \\  first dose ($k$) & \multicolumn{4}{c}{\% (95\% CI)}  \\ \hline
		1 & 85 (71, 93) & 80 (68, 87) & 71 (48, 84) & 60 (-0, 84) \\ 
		7 & 87 (75, 93) & 82 (73, 88) & 75 (56, 86) & 68 (22, 86) \\ 
		14 & 87 (76, 93) & 82 (73, 88) & 77 (60, 87) & 72 (38, 88) \\ 
		21 & 86 (75, 92) & 81 (73, 87) & 77 (62, 86) & 73 (45, 87) \\ 
		28 & 84 (73, 90) & 79 (70, 85) & 72 (57, 81) & 60 (30, 78) \\ 
		34 & 82 (71, 89) & 74 (65, 81) & 59 (41, 71) & 25 (-21, 53) \\ 
		\hline
	\end{tabular}
\end{table}

\subsubsection{Sensitivity analyses for the outcome hazard model}
\label{Sens_OCmodel}

	Two sensitivity analyses to the outcome MSM specification were conducted.  
	First, to assess sensitivity to spline function specifications, substantive information was used to choose knot locations in RCS functions of calendar time and time since vaccination.
	Second, the no-interaction assumption (discussed in Section~\ref{modelSpecification} of the main text) in model~(\ref{calTimeTerms}) was relaxed by including additional product terms in the outcome model.  

\begin{figure}
		\centering
		\includegraphics[width=.9\textwidth]{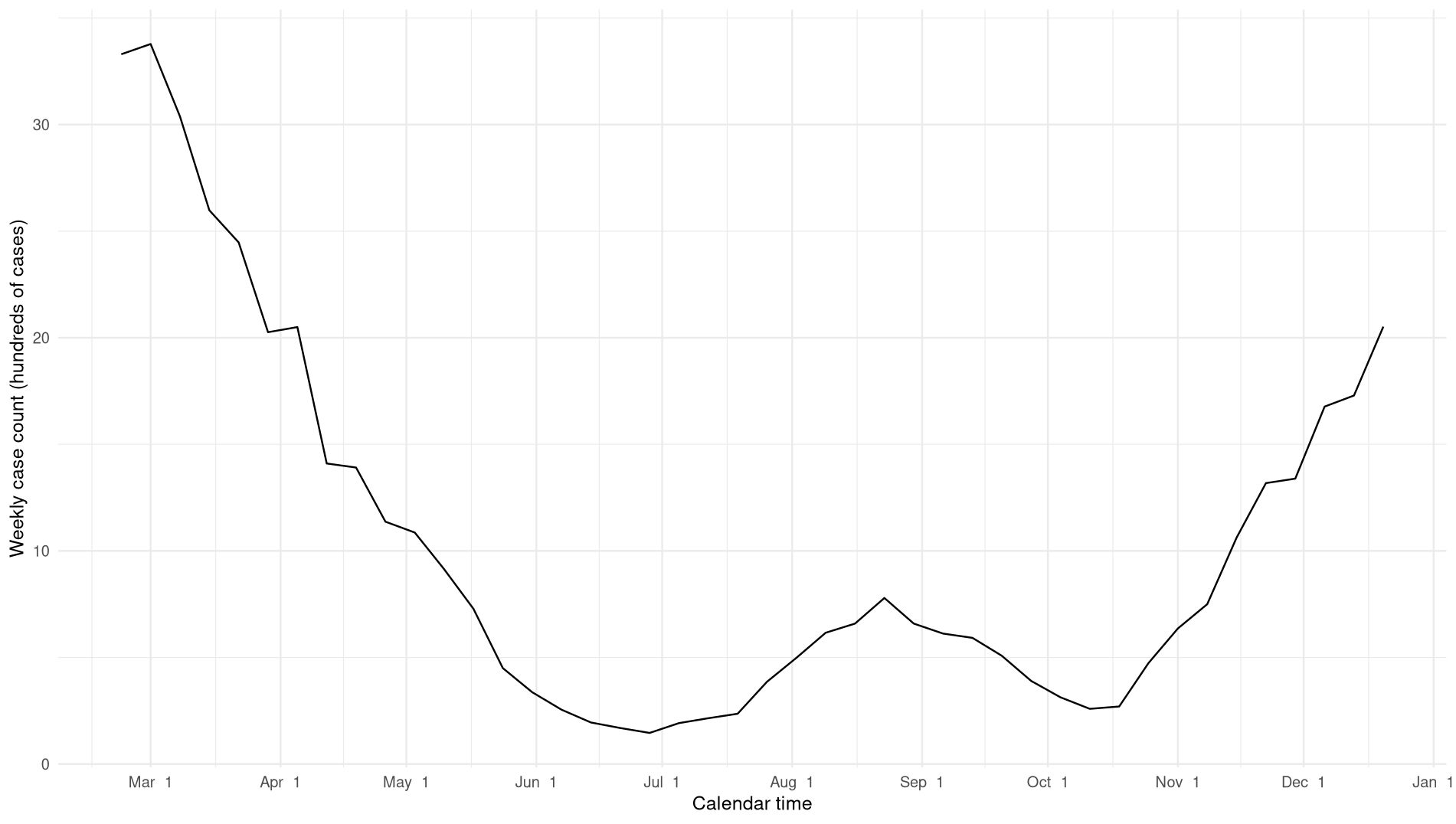}
		\caption{Weekly case counts \citep{ItalyDPC} for SARS-CoV-2 infections in Abruzzo, Italy during the NTE study period.}
		\label{caseCounts}
	\end{figure}

	In the first sensitivity analysis, RCS functions of time were specified as follows.  
	In keeping with epidemiological conditions in Abruzzo over the course of 2021 \citep[][Appendix S2]{ItalyDPC, voc2021, marchetti2022epidemic}, changes in the transformed hazard when unvaccinated were assumed to take separate, linear shapes during the periods February 15-March 14 and November\ 15-December\ 18, and separate, cubic shapes during the periods March 15-June 6, June 7-August 15, and August 16-November\ 14.  
	The first month of the NTE study period saw a steady, near-linear decline in COVID-19 cases (see Figure~\ref{caseCounts}), increasing prevalence of the Alpha variant (see lower half of Figure \ref{deltaPlot} in the main text), and a consistent set of government-imposed social restrictions.  
	During the period March 15-June 6,
	cases declined further, there were multiple circulating SARS-CoV-2 strains, and there was a slight relaxation of government-imposed social restrictions in late April. 
	Background conditions were substantially different from June 8-August 15.  
	Cases were initially low, and in turn, government-imposed social restrictions were substantially lightened for the region. 
	But in late summer, cases began to increase alongside prevalence of the Delta variant.  
	The period August 16-November 14 saw another distinct set of background conditions.  
	Delta was essentially the only circulating variant, case counts initially declined but began to increase in mid-October, and government policies were adapted to relax restrictions for vaccinated individuals \citep{conteduca2022new}.  
	The last month of the NTE study period saw a continuation of (i) the Delta variant wave and (ii) Italian public policy designed to encourage vaccination.  
	Meanwhile, case counts rose approximately linearly \citep[][Appendix S2]{ItalyDPC, voc2021, marchetti2022epidemic}.  
	
	Changes in the increment to the transformed hazard when vaccinated were assumed to take separate, linear shapes during the periods February 15-March 14 and October 18-December 18, and separate, cubic shapes during the periods March 15-May 2, May 3-July 4, and July 5-October 17.  
	The latter three periods reflect distinct patterns in variant prevalence, namely circulation of multiple strains, decline of Alpha and Gamma with increase in Delta, and prominence of Delta to the near-exclusion of other variants, respectively.  
	In turn, it was assumed that changes in the increment in the transformed hazard over calendar time took a different (cubic) shape within each interval.  
	Beginning October\ 15, social restrictions were lightened for vaccinated individuals, suggesting that vaccinated and unvaccinated individuals may have different levels of exposure, and in turn, changes in the increment in the transformed hazard when vaccinated were assumed to have a different trajectory starting from October\ 15.
	
	Changes to the transformed hazard over time since vaccination were assumed to take separate, linear shapes for weeks 1-2 and 36-44, and separate, cubic shapes during weeks 3-5, 6-12, and 13-35 following receipt of the first dose.  
	These assumptions allow the transformed hazard under vaccination to evolve differently prior to receipt of the second dose, in the weeks around the time the second dose is received, in the period after the second dose was received but before vaccine-induced antibody levels are expected to peak \citep{Ebinger}, in the months following said peak, and several months after receipt of the second dose.  

\begin{figure}
	\centering
	\subfloat[][]{\includegraphics[width=.45\textwidth]{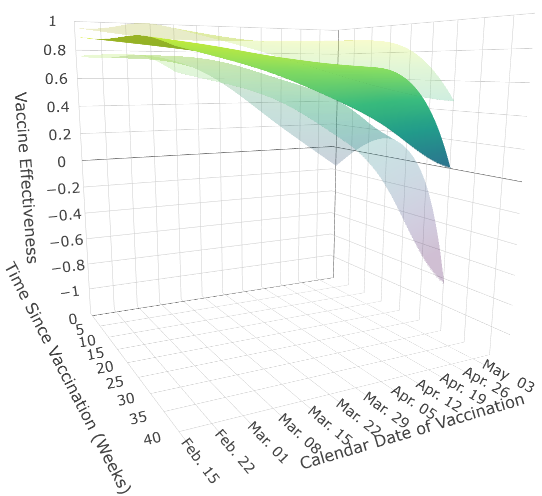}}\quad
	\subfloat[][]{\includegraphics[width=.52\textwidth]{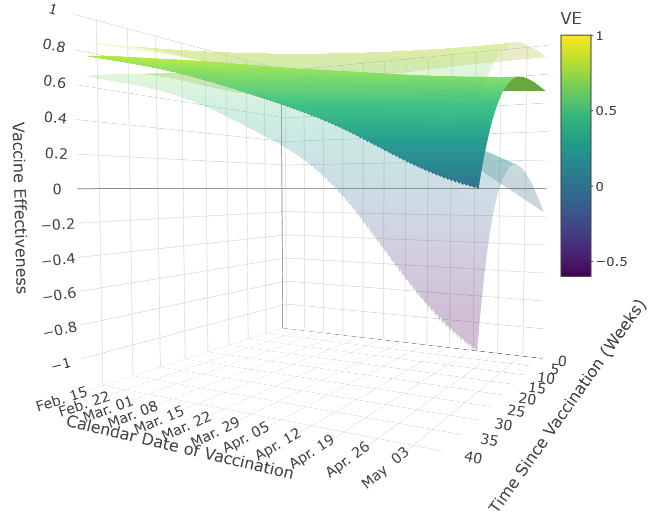}}
	\vspace{8pt}
	\caption{Sensitivity analysis results.  
		Restricted cubic spline functions were specified using published data on time-varying COVID-19 case counts \citep{ItalyDPC}, prevalence of SARS-CoV-2 variants of concern \citep{voc2021}, and public policy \citep[][Appendix S2]{marchetti2022epidemic} in Abruzzo.}
	\label{contourPlots_ext}
\end{figure}

Specifically, the covariate-conditional outcome MSM was specified as ${\lambda}_j^{a}(k \mid \boldsymbol{x})=\text{logit}^{-1}\{\boldsymbol{\gamma} \boldsymbol{f_\gamma}(j,k,a,\boldsymbol{x})\}$, 
	where $\boldsymbol{f_\gamma}(j,k,a,\boldsymbol{x})$=$(1,a$,
	$a\boldsymbol{f_1}(k)^T$,
	$\boldsymbol{f_2}(j+k)^T$, 
	$a\boldsymbol{f_3}(j+k)^T$,
	$\boldsymbol{spl}(x_1-\tilde{x}_1)$, $x_2$,..., $x_8)^T$, 
	$\boldsymbol{f_1}(\cdot)$, 
	$\boldsymbol{f_2}(\cdot)$, and
	$\boldsymbol{f_3}(\cdot)$,
	were RCS functions with knots specified as described in the paragraphs above, 
	and percentiles of $(x_1-\tilde{x}_1)$ were computed in the NTE analysis dataset.
	Point estimates and corresponding pointwise 95\% CIs were largely unchanged relative to those of the primary analysis (Figure~\ref{contourPlots_ext}). 

	The second sensitivity analysis relaxed the no-interaction assumption in model~(\ref{calTimeTerms}).  
	Particularly, the analysis was repeated with additional product terms included in the model.  
	The MSM for the covariate-conditional, counterfactual hazards was specified as 
	$\lambda_j^{a}(m\mid \boldsymbol{x})=\text{logit}^{-1}\{\boldsymbol{\gamma} \boldsymbol{f^*_\gamma}(j,k,a,\boldsymbol{x})\}$, 
	where $\boldsymbol{f^*_\gamma}(j,k,a,\boldsymbol{x})$=$(1$, $\boldsymbol{spl}(j+k)$, $\boldsymbol{spl}(x_1-\tilde{x}_1)$, $x_2$, ..., $x_8$, $a$, $a\boldsymbol{spl}(j+k)$, $a\boldsymbol{spl}(k)$, $a\boldsymbol{spl}(j+k) \times \boldsymbol{spl}(k))^T$.  	
	
\begin{figure}
	\centering
	\subfloat[][]{\includegraphics[width=.46\textwidth]{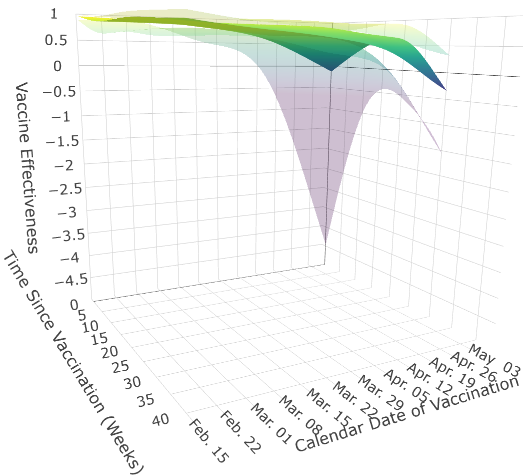}}\quad
	\subfloat[][]{\includegraphics[width=.51\textwidth]{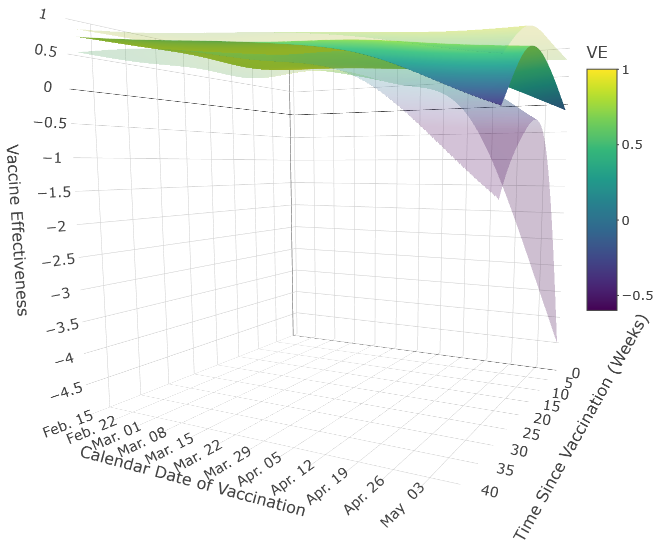}}
	\vspace{8pt}
	\caption{Sensitivity analysis results.  
		The outcome hazard model was specified to include a three-way product term between vaccine regimen assignment, calendar time and time since vaccination.}
	\label{contourPlots_stdPlus}
\end{figure}

	VE point estimates and corresponding pointwise 95\% CIs were largely unchanged relative to those of the primary analysis (Figure~\ref{contourPlots_stdPlus}).  
	However, as trial number increased, 95\% CIs became extremely wide for early and late time points, suggesting that observed events may be too sparse in the Abruzzo data to allow for precise estimation of these additional parameters.  	

\subsubsection{Sensitivity analysis to the exclusion of a covariate}
\label{Sens_std}
	The standardized VE estimand $VE^s_j(k)$ is defined in terms of $\boldmath{X}$, the set of covariates that satisfy identifiability assumptions.  
	If identifiability assumptions are not satisfied for the set of covariates $\boldmath{X}$ used in the analysis, then the proposed standardized VE estimator may exhibit bias. 	
	The proposed standardization procedure ensures that changes in VE estimates across calendar time cannot be attributed to differences in observed covariate distributions across the trial-eligible populations. 
	If additional unmeasured covariates exist that are strong predictors of the outcome conditional on the observed covariates, then heterogeneity in VE estimates across trials may be due to differences in the distribution of these unmeasured covariates. 
	
	To assess the extent to which exclusion of a covariate may influence the results, the full analysis was repeated eight times, each time with one covariate excluded from the analysis (i.e., age, sex, hypertension, diabetes, cardiovascular disease, chronic obstructive pulmonary disease, kidney disease, or cancer).  
	Specifically, for each analysis, the covariate vector $\boldsymbol{X_{-c}}$ was formed by removing the $c$th covariate.  
	The VU model, loss to follow up model, and covariate-conditional outcome hazard MSM were specified as described in Sections~\ref{appliedCensModelSpec} and~\ref{appliedOcModelSpec} using only the variables in $\boldsymbol{X_{-c}}$, and the estimator $\widehat{VE}^s_j(k)$ in~(\ref{eEstimator}) was computed conditional on $\boldsymbol{X_{-c}}$ rather than $\boldsymbol{X}$.  
	Overall, the results were robust to the exclusion of each covariate (Tables~\ref{sens_lou1}-\ref{sens_lou8}).

\begin{table}
	\centering \renewcommand{\arraystretch}{1}
	\caption{Sensitivity analysis results for selected emulated trials and weeks since first dose.
		The covariate age was omitted from the analysis.} 
	\label{sens_lou1}
	\begin{tabular}{lllll}
		\hline
		Weeks since & \multicolumn{1}{c}{$\widehat{VE}_0(k )$} & \multicolumn{1}{c}{$\widehat{VE}_3(k )$} & \multicolumn{1}{c}{$\widehat{VE}_6(k )$} & \multicolumn{1}{c}{$\widehat{VE}_9(k )$} \\  first dose ($k$) & \multicolumn{4}{c}{\% (95\% CI)}  \\ \hline
		1 & 87 (74, 93) & 83 (73, 89) & 77 (59, 87) & 69 (25, 87) \\ 
		7 & 88 (78, 94) & 84 (77, 90) & 79 (65, 88) & 74 (40, 89) \\ 
		14 & 88 (79, 94) & 85 (78, 90) & 81 (68, 89) & 77 (52, 89) \\ 
		21 & 88 (79, 93) & 84 (77, 89) & 81 (69, 88) & 77 (55, 89) \\ 
		28 & 86 (78, 92) & 82 (75, 87) & 76 (65, 84) & 65 (41, 80) \\ 
		34 & 85 (77, 91) & 79 (71, 84) & 64 (50, 74) & 31 (-10, 57) \\ 
		\hline
	\end{tabular}
\end{table}
\begin{table}
	\centering \renewcommand{\arraystretch}{1}
	\caption{Sensitivity analysis results for selected emulated trials and weeks since first dose.
		The covariate sex was omitted from the analysis.} 
	\label{sens_lou2}
	\begin{tabular}{lllll}
		\hline
		Weeks since & \multicolumn{1}{c}{$\widehat{VE}_0(k )$} & \multicolumn{1}{c}{$\widehat{VE}_3(k )$} & \multicolumn{1}{c}{$\widehat{VE}_6(k )$} & \multicolumn{1}{c}{$\widehat{VE}_9(k )$} \\  first dose ($k$) & \multicolumn{4}{c}{\% (95\% CI)}  \\ \hline
		1 & 86 (72, 93) & 80 (68, 87) & 71 (48, 84) & 59 (-3, 84) \\ 
		7 & 87 (76, 93) & 82 (73, 88) & 75 (55, 86) & 67 (20, 86) \\ 
		14 & 87 (77, 93) & 82 (73, 88) & 77 (60, 86) & 72 (37, 88) \\ 
		21 & 86 (75, 92) & 81 (73, 87) & 77 (62, 86) & 73 (44, 87) \\ 
		28 & 84 (73, 90) & 79 (70, 85) & 72 (57, 81) & 61 (30, 78) \\ 
		34 & 82 (71, 89) & 74 (64, 81) & 59 (41, 71) & 26 (-20, 54) \\ 
		\hline
	\end{tabular}
\end{table}
\begin{table}
	\centering  \renewcommand{\arraystretch}{1}
	\caption{Sensitivity analysis results for selected emulated trials and weeks since first dose.
		The covariate cancer was omitted from the analysis.} 
	\label{sens_lou3}
	\begin{tabular}{lllll}
		\hline
		Weeks since & \multicolumn{1}{c}{$\widehat{VE}_0(k )$} & \multicolumn{1}{c}{$\widehat{VE}_3(k )$} & \multicolumn{1}{c}{$\widehat{VE}_6(k )$} & \multicolumn{1}{c}{$\widehat{VE}_9(k )$} \\  first dose ($k$) & \multicolumn{4}{c}{\% (95\% CI)}  \\ \hline
		1 & 86 (72, 93) & 80 (68, 87) & 71 (48, 84) & 60 (-2, 84) \\ 
		7 & 87 (76, 93) & 82 (73, 88) & 75 (55, 86) & 67 (21, 86) \\ 
		14 & 87 (76, 93) & 82 (73, 88) & 77 (60, 86) & 72 (37, 88) \\ 
		21 & 86 (75, 92) & 81 (72, 87) & 77 (62, 86) & 73 (44, 87) \\ 
		28 & 84 (73, 90) & 79 (70, 85) & 72 (57, 81) & 60 (30, 78) \\ 
		34 & 82 (71, 89) & 74 (64, 81) & 59 (41, 71) & 25 (-21, 54) \\ 
		\hline
	\end{tabular}
\end{table}
\begin{table}
	\centering \renewcommand{\arraystretch}{1}
	\caption{Sensitivity analysis results for selected emulated trials and weeks since first dose.
		The covariate chronic obstructive pulmonary disease was omitted from the analysis.} 
	\label{sens_lou4}
	\begin{tabular}{lllll}
		\hline
		Weeks since & \multicolumn{1}{c}{$\widehat{VE}_0(k )$} & \multicolumn{1}{c}{$\widehat{VE}_3(k )$} & \multicolumn{1}{c}{$\widehat{VE}_6(k )$} & \multicolumn{1}{c}{$\widehat{VE}_9(k )$} \\  first dose ($k$) & \multicolumn{4}{c}{\% (95\% CI)}  \\ \hline
		1 & 86 (72, 93) & 80 (68, 87) & 71 (48, 84) & 60 (-3, 85) \\ 
		7 & 87 (76, 93) & 82 (73, 88) & 75 (56, 86) & 68 (21, 87) \\ 
		14 & 87 (76, 93) & 82 (74, 88) & 77 (60, 87) & 73 (38, 88) \\ 
		21 & 86 (75, 92) & 81 (73, 87) & 77 (62, 86) & 74 (45, 88) \\ 
		28 & 84 (73, 90) & 79 (70, 85) & 72 (57, 82) & 61 (31, 78) \\ 
		34 & 82 (71, 89) & 75 (65, 82) & 60 (42, 72) & 26 (-19, 55) \\ 
		\hline
	\end{tabular}
\end{table}
\begin{table}
	\centering  \renewcommand{\arraystretch}{1}
	\caption{Sensitivity analysis results for selected emulated trials and weeks since first dose.
		The covariate cardiovascular disease was omitted from the analysis.} 
	\label{sens_lou5}
	\begin{tabular}{lllll}
		\hline
		Weeks since & \multicolumn{1}{c}{$\widehat{VE}_0(k )$} & \multicolumn{1}{c}{$\widehat{VE}_3(k )$} & \multicolumn{1}{c}{$\widehat{VE}_6(k )$} & \multicolumn{1}{c}{$\widehat{VE}_9(k )$} \\  first dose ($k$) & \multicolumn{4}{c}{\% (95\% CI)}  \\ \hline
		1 & 87 (73, 93) & 81 (70, 88) & 72 (49, 85) & 60 (-4, 85) \\ 
		7 & 88 (77, 94) & 83 (74, 88) & 75 (56, 86) & 67 (20, 87) \\ 
		14 & 88 (78, 94) & 83 (75, 89) & 78 (61, 87) & 73 (37, 88) \\ 
		21 & 87 (77, 92) & 82 (74, 88) & 78 (63, 86) & 74 (45, 88) \\ 
		28 & 85 (75, 91) & 80 (71, 86) & 73 (59, 82) & 62 (33, 79) \\ 
		34 & 83 (73, 90) & 76 (67, 82) & 61 (44, 73) & 29 (-14, 56) \\ 
		\hline
	\end{tabular}
\end{table}
\begin{table}
	\centering  \renewcommand{\arraystretch}{1}
	\caption{Sensitivity analysis results for selected emulated trials and weeks since first dose.
		The covariate diabetes was omitted from the analysis.} 
	\label{sens_lou6}
	\begin{tabular}{lllll}
		\hline
		Weeks since & \multicolumn{1}{c}{$\widehat{VE}_0(k )$} & \multicolumn{1}{c}{$\widehat{VE}_3(k )$} & \multicolumn{1}{c}{$\widehat{VE}_6(k )$} & \multicolumn{1}{c}{$\widehat{VE}_9(k )$} \\  first dose ($k$) & \multicolumn{4}{c}{\% (95\% CI)}  \\ \hline
		1 & 86 (72, 93) & 80 (68, 87) & 71 (49, 84) & 60 (-1, 84) \\ 
		7 & 87 (76, 93) & 82 (73, 88) & 75 (56, 86) & 67 (22, 86) \\ 
		14 & 87 (77, 93) & 82 (74, 88) & 77 (60, 87) & 72 (38, 88) \\ 
		21 & 86 (75, 92) & 81 (73, 87) & 77 (62, 86) & 73 (45, 87) \\ 
		28 & 84 (73, 90) & 79 (70, 85) & 72 (57, 81) & 61 (30, 78) \\ 
		34 & 82 (72, 89) & 74 (65, 81) & 59 (41, 71) & 25 (-21, 54) \\ 
		\hline
	\end{tabular}
\end{table}
\begin{table}
	\centering  \renewcommand{\arraystretch}{1}
	\caption{Sensitivity analysis results for selected emulated trials and weeks since first dose.
		The covariate hypertension was omitted from the analysis.} 
	\label{sens_lou7}
	\begin{tabular}{lllll}
		\hline
		Weeks since & \multicolumn{1}{c}{$\widehat{VE}_0(k )$} & \multicolumn{1}{c}{$\widehat{VE}_3(k )$} & \multicolumn{1}{c}{$\widehat{VE}_6(k )$} & \multicolumn{1}{c}{$\widehat{VE}_9(k )$} \\  first dose ($k$) & \multicolumn{4}{c}{\% (95\% CI)}  \\ \hline
		1 & 86 (72, 93) & 80 (68, 87) & 71 (47, 84) & 59 (-7, 84) \\ 
		7 & 87 (76, 93) & 82 (73, 88) & 74 (55, 86) & 66 (17, 86) \\ 
		14 & 87 (77, 93) & 82 (74, 88) & 77 (59, 86) & 72 (35, 88) \\ 
		21 & 86 (75, 92) & 81 (73, 87) & 76 (61, 86) & 73 (42, 87) \\ 
		28 & 84 (73, 90) & 79 (70, 85) & 71 (56, 81) & 60 (28, 78) \\ 
		34 & 82 (72, 89) & 74 (65, 81) & 59 (41, 71) & 25 (-22, 54) \\ 
		\hline
	\end{tabular}
\end{table}
\begin{table}
	\centering  \renewcommand{\arraystretch}{1}
	\caption{Sensitivity analysis results for selected emulated trials and weeks since first dose.
		The covariate kidney disease was omitted from the analysis.} 
	\label{sens_lou8}
	\begin{tabular}{lllll}
		\hline
		Weeks since & \multicolumn{1}{c}{$\widehat{VE}_0(k )$} & \multicolumn{1}{c}{$\widehat{VE}_3(k )$} & \multicolumn{1}{c}{$\widehat{VE}_6(k )$} & \multicolumn{1}{c}{$\widehat{VE}_9(k )$} \\  first dose ($k$) & \multicolumn{4}{c}{\% (95\% CI)}  \\ \hline
		1 & 86 (73, 93) & 80 (69, 87) & 71 (47, 84) & 59 (-6, 84) \\ 
		7 & 88 (77, 93) & 82 (73, 88) & 75 (55, 86) & 66 (19, 86) \\ 
		14 & 88 (77, 93) & 83 (74, 88) & 77 (60, 87) & 72 (36, 88) \\ 
		21 & 86 (76, 92) & 82 (73, 87) & 77 (62, 86) & 73 (44, 87) \\ 
		28 & 84 (74, 91) & 79 (70, 85) & 72 (57, 81) & 59 (27, 77) \\ 
		34 & 83 (72, 89) & 75 (65, 82) & 59 (40, 71) & 23 (-26, 53) \\ 
		\hline
	\end{tabular}
\end{table}
\section{Unstandardized data analysis results}
\label{dataAnalysisRes_un}
\begin{figure}
	\centering
	\subfloat[][]{\includegraphics[width=.45\textwidth]{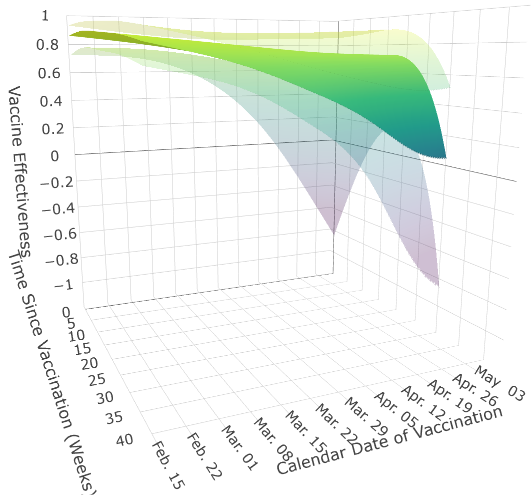}}\quad
	\subfloat[][]{\includegraphics[width=.52\textwidth]{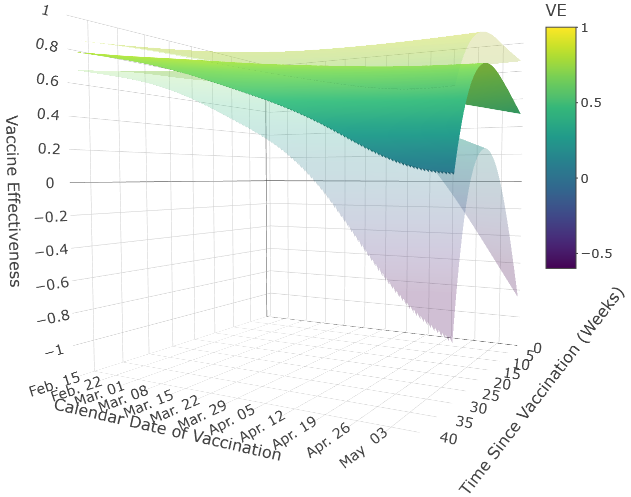}}
	\vspace{8pt}
	\caption{Unstandardized VE estimates against severe COVID-19 or COVID-19-related death among Abruzzo residents aged 80 years or older between February 15, 2021 and December 18, 2021.  
		The opaque surfaces depict the unstandardized VE point estimates and the transparent surfaces the corresponding 95\% Wald CIs.  
		Panels (a) and (b) display the same surface from two vantage points.}
	\label{contourPlots_supp}
\end{figure}

The Abruzzo data analysis was repeated without standardization using the method detailed in \ref{webA} and Section~\ref{methods} of the main text.  
The models for vaccine uptake and censoring due to loss to follow up were identical to those described in \ref{dataAnalysisDetails} above.  
The MSM for the outcome hazards was identical to (\ref{appliedOcModel}) above, except that baseline covariate terms were not included.  
VE point estimates and 95\% CIs are presented in Figure~\ref{contourPlots_supp} and Table~\ref{resTable_un}.  
The generalized Wald test of (\ref{Hnaught}) in the main text (with $K_J=33)$ yielded a one-sided $P$-value of $0.06$, 
providing some evidence against the null that $VE_j(k)$ is homogeneous across trials.
Overall, results were highly similar to those obtained using the standardization method.

\begin{table}
	\centering \renewcommand{\arraystretch}{1}
	\caption{Unstandardized VE estimates and 95\% pointwise confidence intervals (CIs) against severe COVID-19 or COVID-19-related death among Abruzzo residents aged 80 years or older ($n=110{,}623$) between February 15, 2021 and December 18, 2021.  
		Results are presented for selected emulated trials and weeks since first dose.
		The starting date of each trial was as follows: trial $j=0$, February 15, 2021; trial $j=3$, March 8, 2021; trial $j=6$, March 29, 2021; and trial $j=9$, April 19, 2021.} 
	\label{resTable_un}
	\begin{tabular}{lllll}
		\hline
		Weeks since & \multicolumn{1}{c}{$\widehat{VE}_0(k )$} & \multicolumn{1}{c}{$\widehat{VE}_3(k )$} & \multicolumn{1}{c}{$\widehat{VE}_6(k )$} & \multicolumn{1}{c}{$\widehat{VE}_9(k )$} \\  first dose ($k$) & \multicolumn{4}{c}{\% (95\% CI)}  \\ \hline
		1 & 86 (72, 93) & 80 (69, 87) & 71 (47, 84) & 59 (-7, 85) \\ 
		7 & 88 (76, 94) & 82 (73, 88) & 75 (56, 86) & 67 (18, 87) \\ 
		14 & 88 (77, 93) & 83 (74, 89) & 77 (61, 87) & 72 (36, 88) \\ 
		21 & 86 (76, 92) & 82 (74, 88) & 77 (62, 86) & 73 (43, 87) \\ 
		28 & 85 (74, 91) & 80 (71, 85) & 72 (57, 82) & 60 (27, 78) \\ 
		34 & 83 (73, 90) & 76 (66, 82) & 60 (42, 72) & 24 (-24, 53) \\ 
		\hline
	\end{tabular}
\end{table}

\section{Comparison with Cox model approach}
\label{CoxModel}
A common approach to estimating COVID-19 vaccine effectiveness entails fitting a Cox proportional hazards model and estimating VE by one minus the estimated hazard ratio associated with vaccination  \citep{fintzi2021assessing,tartof2021effectiveness,lin2022durability,Lin,lin2022evaluating,lin2022reliably,shrestha2024effectiveness}.  
	However, causal interpretation of hazard ratio-based VE parameters is difficult, in part because at later study time points the hazard ratio may not contrast the same population under two hypothetical treatment conditions \citep{MartinussenHazards}.  
	For example, if the treatment has a protective effect, then the active treatment group may be more frail at later time points because the frailest people will have been already removed from the comparator group due to experiencing events \citep{HernanAndRobins}.  
	\citet{HernanHazards} refers to such depletion of susceptibles as an example of the ``built-in selection bias'' of the hazard ratio.  
	Because the Cox model is commonly used to analyze observational COVID-19 vaccine data, a Cox-model-based analysis of the Abruzzo data is included below.  
	Results are compared with those obtained using the proposed NTE methods.  

\subsection{Method}
\label{CoxModelMethod}
	Let $t$ index (continuous) calendar time, measured in days from February 15, 2021 with $t=\tau$ corresponding to December 18, 2021.  
	Recall that $T$ is the (possibly right censored) event time, $U$ is the dropout time, $T^*=\text{min}(T, U)$, $\Delta=I(T<U)$, and $V_1$ is the time of first COVID-19 vaccine dose.  
	In this section, each of these variables is measured in days from February 15, 2021.  
	Following \citet{Lin}, individuals are followed until they experience the event or until time $\tau$, whichever occurs first.  
	In this section, the dropout variable $U$ corresponds to the time of earliest treatment noncompliance or the administrative censoring time $\tau+1$, whichever occurs first.   
	Treatment noncompliance can occur in the following ways: 
	(i) for vaccinated and unvaccinated individuals, receipt of any non-mRNA COVID-19 vaccine dose, 
	(ii) for individuals who received a first dose of mRNA COVID-19 vaccine, failure to receive the second dose within the recommended time window, and 
	(iii) for individuals who received two doses of mRNA COVID-19 vaccine, receipt of a third COVID-19 vaccine dose.  
	The date of (ii) was defined according to CDC guidelines for timing of the second dose of the original monovalent Pfizer-BioNTech and Moderna vaccines \citep{CDC}.
	
	Assume the hazard of $T$, conditional on $V_1$ and $\boldsymbol{X}$, takes the following form
	\begin{equation}
		\label{eq:CoxModel}
		\lambda(t \mid  V_1, \boldsymbol{X})= \lambda_0(t)\text{exp}[
		\boldsymbol{\zeta}\boldsymbol{f}_{\zeta}(\boldsymbol{X}) + 
		\{\boldsymbol{\eta_1}\boldsymbol{f_{\eta_1}}(t) +
		\boldsymbol{\eta_2}\boldsymbol{f_{\eta_2}}(t-V_1)\} I_t]
	\end{equation}
	where 
	$\lambda_0$ is an arbitrary baseline hazard function, 
	$(\boldsymbol{\zeta}, \boldsymbol{\eta_1}, \boldsymbol{\eta_2})$ are unknown parameters, 
	$\boldsymbol{f}_{\zeta}(\boldsymbol{X})$ is a column vector of user-specified functions of covariates, 
	$\boldsymbol{f_{\eta_1}}(\cdot)$ and $\boldsymbol{f_{\eta_2}}(\cdot)$ are column vectors of user-specified functions of (continuous) time,  
	and $I_t=I(V_1 < t)$ is a time-varying treatment indicator.  
	Under model~(\ref{eq:CoxModel}), the baseline hazard may vary over calendar time, and the vaccine effect may vary over both calendar time and time since vaccination.  
	
	The partial likelihood for $(\boldsymbol{\zeta}, \boldsymbol{\eta_1}, \boldsymbol{\eta_2})$ (with no tied event times) is 
	\begin{equation*}
		\label{partialLikelihood}
		PL(\boldsymbol{\zeta}, \boldsymbol{\eta_1},\boldsymbol{\eta_2})=\prod_{i=1}^{N}
		\bigg(
		\dfrac{\exp[\boldsymbol{\zeta}\boldsymbol{f}_{\zeta}(\boldsymbol{X_i}) + 
			\{\boldsymbol{\eta_1}\boldsymbol{f_{\eta_1}}(T_i) +
			\boldsymbol{\eta_2}\boldsymbol{f_{\eta_2}}(T_i-V_{1i})\}I_{ti}]}
		{\sum_{i^\prime=1}^N I(T^*_{i^\prime} \geq T^*_{i})
			\exp[\boldsymbol{\zeta}\boldsymbol{f}_{\zeta}(\boldsymbol{X_{i^\prime}}) + 
			\{\boldsymbol{\eta_1}\boldsymbol{f_{\eta_1}}(T_{i^\prime}) +
			\boldsymbol{\eta_2}\boldsymbol{f_{\eta_2}}(T_{i^\prime}-V_{1i^\prime})\}I_{ti^\prime}]}
		\bigg)^{\Delta_i},
	\end{equation*}
	and is maximized using the Newton-Raphson algorithm.
	Let $(\boldsymbol{\hat{\eta}_1}, \boldsymbol{\hat{\eta}_2})$ denote the maximum partial likelihood estimator of $(\boldsymbol{\eta_1},\boldsymbol{\eta_2})$.  
	Then a Cox-model-based estimator of VE $s$ days after vaccination for those vaccinated at calendar time $v$ is given by 
	$1-\exp\{\boldsymbol{\hat{\eta}_1}\boldsymbol{f_{\eta_1}}(v+s)+\boldsymbol{\hat{\eta}_2}\boldsymbol{f_{\eta_2}}(s)\}$.
	Endpoints of Wald-type, pointwise $(1-\iota)100\%$ CIs are given by $1-\exp\big\{
	(\boldsymbol{\hat{\eta}_1}, \boldsymbol{\hat{\eta}_2})\boldsymbol{f_{v,s}}
	\pm \Phi_{1-\iota/2}\sqrt{\boldsymbol{f_{v,s}} \hat{V}(\boldsymbol{\hat{\eta}_1},\boldsymbol{\hat{\eta}_2})\boldsymbol{f^T_{v,s}} }\big\}$, where $\boldsymbol{f_{v,s}}=(\boldsymbol{f_{\eta_1}}(s+v)^T,\boldsymbol{f_{\eta_2}}(s)^T)^T$.
	
	For the analysis presented below, model (\ref{eq:CoxModel}) was specified with $\boldsymbol{f}_{\zeta}(\boldsymbol{x})=(\boldsymbol{spl}(x_1-\tilde{x}_1), x_2, ..., x_8)^T$, and  	
	$\boldsymbol{f_{\eta_1}}$ and $\boldsymbol{f_{\eta_2}}$ as RCS functions with knots at the 10th, 50th, and 90th percentiles of the observed event times and observed event times among the vaccinated, respectively.  
	Ties were handled using the Efron approximation.  

\subsection{Analysis}
\label{CoxModelAnalysis}
	There were a total of 310 observed events during unvaccinated person-time and 66 observed events during vaccinated person-time.  
	The 66 vaccinated events were approximately evenly spread out across the study period.  
	On the other hand, approximately 90\% of unvaccinated events occurred before June 1, 2021.  
	
\begin{figure}
	\centering
	\subfloat[][]{\includegraphics[width=.45\textwidth]{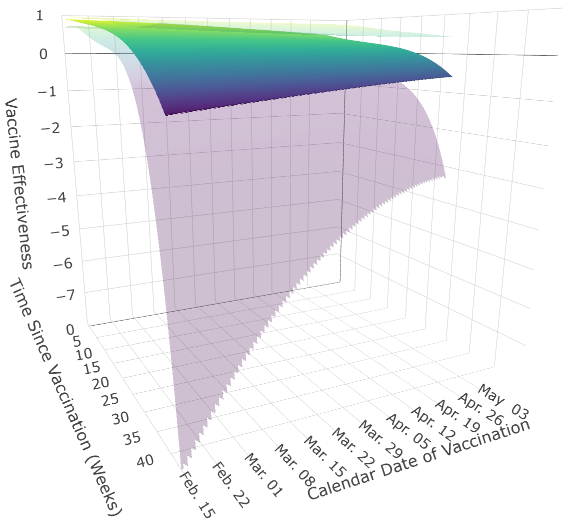}}\quad
	\subfloat[][]{\includegraphics[width=.52\textwidth]{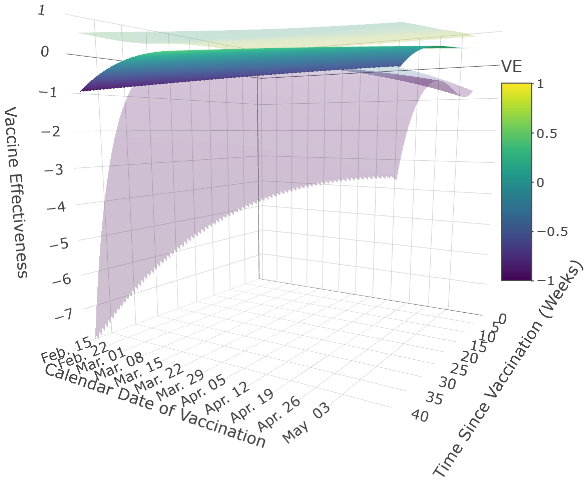}}
	\vspace{8pt}
	\caption{Cox model estimates of VE against severe COVID-19 or COVID-19-related death among Abruzzo residents aged 80 years or older between February 15, 2021 and December 18, 2021.  
		The opaque surfaces depict VE estimates and the transparent surfaces the corresponding 95\% Wald CIs.  
		Panels (a) and (b) display the same surface from two vantage points.}
	\label{contourPlots_Cox}
\end{figure}

\begin{figure}
	\centering
	\subfloat[][]{\includegraphics[width=.48\textwidth]{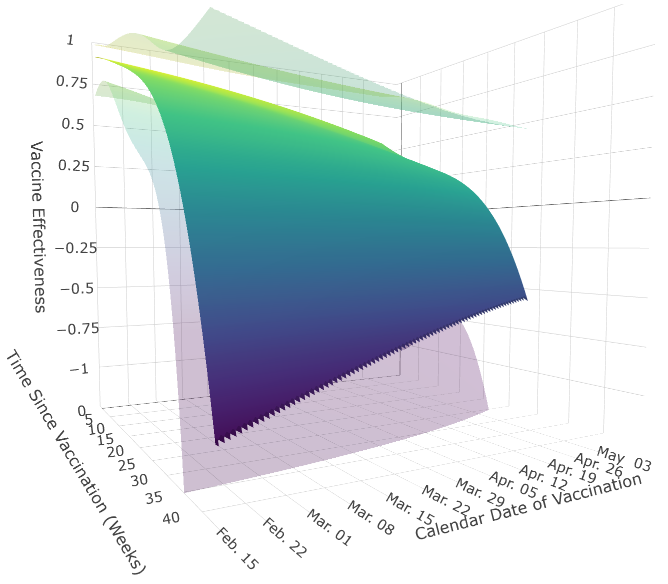}}\quad
	\subfloat[][]{\includegraphics[width=.49\textwidth]{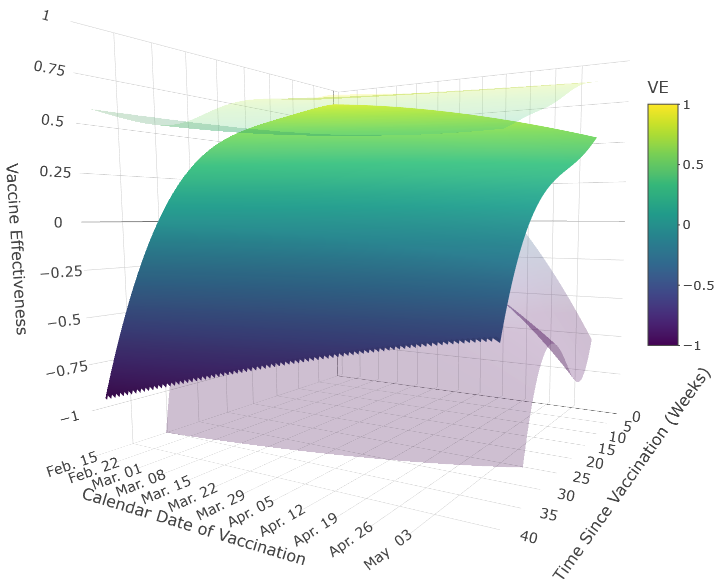}}
	\vspace{8pt}
	\caption{Detail of Figure~\ref{contourPlots_Cox}, with magnified VE-axis region $[-1,1]$.
		Panels (a) and (b) display the same surface from two vantage points.}
	\label{contourPlots_Cox_point}
\end{figure}

	Figure~\ref{contourPlots_Cox} displays the Cox model estimated VE surface and corresponding pointwise 95\% CIs.  
	The extremely wide confidence intervals at later calendar times may plausibly be explained by sparsity of observed events between June and December 2021.  
	To ease comparison with Figure~\ref{contourPlots} in the main text, a detail of Figure~\ref{contourPlots_Cox}, with magnified VE-axis region $[-1,1]$ is shown in Figure~\ref{contourPlots_Cox_point}.  
	
	Cox model estimates of VE one week after vaccination were fairly similar to those obtained using the NTE approach. 
	However, the Cox model estimates declined noticeably faster over time since vaccination in each of these trials.  
	This result is consistent with the ``built-in selection bias'' mentioned above.  
	As in the NTE analysis, (i) for most fixed times since vaccination, VE point estimates decreased with calendar time, and (ii) for each fixed time since vaccination, CI widths tended to increase with calendar time.  
	In the Cox model analysis, CIs were substantially wider than those of the NTE analysis at later times since vaccination.  

\section{Illustrative pseudocode}
\label{pseudocode}	
	The complete estimation procedure for the standardized analysis is summarized in Algorithm~\ref{algo}.
	Recall that the analytic cohort comprises the set of individuals in the observational database who meet eligibility criteria for trial 0 (i.e., on February 15, 2021).  
	The number of trials to be emulated is chosen by the analyst.  
	Step~\ref{censoringModels} can be completed using standard software (e.g., \texttt{glm} in \texttt{R}) and the analytic cohort dataset (in long format).  
	Construction of the NTE analysis dataset is summarized by steps~\ref{startDerive}-\ref{endDerive} and illustrated via a worked example in Section~\ref{NTEdsConstruction}.  
	Step~\ref{ocModel} can be accomplished using standard software (e.g., \texttt{glm} in \texttt{R}) and the NTE analysis dataset (in long format).  
	Steps~\ref{startInd}-\ref{endInd} entail computing predicted conditional counterfactual hazards and risks under $a=0,1$ (based on estimates of $\boldsymbol{\gamma}$ obtained in step~\ref{ocModel}) for each individual in the analytic cohort at each time point in the NTE analysis.  
	Step~\ref{penny} can be completed using standard software (e.g., \texttt{lm} in \texttt{R}).
	Finally, step~\ref{geexDeli} can be accomplished using M-estimation software, e.g., the R package \texttt{geex} \citep{Saul} or the Python library \texttt{delicatessen} \citep{Zivich}.

\newpage
\begin{algorithm}
	\caption{Estimation procedure for standardized analysis}
		\label{algo}
	{\fontsize{12}{11}\selectfont
		\textbf{Input:} Observed data $(T^*_i, \Delta_i, \boldsymbol{X_i}, \overline{B}_{\tau i})_{i=1}^n$ for the analytic cohort; 
		choice of $J$, the number of trials to be emulated.
		\begin{algorithmic}[1]
			\State \label{censoringModels} Obtain solutions $\boldsymbol{\hat\xi}$ and $\boldsymbol{\hat\kappa}$ to the vector estimating equations $\sum_{i=1}^n\boldsymbol{\psi_{\boldsymbol{\xi}}}(\boldsymbol{O_i}; \boldsymbol{\xi})=\boldsymbol{0}$ and $\sum_{i=1}^n\boldsymbol{\psi_{\boldsymbol{\kappa}}}(\boldsymbol{O_i}; \boldsymbol{\kappa})=\boldsymbol{0}$.  
			\For{$i \in \{1,...,n\}$} \label{startDerive}
			\For{$(j,k) \in \mathcal{W}$}		
			\State Derive eligibility indicators $E_{ji}$ and trial-specific variables $A_{ji}, C_{ji}(k), R_{ji}(k)$ according to specifications in the target trial protocol.
			\State Construct estimated IP weights $W_{ji}(k, A_{ji}; \boldsymbol{\hat\kappa}, \boldsymbol{\hat\xi})$. 
			\EndFor
			\EndFor  \label{endDerive}
			\State \label{ocModel} Obtain the solution $\boldsymbol{\hat\gamma}$ to the vector estimating equation
			\Statex $\sum_{i=1}^{n} \sum_{j=0}^{J}  \sum_{k=1}^{K_j}  
			W_{ji}(k, A_{ji}; \boldsymbol{\hat\kappa}, \boldsymbol{\hat\xi})
			\{Y_{ji}(k)- \lambda_j^{A_{ji}}(k \mid \boldsymbol{X_i} ;\boldsymbol{\gamma})\}
			\boldsymbol{f}_{\boldsymbol{\gamma}}(j, k,A_{ji}, \boldsymbol{X_i})=\boldsymbol{0}$. 
			\For{$(j,k) \in \mathcal{W}$}		 \label{startInd}
			\For{$i \in \{1,...,n\}$}        \label{startInd_sorta}
			\For{$a \in \{0,1\}$}
			\State Compute $\hat{E}(Y_{j}^a(k)|E_{j}=1,\boldsymbol{X}=\boldsymbol{x_i})$ based on (\ref{condRisk}). \label{lineRef}
			\EndFor
			\EndFor	\label{endInd_sorta}
			\State Using estimates obtained in lines~\ref{startInd_sorta}-\ref{endInd_sorta}, calculate $\widehat{VE}^s_j(k)$ via (\ref{eEstimator}).
			\EndFor	 \label{endInd}
			\State Compute $\hat{\beta}$ according to (\ref{leastSquares}) with $\widehat{VE}_j(k)$ replaced by $\widehat{VE}^s_j(k)$. \label{penny}
			\State Obtain sandwich variance estimates and then construct the test statistic $U_{\beta}$ and pointwise $(1-\upsilon)100\%$ CIs for ${VE}^s_j(k)$ for $(j,k) \in \mathcal{W}$, as described in Sections~\ref{TEH} and~\ref{inference} of the main text, respectively. \label{geexDeli}
		\end{algorithmic}
		\textbf{Output:} Estimates of ${VE}^s_j(k)$ for $(j,k) \in \mathcal{W}$ with corresponding pointwise $(1-\upsilon)100\%$ CIs; test statistic $U_{\beta}$ and corresponding $P-$value.}
\end{algorithm}